\documentclass{jfm}
\usepackage{graphicx}
\usepackage{epstopdf,epsfig}
\usepackage{amssymb}
\usepackage{amsmath}
\usepackage{subfig}
\usepackage{color,soul}
\usepackage{natbib}

 \def\vector#1{\mbox{\boldmath $#1$}}
 
 \def\tensor#1{\mathcal #1}

\begin{document}

\newtheorem{lemma}{Lemma}
\newtheorem{corollary}{Corollary}

\shorttitle{Bubble cloud dynamics} 
\shortauthor{K. Maeda and T. Colonius} 

\title{Bubble cloud dynamics in an ultrasound field}

\author
 {
 Kazuki Maeda
  \corresp{\email{maeda@caltech.edu}},
  Tim Colonius
  }

\affiliation
{
Division of Engineering and Applied Science, California Institute of Technology, Pasadena, CA 91125, USA
}
\maketitle

\begin{abstract}
The dynamics of bubble clouds induced by high-intensity focused ultrasound are investigated in a regime where the cloud size is similar to the ultrasound wavelength.
High-speed images show that the cloud is asymmetrical; the bubbles nearest the source grow to a larger radius than the distal ones.
Similar structures of bubble clouds are observed in numerical simulations that mimic the laboratory experiment.
To elucidate the structure, a parametric study is conducted for plane ultrasound waves with various amplitudes and diffuse clouds with different initial void fractions.
Based on an analysis of the kinetic energy of liquid induced by bubble oscillations, a new scaling parameter is introduced to characterize the dynamics. The new parameter generalizes the cloud interaction parameter originally introduced by \citet{dAgostino89}.
The dynamic interaction parameter controls the energy localization and consequent anisoptropy of the cloud.
Moreover, the amplitude of the far-field, bubble-scattered acoustics is likewise correlated with the proposed parameter. Findings of the present study not only shed light on the physics of cloud cavitation, but may also be of use to quantification of the effects of cavitation on outcomes of ultrasound therapies including HIFU-based lithotripsy.
\end{abstract}

\section{Introduction}
\label{section:intro}
The dynamics of cavitation bubble clouds excited in an intense ultrasound field are of critical importance for the safety and efficacy of lithotripsy and high-intensity focused ultrasound (HIFU).
In such therapy, cavitation bubbles can be formed in the human body during the passage of the tensile part of ultrasound pulses. Bubbles can scatter and absorb subsequent pulses, and the violent collapse of bubbles can cause cavitation damage \citep{Coleman87,Pishchalnikov03,Matsumoto05b,McAteer05,Ikeda06,Bailey06,Stride10,Miller12}.
Due to the short time scale and three-dimensional nature of cloud cavitation, precise measurement of individual bubbles has been challenging.
Numerical simulations using mixture-averaging approaches \citep{Van68,Biesheuvel84} have remained central tools for quantification of the dynamics of bubble clouds.

Early studies of bubble cloud dynamics focused on assessment of cavitation noise and erosion on materials.
\citet{Morch80,Morch82} theoretically modeled the inward-propagating collapse of spherical bubble clusters and quantified the resulting collapse pressure.
\citet{Omta87} studied acoustic emission from the spherical bubble cloud excited by step change of the pressure in the surrounding liquid.
\citet{dAgostino89} formulated the linearized dynamics of monodisperse, spherical bubble clouds under weak, long wavelength pressure excitation and identified that the cloud interaction parameter, $B=\beta R_c^2/R_{b0}^2$, dictates the linear dynamics of the cloud, where $\beta$ is the void fraction, $R_c$ and $R_{b0}$ are the initial radius of the cloud and the bubbles, respectively. \citet{Wang94,Wang99} extended the study to the nonlinear regime, further characterizing the strong collapse of bubble clouds accompanied by a shockwave.
\citet{Shimada00} used a similar approach to assess the effect of the polydispersity of nuclei on the nonlinear dynamics of spherical bubble clouds.

Later, numerical studies of cavitation have gained interest for medical applications. \citet{Tanguay03} extended the mixture-averaging approach to simulate and characterize the dynamics of cavitation bubble clouds induced in extra-corporeal shock wave lithotripsy (ESWL).
\citet{Matsumoto05} extended the method of \citet{Shimada00} to quantify amplifications in the pressure due to bubble cloud collapse under excitation by resonant HIFU waves and discussed applications of the collapse energy to kidney stone comminution as an alternative method of ESWL.

In experiments, \citet{Reisman98} used high-speed imaging to observe cloud cavitation collapse on a finite-span hydrofoil, and analyzed acoustic signals from the cloud collapse, and associated the results with the inward propagating shockwave predicted in the aforementioned studies.
\citet{Arora07} observed collapse in bubble clouds of controlled nuclei concentration, and observed an inward-propagating collapse with high nuclei concentrations. \citet{Lu13} studied the spatial distribution and the translational motions of bubble clouds in standing acoustic waves in a time scale longer than the collapse.

The aforementioned theoretical studies focus on bubble clouds in an otherwise incompressible liquid so that the wavelength of the pressure excitation is much larger than the size of the cloud.
In practical conditions of ultrasound therapies, however, the scale separation invoked above does not hold.
In fact, \citet{Maeda15} observed bubble clouds with a size of $O(1)$ mm {\it{in vitro}} during the passage of a strong ultrasound wave with a wavelength as short as the cloud size.

\begin{figure}
  \center
  \includegraphics[width=130mm,trim=0 0 0 0, clip]{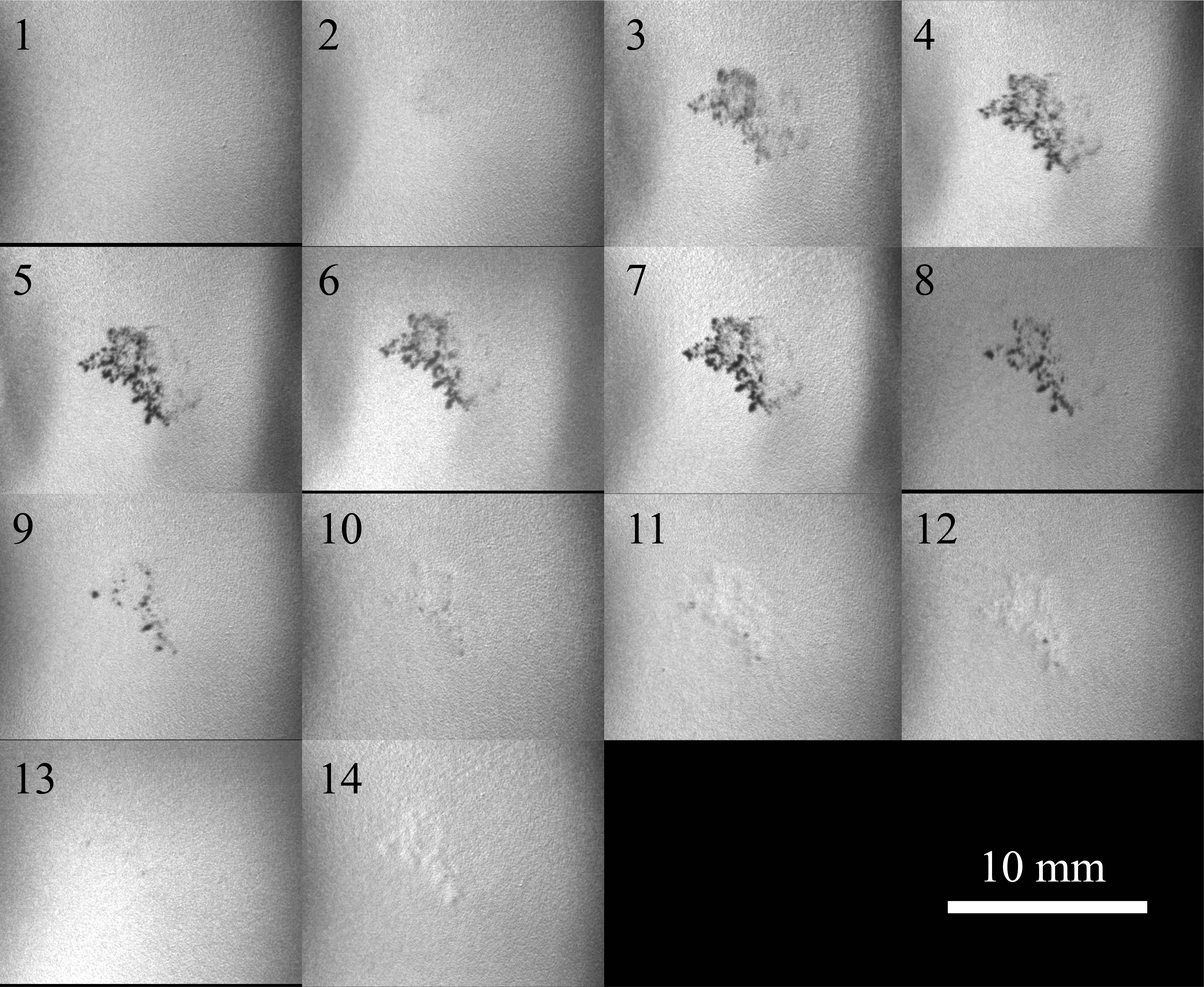}
  \caption{High-speed images showing evolution of a representative bubble cloud excited by a focused ultrasound wave. The gray shadows with a shape of the bubble cloud present in the 11th through 14th frames are an artifact of the imaging system.}
   \label{fig:imacon} 
\end{figure}
As a representative example of such bubble clouds, in figure \ref{fig:imacon} we show the evolution of an isolated bubble cloud nucleated in a pulse of focused ultrasound consisting of 10 wavelengths with a carrier frequency of $f=335$ kHz, thus a wavelength of $\lambda=4.4$ mm, generated by a medical transducer in water.
The experiment, extended from a setup documented in \citet{Maeda15}, is designed to characterize the dynamics of bubble clouds in a recently proposed HIFU-based lithotripsy, {\itshape{burst wave lithotripsy}} \citep{Maxwell15}. Details of the experimental setup are described in appendix A.
The cloud is growing up through the 10th frame, as the ultrasound wave is propagating through the focal region, and then decays in the subsequent frames.
The bubble cloud occupies an approximately sphereical volume with $R_c\approx 2.5$ mm.
The size of the bubble cloud is thus at the same scale as the wavelength of the incident wave.
Notably, the cloud possesses an anisotropic structure in that the proximal bubbles grow to a larger radius than the distal bubbles.
These dynamics are significantly different from bubble clouds in the long wavelength regime.

Advanced interface capturing methods are capable of simulating detailed dynamics of each bubble in a cloud in a compressible liquid at fine spatial scales, and have been applied to bubble cloud collapse in a free field by \citet{Rossinelli13,Rasthofer17} and near a wall by \citet{Tiwari15}.
Yet, such methods are still computationally intensive and applications are limited to the dynamics within a short time scale, typically that of a single cycle of bubble collapse. For more complex problems, modeling assumptions have to be made to reduce the computational cost. In this spirit, we have recently developed a method that enables simulation of cloud cavitation in an intense ultrasound field with a fine spatial resolution without a constraint of the scale separation \citep{Maeda17b}. The method solves mixture-averaged equations using an Eulerian-Lagrangian approach \citep{Kameda96,Fuster11,Ma18}. The bubbly-mixture is discretized on an Eulerian grid, while the individual bubbles are tracked on Lagrangian points and their radial dynamics are solved using the Keller-Miksis equation. By doing so, the dynamics of each bubble and the bubble-scattered pressure wave are accurately captured with a reasonable computational expense.

\begin{figure}
  \center
  \subfloat[$t^*=0$]{\includegraphics[width=32mm,trim=32 0 32 0, clip]{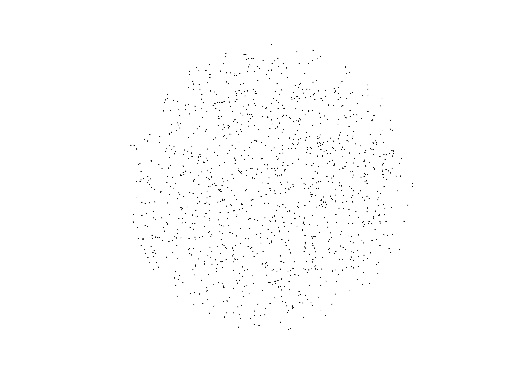}}
  \subfloat[$t^*=3.8$]{\includegraphics[width=32mm,trim=32 0 32 0, clip]{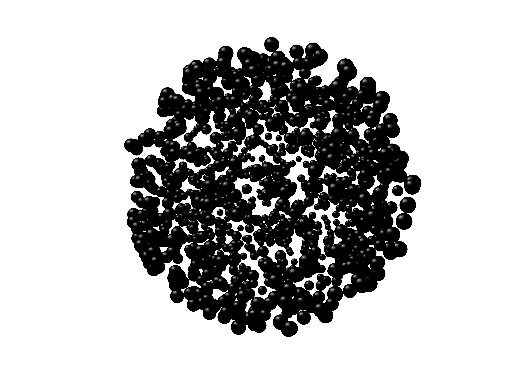}}
  \subfloat[$t^*=6.9$]{\includegraphics[width=32mm,trim=32 0 32 0, clip]{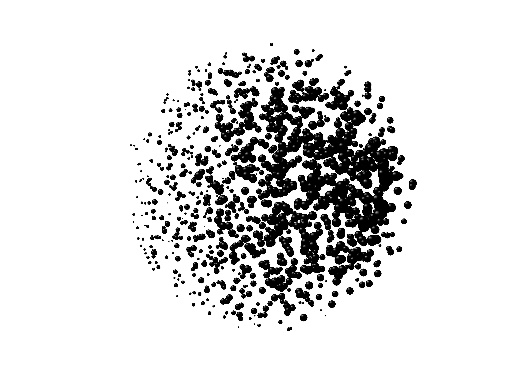}}
  \subfloat[$t^*=7.3$]{\includegraphics[width=32mm,trim=32 0 32 0, clip]{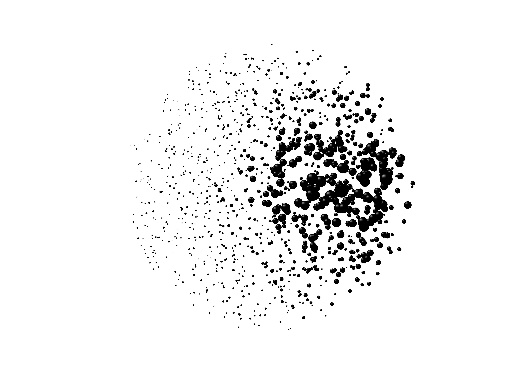}}
  \caption{Evolution of a bubble cloud excited by a single cycle of sinusoidal wave with a frequency of $f=50$ kHz. $t^*$ denotes non-dimensional time $t^*=tf$.}
   \label{fig:snap_f50} 
\end{figure}
Our simulation successfully reproduces the aforementioned inward-propagating spherical cloud collapse in the long wavelength regime.
Figure \ref{fig:snap_f50} shows the evolution of bubbles with an initial radius of 10 $\mu$m that are randomly distributed within a sphere of a radius 2.5 mm and excited by a single cycle of a planer sinusoidal wave with a frequency of 50 kHz, thus a wavelength of 30 mm, and an amplitude of 1 MPa.
The wavelength is much longer than the cloud size and this effectively models the scale separation.
\begin{figure}
  \center
  \includegraphics[width=70mm]{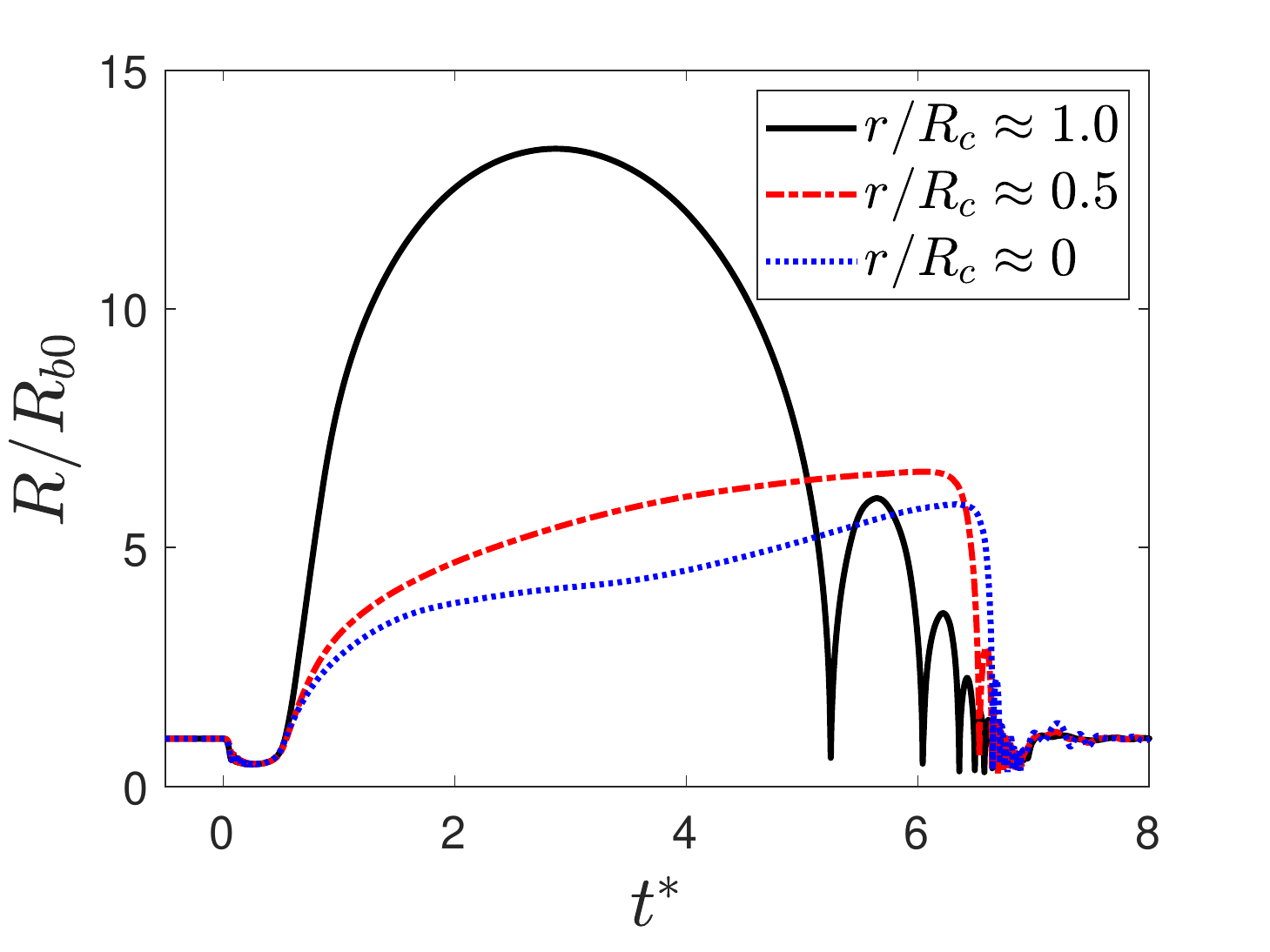}
  \caption{Evolution of the normalized radii of three bubbles at distinct locations in the cloud. $r$ denotes the distance from the cloud center.}
   \label{fig:Brennenb}
\end{figure}
Figure \ref{fig:Brennenb} shows the evolution of the radii of representative bubbles that are initially located at three distinct radial coordinates of the cloud: cloud center; mid-point between the center and the cloud periphery; periphery.
The peripheral bubble grows to a larger maximum radius and collapse faster than the other bubbles, while the inner bubbles are subsequently collapsed during the arrival of the inward propagating bubbly shockwave.
The result qualitatively reproduces the numerical simulation of \citet{Wang94,Wang99} as well as the experimental observation of \citet{Arora07}.

Motivated by the high-speed imaging and the preliminary simulation, in this paper we aim to use numerical simulations to provide a first insight into the bubble clouds excited by HIFU where the cloud radius is commensurate with the wavelength.

The paper is organized as follows.
$\S$ \ref{section:model} provides a summary of the modeling and numerical methods.
In $\S$ \ref{section:metrics} we introduce metrics to quantify the dynamics of bubble clouds, including the cloud interaction parameter introduced by \citet{dAgostino89}, and moments of the volume and the kinetic energy.
In $\S$ \ref{section:focused} we simulate the dynamics of bubble clouds excited by a focused ultrasound wave with various polydispersities and populations of nuclei in a setup that mimics the experimental condition. We quantitatively compare results with the experimental high-speed images shown in figure \ref{fig:imacon} and evaluate the anisotropic structure.
To further elucidate the dynamics in more generalized conditions, in $\S$ \ref{section:parametric} we conduct a parametric study of bubble clouds excited by a plane ultrasound wave, varying the nuclei populations and the amplitudes of the wave.
In $\S$ \ref{subsec:anis} we quantitatively analyze the anisotropic structure, and in $\S$ \ref{section:dynamicB} we propose a new scaling parameter to characterize the dynamics of the clouds by generalizing the cloud interaction parameter of \cite{dAgostino89}.
In $\S$ \ref{section:moment} we collapse the moments of bubble-induced kinetic energy in terms of the proposed parameter and identify the mechanisms by which energy is localized in the proximal side of the cloud.
In $\S$ \ref{section:acoustics} the amplitude and directionality of the scattered acoustic field are evaluated and collapsed by the proposed parameter. The energy localization and the scattered acoustics are directly correlated.
In $\S$ \ref{section:lith} we discuss implications of the numerical results to the effects of cloud cavitation on outcomes of HIFU-based lithotripsy.
In $\S$ \ref{section:conc} we state conclusions.

\section{Model formulation}
\label{section:model}
\subsection{Bubble cloud dynamics}
\label{section:gov_eqn}
Here we briefly summarize the physical model and numerical method for simulation of cloud cavitation employed in the present study. Further details are available in \citet{Maeda17}.
We describe the dynamics of bubbly-mixture using volume-averaged equations of motion \citep{Caflisch85,Commander89}:
\begin{align}
\frac{\partial \overline\rho}{\partial
t}+\nabla\cdot(\overline{\rho\vector{u}})
&=0,\label{eqn:vaeqn_ma}\\
\frac{\partial (\overline{\rho\vector{u}})}{\partial t} + \nabla\cdot
(\overline{\rho\vector{u}}\otimes\overline{\vector{u}}+p\tensor{I}
-\tensor{T})
&=0,\label{eqn:vaeqn_mo}\\
\frac{\partial \overline{E}}{\partial t} + \nabla\cdot
\left((\overline{E}+p)\overline{\vector{u}}
-\tensor{T}\cdot\overline{\vector{u}}\right)
&=0\label{eqn:vaeqn_e},
\end{align}
where $\rho$ is the density, $\vector{u}=(u,v,w)^{\mathrm{T}}$ is the velocity,
$p$ is the pressure and $E$ is the total energy, respectively.
$\overline{(\cdot)}$ denotes the volume averaging operator:
$\overline{(\cdot)}=(1-\beta)(\cdot)_l+\beta(\cdot)_g$,
where $\beta\in[0,1)$ is the volume fraction of gas (void fraction), and
subscripts $l$ and $g$ denote the liquid and gas phase, respectively.
$\tensor{T}$ is the effective viscous stress tensor of the mixture, that we approximate as that of the liquid phase: $\tensor{T}\approx\tensor{T}_l$.
We invoke two approximations valid at the limit of low void fraction: the density of the mixture is approximated by that of the liquid:
$\overline\rho\approx(1-\beta)\rho_l$; the slip velocity between the two phases is zero:
$\overline{\vector{u}}\approx\vector{u}_l=\vector{u}_g$.

Equations
(\ref{eqn:vaeqn_ma}-\ref{eqn:vaeqn_e}) can be rewritten
as conservation equations in terms of the mass, momentum and energy of the liquid
with source terms, as an inhomogeneous hyperbolic system:
\begin{align}
\frac{\partial \rho_l}{\partial t} + \nabla\cdot (\rho_l
\vector{u}_l)
&=\frac{\rho_l}{1-\beta}\left[\frac{\partial
\beta}{\partial t}+\vector{u}_l\cdot\nabla\beta\right],\label{eqn:vaeqn2_ma}\\
\frac{\partial (\rho_l\vector{u}_l)}{\partial t} +
\nabla\cdot (\rho_l\vector{u}_l\otimes \vector{u}_l+p\tensor{I}-\tensor{T}_l)
&=\frac{\rho_l\vector{u}}{1-\beta}\left[\frac{\partial
\beta}{\partial
t}+\vector{u}_l\cdot\nabla\beta\right]
-\frac{\beta\nabla\cdot(p\tensor{I}-\tensor{T}_l)}{1-\beta},\label{eqn:vaeqn2_mo}\\
\frac{\partial E_l}{\partial t} + \nabla\cdot
\left((E_l+p)\vector{u}_l-\tensor{T}_l\cdot\vector{u}_l\right)
&=\frac{E_l}{1-\beta}\left[\frac{\partial
\beta}{\partial
t}+\vector{u}_l\cdot\nabla\beta\right]
-\frac{\beta\nabla\cdot(p\vector{u}_l-\tensor{T}_l\cdot\vector{u}_l)}{1-\beta}.
\label{eqn:vaeqn2_e}
\end{align}
For a thermodynamic closure for the liquid, we employ stiffened gas equation of state:
\begin{equation}
p=(\gamma-1)\rho\varepsilon-\gamma\pi_\infty,
\label{eqn:stiff_gas}
\end{equation}
where $\varepsilon$ is the internal energy of liquid, $\gamma$ is the specific
heat ratio, and $\pi_\infty$ is the stiffness, respectively.
In the present study we use $\gamma=7.1$ and $\pi_\infty=3.06 \times 10^8$ Pa for water.

At the limit of small change in the density of liquid, the equation of state can
be linearized as
\begin{equation}
p=p_0+c_0^2(\rho-\rho_0)
\label{eqn:lin_eqs},
\end{equation}
where
\begin{equation}
c=\sqrt{\gamma(p+\pi_\infty)/\rho}
\label{eqn:c}
\end{equation}
is the speed of sound in liquid and the subscript 0 denotes reference states.
With $\rho_0=1000$ m$^3$kg$^{-1}$, we recover an ambient speed of sound in water, $c_0=1475$ ms$^{-3}$.

To model the gas phase, we employ a Lagrangian point-bubble approach, in that the gas phase is modeled as spherical, radially oscillating cavities consisted of a non-condensible gas and liquid vapor.
The center of $n$th bubble ($n\in\mathbb{Z}:n\in[1, N]$), with a radius of
$R_n$ and a radial velocity of $\dot{R}_n$, is initially defined at the coordinate
$\vector{x}_n$ and tracked as Lagrangian points during simulations.
To define the continuous field of the void fraction in the mixture at coordinate
$\vector{x}$, we smear the volume of bubble using a regularization kernel $\delta$:
\begin{equation}
\beta(\vector{x})=\sum^N_{n=1}V_n(R_n)\delta(d_n, h),
\label{eqn:beta}
\end{equation}
where $V_n$ is the volume of bubble $n$, $V_n=4\pi/3 R_n^3$, and $d_n$ is the distance of the coordinate $\vector{x}$ from the center of the bubble, $d_n=|\vector{x}-\vector{x}_n|$.
We discretize equations (2.4)-(2.6) on an axi-symmetric grid.
The Lagrangian bubbles are distributed in three-dimensional space.
The kernel $\delta$ maps the volume of bubbles onto the void fraction field defined on axi-symmetric coordinates.
The bubbles, while three dimensional, are forced by the axisymmetric pressure field, which is appropriate to obtain spatially-averaged quantities of the bubbles addressed in the present study \citep{Maeda17}.
Spatial integration is realized using a fifth-order finite volume Weighted Essentially Non-Oscillatory (WENO) scheme \citep{Coralic14}.
A 4th/5th order Runge-Kutta-Cash-Karp (RKCK) algorithm \citep{Cash90} is employed for time integration of solutions.

To model the dynamics of volumetric oscillations of bubbles, we employ the Keller-Miksis equation \citep{Keller80}:
\begin{equation}
\left(R_n\left(1-\frac{\dot{R}_n}{c}\right)\right)\ddot{R}_n
+
\frac{3}{2}\dot{R}^2\left(1-\frac{\dot{R}_n}{3c}\right)
=
\frac{p_n-p_\infty}{\rho}
\left(1+\frac{\dot{R}_n}{c}\right)+\frac{R_n\dot{p}_n}{\rho c},\label{eqn:KM}
\end{equation}
\begin{equation}
p_{n}=p_{Bn}-\frac{4\mu_l\dot{R}_n}{R_n}-\frac{2\sigma_s}{R_n},\label{eqn:pb}
\end{equation}
where $p_n$ is the pressure at the bubble wall,
$p_{Bn}$ is the pressure inside the bubble,
$\sigma_s$ is the surface tension,
and $p_\infty$ is the component of the pressure that forces the radial oscillations of the bubble. We use a reduced-order model \citep{Preston07} to account for heat and mass transfer across the bubble-liquid interface.
The reduced-order model formulates $\dot{p}_n$ and the vapor mass in the bubble $\dot{m}_{Vn}$ as
\begin{align}
\dot{p}_{Bn}
&=\textsf{func}
[R_n,\dot{R}_n,{m}_{Vn}]\label{eqn:preston_pdot}\\
\dot{m}_{Vn}
&=
\textsf{func}[R_n,{m}_{Vn}]\label{eqn:preston_mdot}.
\end{align}
Overall, equations (\ref{eqn:KM}-\ref{eqn:preston_mdot}) consist a system of
ODEs in terms of $[R_n,\dot{R}_n,p_{Bn},m_{Vn}]$, that can be integrated given initial conditions and $p_\infty$.

\subsection{Acoustic source}
In simulations we excite volumetric oscillations of bubbles using plane and focused pressure waves.
In order to generate the waves in the computational domain, we utilize a source-term approach introduced by \cite{Maeda17}.
The method can generate uni-directional acoustic waves from an arbitrarily chosen source surface, by forcing the mass, momentum and energy equations (\ref{eqn:vaeqn2_ma}-\ref{eqn:vaeqn2_e}) in a thin volume enclosing the surface..

\section{Theory and scaling for the dynamics of bubble clouds}
\label{section:metrics}
\subsection{Cloud interaction parameter}
\citet{dAgostino89} (hereafter DB) studied the linear response of monodisperse, spherical bubble clouds subjected to harmonic, long wavelength pressure excitation.
DB deduced that the response of the bubble cloud with a low void fraction is characterized by a non-dimensional parameter,
\begin{equation}
B_0=\frac{\beta_0R^2_c}{R^2_{b0}},\label{eqn:DB}
\end{equation}
termed as the {\itshape{cloud interaction parameter}}.
DB found that, when $B_0\ll 1$, the effect of inter-bubble interaction is weak and each bubble in the cloud behaves like a single, isolated bubble. When $B_0\gg 1$, inter-bubble interactions cause bubbles to oscillate coherently at a lower frequency than an isolated single bubble.
\citet{Wang99} simulated the dynamics of a spherical bubble cloud with various values of $B_0$ in nonlinear regime.

The cloud interaction parameter can be interpreted in different ways.
Substituting $\beta=N_bR^3_{b0}/R^3_c$ into equation \ref{eqn:DB}, $B_0$ can be rewritten as
\begin{equation}
B_0=\frac{N_bR^3_{b0}}{R^3_c}\frac{R^2_c}{R^2_{b0}}=\frac{N_bR_{b0}}{R_c},
\end{equation}
where $N_b$ is the number of bubbles in the cloud.
We notice that this scaling parameter can be independently derived from the Lagrangian mechanics of spherical bubbles under mutual interactions.
The global kinetic energy of potential flow of an incompressible liquid induced by volumetric oscillations of $N_b$ spherical bubbles can be expressed using a multipole expansion \citep{Takahira94,Ilinskii07} as
\begin{equation}
K = 2\pi\rho\left[\sum^{N_{b}}_iR_i^3\dot{R}_i^2+\sum^{N_{b}}_i\sum^{N_{b}}_j\frac{R_i^2R_j^2\dot{R}_i\dot{R}_j}{r_{i,j}}+O\left(\frac{R^7\dot{R}^2}{r^4}\right)\right],
\label{eqn:mRP}
\end{equation}
where $r_{i,j}$ is the distance between the centers of bubble $i$ and bubble $j$.
The first term in the bracket represents the kinetic energy induced by direct contributions from each bubble and the second term represents the energy induced by the inter-bubble interactions.
When bubbles have an approximately uniform size distribution and experience simultaneous change in pressure, we can assume that each bubble takes the same characteristic radius and the velocity, $R$ and $\dot{R}$.
The characteristic inter-bubble distance can be scaled as $r\sim R_c$.
Then $K$ can be scaled as
\begin{equation}
K \sim N_b R^3\dot{R}^2\left(1+\frac{N_{b}R}{R_c}\right).
\end{equation}
In the limit of small amplitude oscillations we have $R\approx R_{b0}$, and therefore we obtain
\begin{equation}
K \sim N_b R_{b0}^3\dot{R}^2\left(1+B_0\right).
\end{equation}
We see that the interaction parameter dictates the kinetic energy induced by bubbles.
With $B_0=0$ the kinetic energy is that of an isolated bubble, while with $B_0>1$ there is an additional contribution from the inter-bubble interactions.
Based on equation (33), an extended R-P equation for the dynamics of the bubbles can be derived \citep{Takahira94,Doinikov04,Bremond06,Ilinskii07,Zeravcic11}.
In fact, the scaling of kinetic energy in terms of $N_bR_b/R_c$ was mentioned by \citet{Ilinskii07}, but was not associated with the parameter derived by DB.

In what follows we specify the size distribution of nuclei to be log-normal distribution given by ${\mathrm{ln}}(R_b/R_{b,ref})\sim N(0,\sigma^2)$.
Therefore we employ the following expression for $B_0$:
\begin{equation}
B_0\approx\frac{N_bR_{b0,ref}}{R_c}.
\end{equation}

\subsection{Moments}
In order to quantify the anisotropic structure and associated bubble dynamics, we use the moments of either bubble volume or kinetic energy of the liquid, both measured with respect to the initial center of the cloud (hereafter denoted as the moment of volume and moment of kinetic energy, respectively).
The $n$-th moments of the bubble volume and the bubble-induced kinetic energy of liquid are thus respectively defined as:
\begin{equation}
\mu_{Vcn}=\frac{\sum_{bubble}\frac{4\pi}{3}R^3_b\left(\frac{x_b}{R_c}\right)^n}{\sum_{bubble}\frac{4\pi}{3}R^3_b}
\hspace{3mm}and \hspace{3mm}
\mu_{Kcn}=\frac{\sum_{bubble}2\pi\rho R^3_b\dot{R}^2_b\left(\frac{x_b}{R_c}\right)^n}{\sum_{bubble}2\pi\rho R^3_b\dot{R}^2_b}.
\end{equation}
We will treat the first moment ($n=1$), unless otherwise noted.
The moments are normalized to vary within the range of $[-1, 1]$. In an extreme case, when monodisperse bubbles are distributed in a left hemisphere ($-x$) and oscillate with the same radial velocity, the 1st moments satisfy $\mu_{Vcn}=\mu_{Kcn}=-0.375$. Therefore, moments smaller than this value indicate a large bias in the volume or kinetic energy toward the proximal side of the cloud.

\section{Cloud cavitation in a focused ultrasound wave}
\label{section:focused}
\subsection{Setup}
In order to investigate the dynamics of bubble clouds, we conduct numerical simulations that mimics the laboratory setup.
\begin{figure}
  \center
  \includegraphics[width=140mm,trim=0 0 0 0, clip]{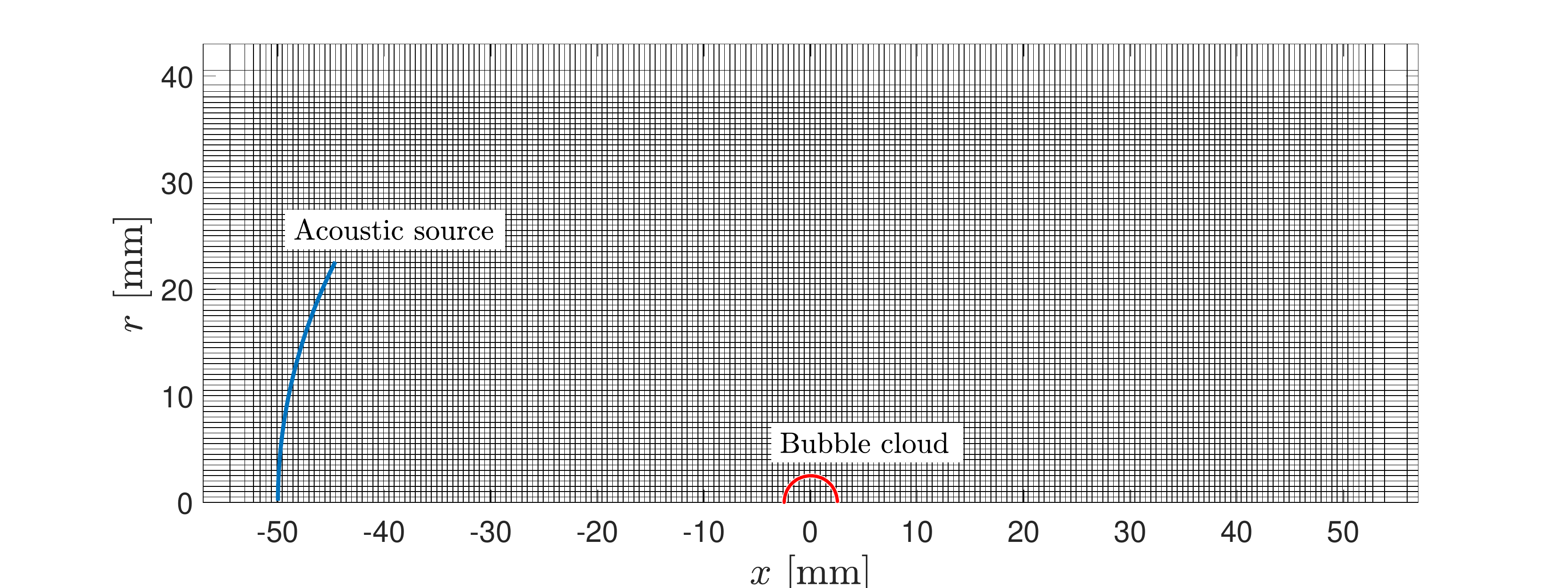}
  \caption{Schematic of the numerical setup. Every fifth point is plotted in the computational mesh.}
   \label{fig:focus_setup} 
\end{figure}
Figure \ref{fig:focus_setup} shows the schematic of numerical setup.
The size of the simulation domain is $500\times250$ mm, which has been verified to be sufficiently large to effectively mimic free space.
For the initial condition, we randomly distribute bubble nuclei in a spherical region of with radius 2.5 mm with its center located at the origin of $x-r$ axi-symmetric coordinates.
The grid size is uniform near the region of bubble cloud with a characteristic grid size of 100 $\mu$m. Symmetry boundary condition is used on the axis of symmetry.
The grid is smoothly stretched toward the other domain boundaries, where characteristic boundary conditions are used to reduce spurious reflections of waves.
The transducer used in the experiment is modeled by an acoustic source uniformly distributed on a portion of spherical surface with an aperture of 30 mm and a radius of 50 mm concentric with the bubble cloud.
The axis of the spherical surface is aligned with the axis of symmetry of the coordinates.

\begin{figure}
  \center
  \includegraphics[width=80mm,trim=0 0 0 0, clip]{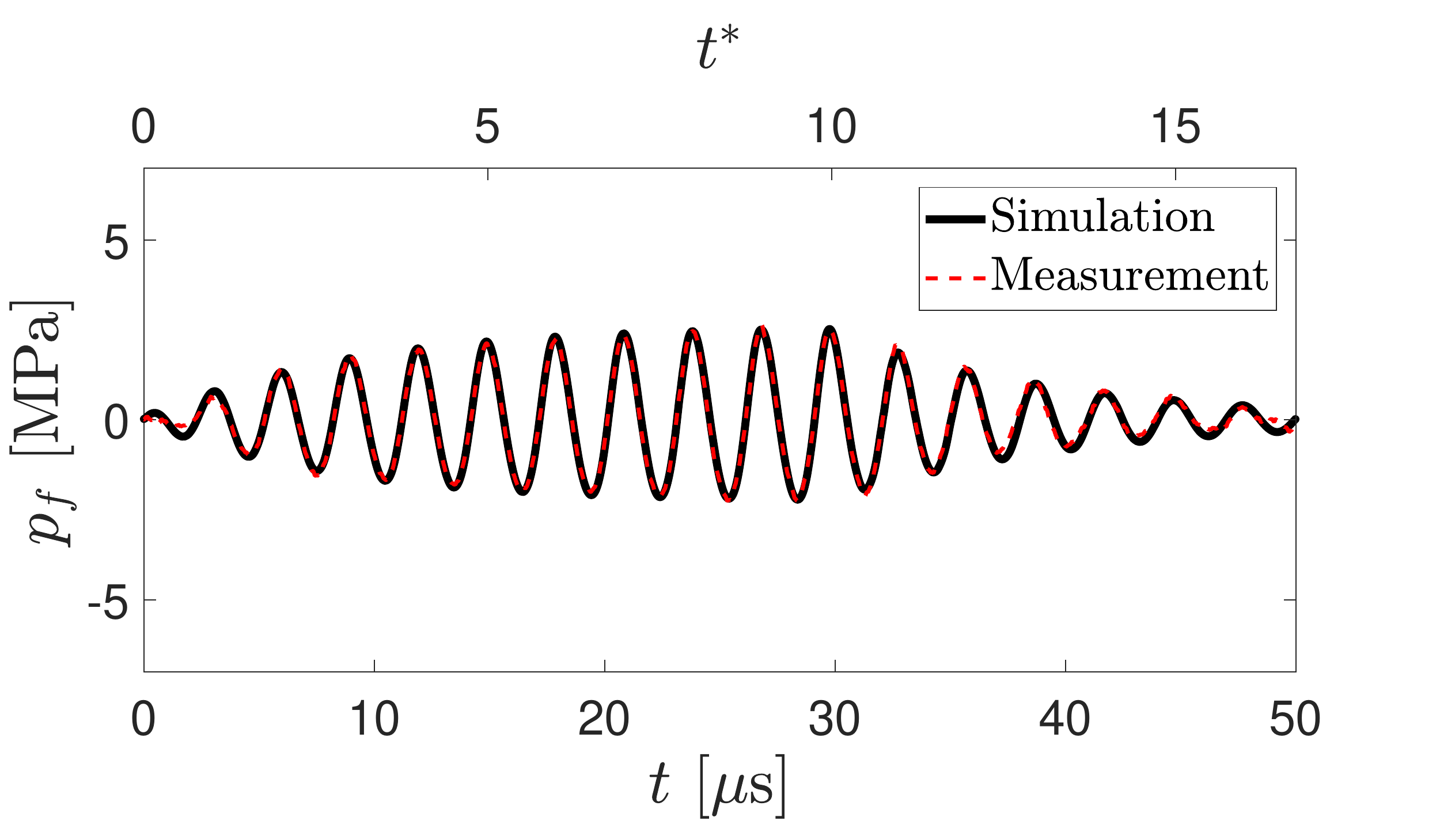}
  \caption{Focal pressure evolution of the modeled transducer and that of the experimental measurement.}
   \label{fig:waveform} 
\end{figure}
The modeled acoustic source is calibrated by comparing the focal pressure evolution with an experimental measurement with a low input voltage, producing the good agreement shown in the figure \ref{fig:waveform}.
Details of the calibration are described in appendix A.
In the simulations of bubble clouds, the peak maximum and negative amplitudes are adjusted to 6.0 and -4.5 MPa, respectively.

\begin{table}
 \begin{center}
  \begin{tabular}{cccc}
    Run & $B_0$  & $\sigma$ \\[3pt]
     F1 & 0.625  & 0\\
     F2 & 1.25  & 0\\
     F3 & 2.5  & 0\\ 
     F4 & 5.0  & 0\\
     F5 & 0.625  & 0.7\\
     F6 & 1.25  & 0.7\\
     F7 & 2.5  & 0.7\\ 
     F8 & 5.0  & 0.7\\
 \end{tabular}
  \caption{List of parameters used 8 runs for simulations of bubble cloud dynamics in the focused ultrasound wave.
  }{\label{tab:pm}}
 \end{center}
\end{table}
The parameters of bubble clouds used in the simulations are summarized on table \ref{tab:pm}.
It is challenging to measure the population and the initial size distribution of nuclei in the experiment.
Therefore, we empirically assess the effects of the nuclei population on the resulting bubble cloud dynamics by varying the value of $B_0$ within a range of $B_0\in[0.625,5]$.
To assess the effect of polydispersity, for each value of $B_0$ we simulate monodisperse and polydisperse clouds.
For the polydisperse case, the initial radii of bubbles follow a log-normal distribution given as ${\mathrm{ln}}(R_{b0}/R_{b,ref})\sim N(0,\sigma^2)$ \citep{Ando11}, where $R_{b,ref}$ is the most probable bubble size, chosen as $R_{b,ref}=10$ $\mu$m. In the monodisperse and polydisperse cases, we use $\sigma=0$ and $0.7$, respectively. $\sigma=0.7$ models highly polydisperse bubble clouds.
This is in order to obtain an upper bound of the variability in the resulting bubble dynamics due to polydispersity. We neglect fission/break-up of bubbles during the simulations.
In order to assess the variability of the bubble cloud dynamics due to the initial reference radius of bubbles, we also simulated monodisperse and polydisperse clouds with $R_{b0,ref}=5$ $\mu$m with various values of $B_0:B_0\in[0.625,5]$.
The results did not show a significant difference from cases with $R_{b0,ref}=10$ $\mu$m, thus they are omitted in this paper.

\subsection{Comparisons with the high-speed image}
\begin{figure}
  \center
  \includegraphics[width=140mm,trim=0 0 0 0, clip]{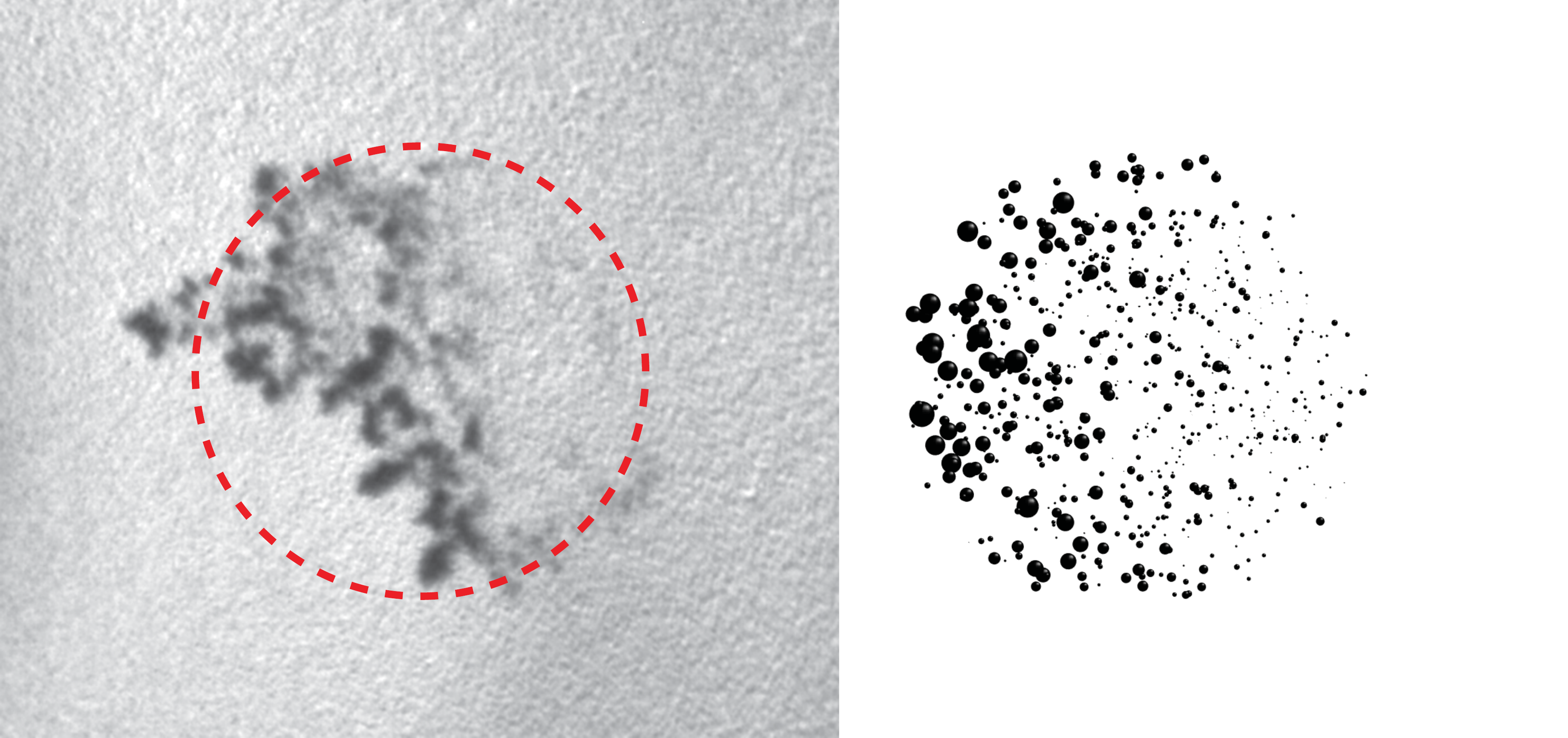}
  \caption{Images of the bubble cloud obtained in the experiment and simulation (F8) at $t^*=13.6$. The red, dotted line shows a circle with radius 2.5 mm. }
   \label{fig:cloud_comp} 
\end{figure}
Figure \ref{fig:cloud_comp} compares the high-speed image (the 7th image of fig \ref{fig:imacon}) and the image of bubbles obtained in run-F7 at $t^*=10$.
A similar anisotropic structure is evident in the simulated cloud; the proximal bubbles are larger than the distal bubbles.

\begin{figure}
  \center
  \subfloat[]{\includegraphics[width=70mm,trim=0 0 0 0, clip]{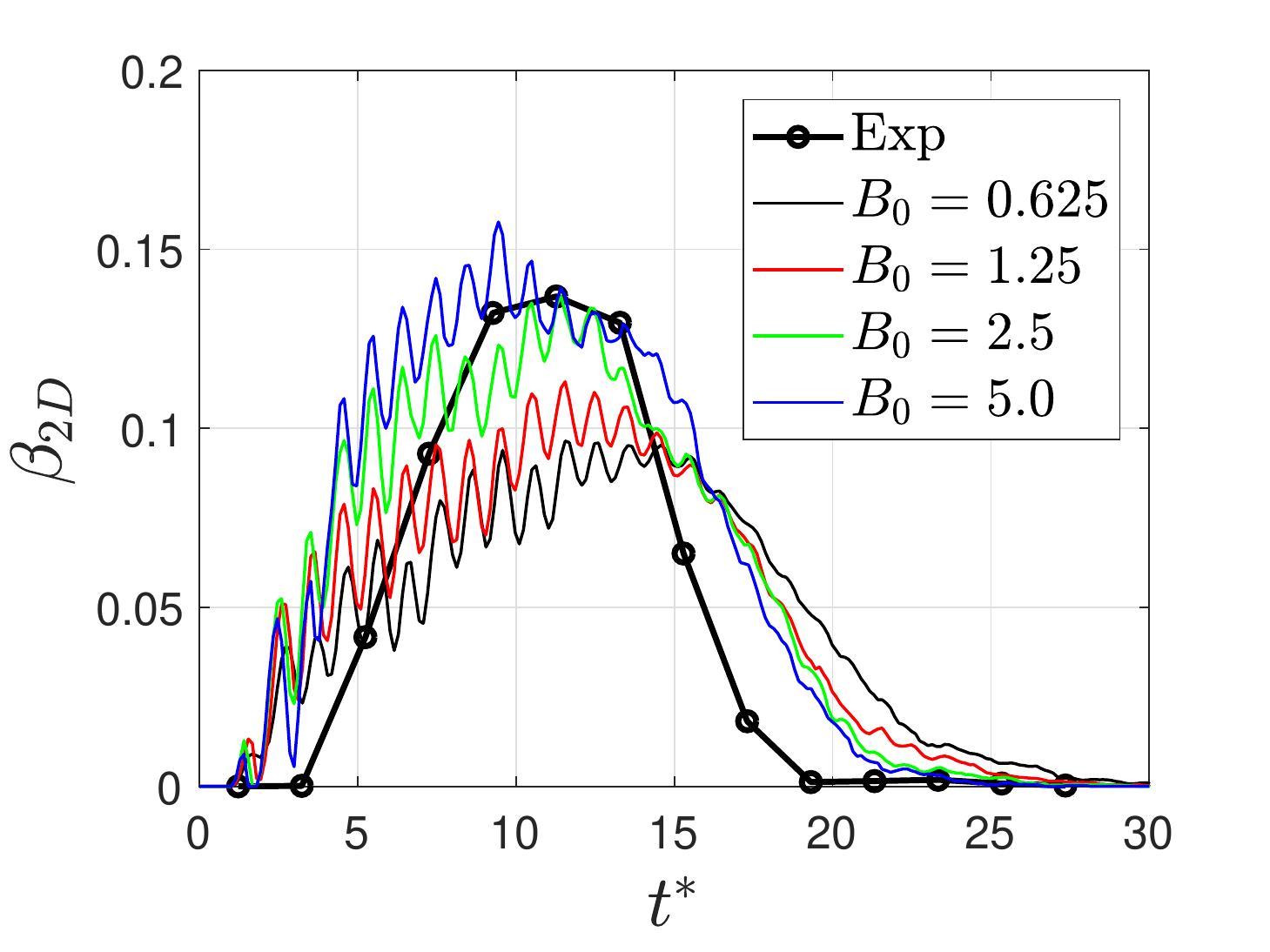}}
  \subfloat[]{\includegraphics[width=70mm,trim=0 0 0 0, clip]{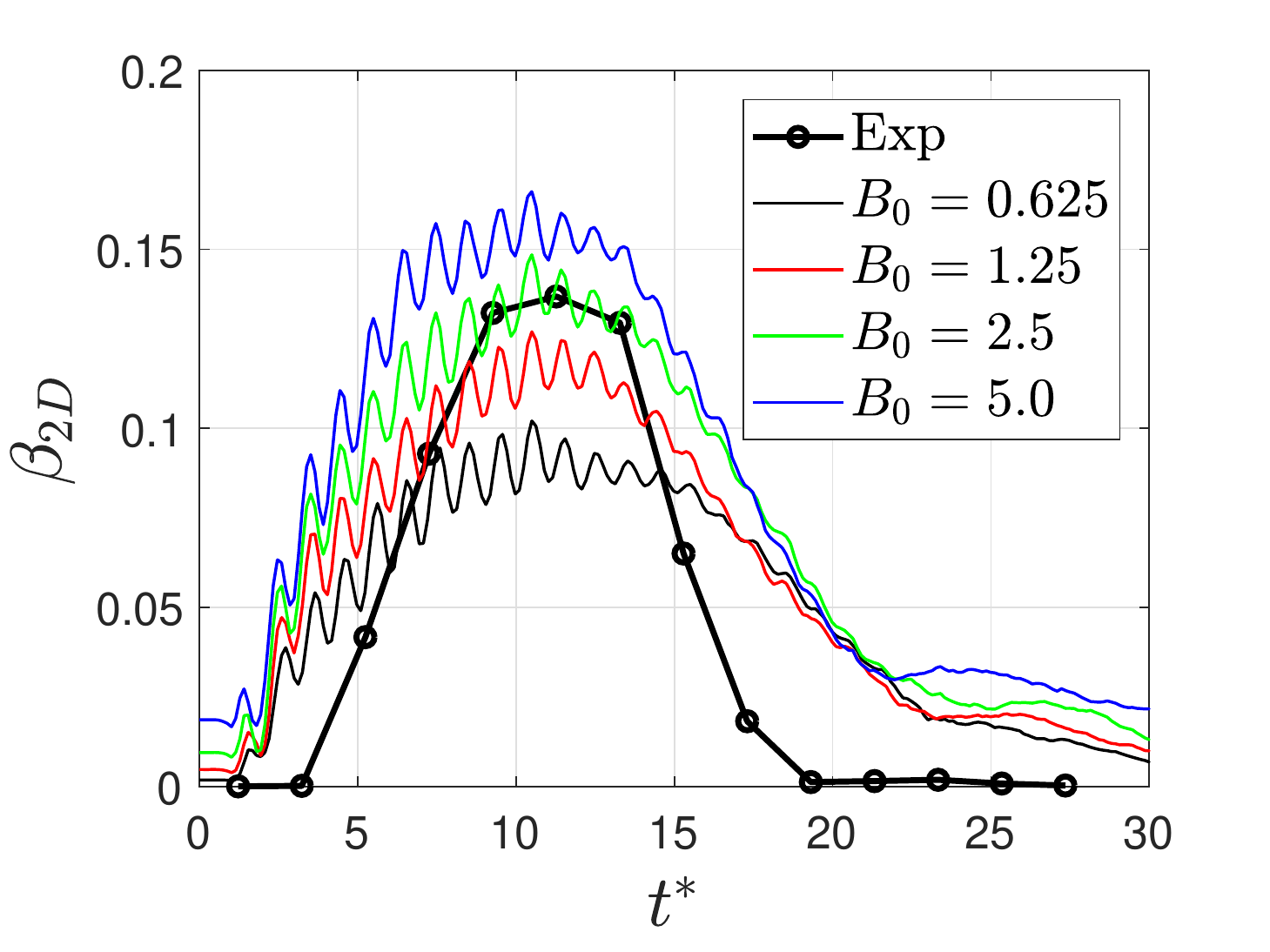}}
  \caption{Comparisons of the evolutions of the two dimensional void fraction of bubbles during the experiment and the simulations.}
   \label{fig:2Devol} 
\end{figure}
Figure \ref{fig:2Devol} (a) and (b) compare the evolutions of the two-dimensional void fraction of bubbles in the experiment, and the simulation of initially monodisperse (run F1-4) and polydisperse (run F5-8) clouds, respectively.
The two-dimensional void fraction is obtained as
\begin{equation}
\beta_{2D}=\frac{A}{\pi R_c^2},
\end{equation}
where $A$ is the area occupied by the bubbles on the two-dimensional images.
In all the cases, the projected area steadily grows and reaches its maximum value within the range of 0.1-0.15 at around $t^*=10-15$ then decays.
Overall, trends in the evolution of the void fraction are similar between the monodisperse and polydisperse clouds, though the polydisperse clouds present slightly higher peak values than the monodisperse clouds with the same values of $B_0$.
The magnitudes of the slope of the void fraction during the growth and the decay are larger in the experiment than the simulation.
The discrepancies could be due to experimental uncertainties, including the size distribution of nuclei in the simulation, non-sphericity of the cloud in the experiment, and the finite resolution and/or the noise of the high-speed images.
Nevertheless, the results confirm that the simulated bubble clouds quantitatively reproduce the experimental observation with reasonable accuracy.

\begin{figure}
  \center
  \subfloat[]{\includegraphics[width=70mm,trim=0 0 0 0, clip]{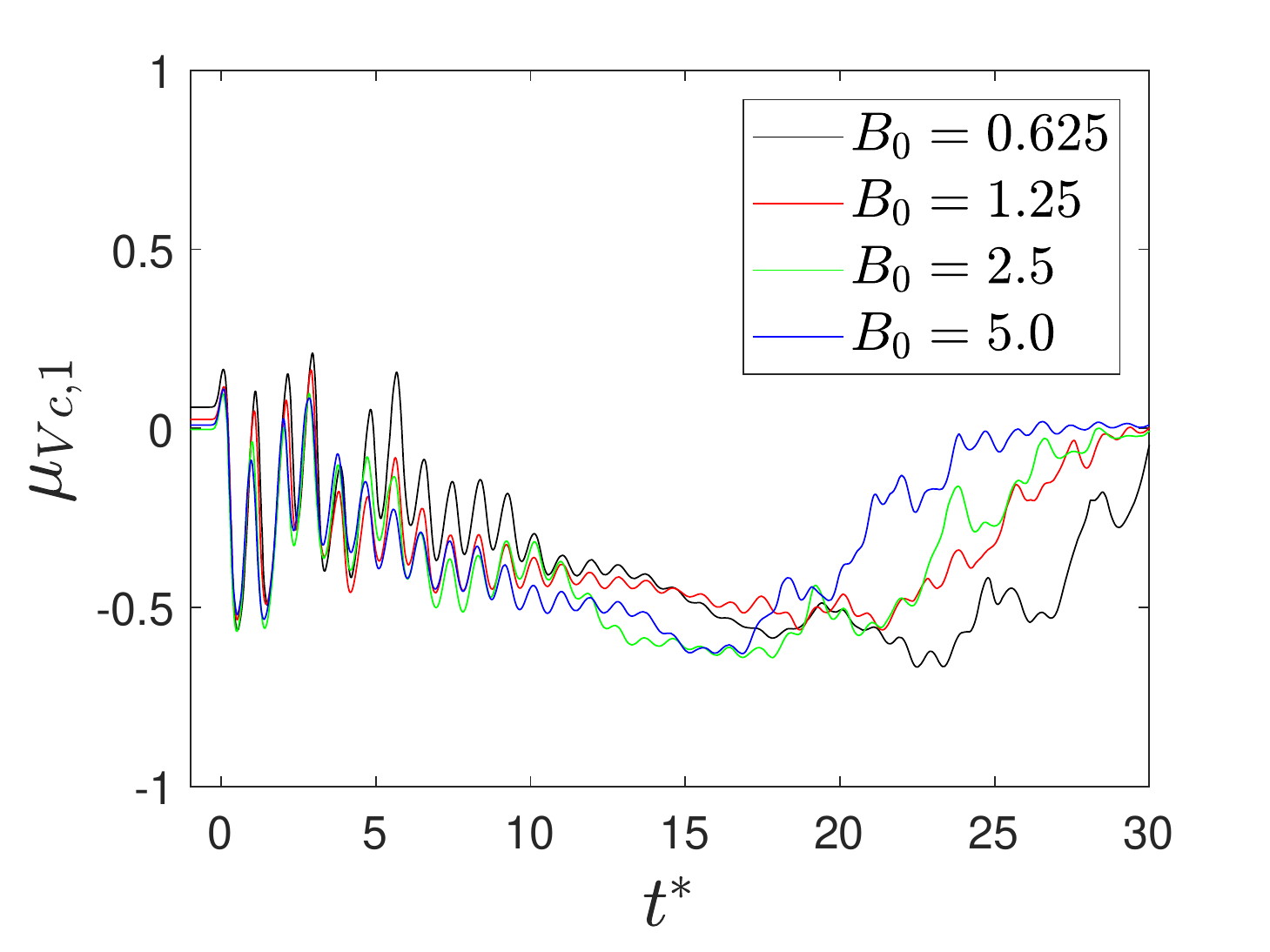}}
  \subfloat[]{\includegraphics[width=70mm,trim=0 0 0 0, clip]{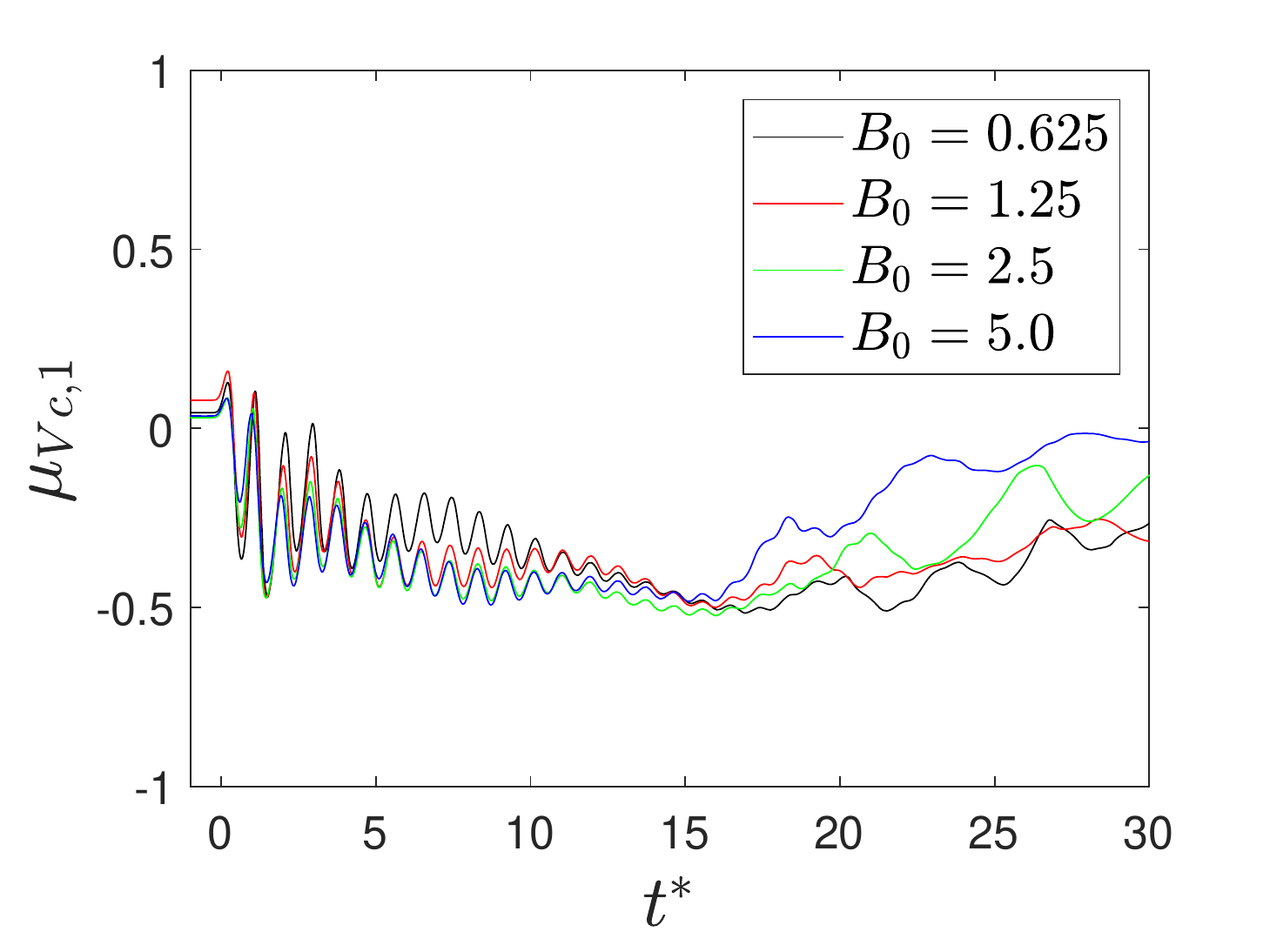}}\\
  \subfloat[]{\includegraphics[width=70mm,trim=0 0 0 0, clip]{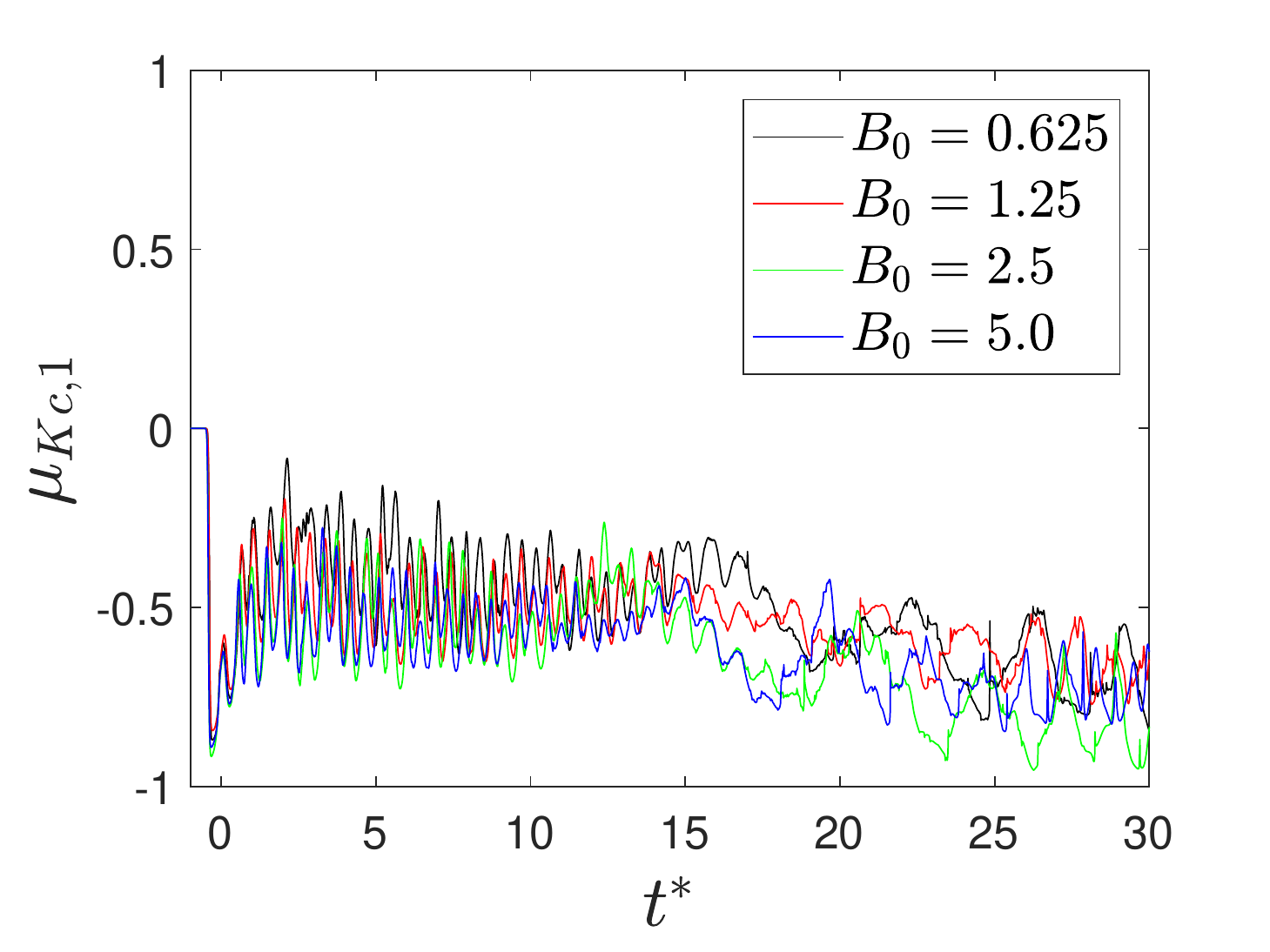}}
  \subfloat[]{\includegraphics[width=70mm,trim=0 0 0 0, clip]{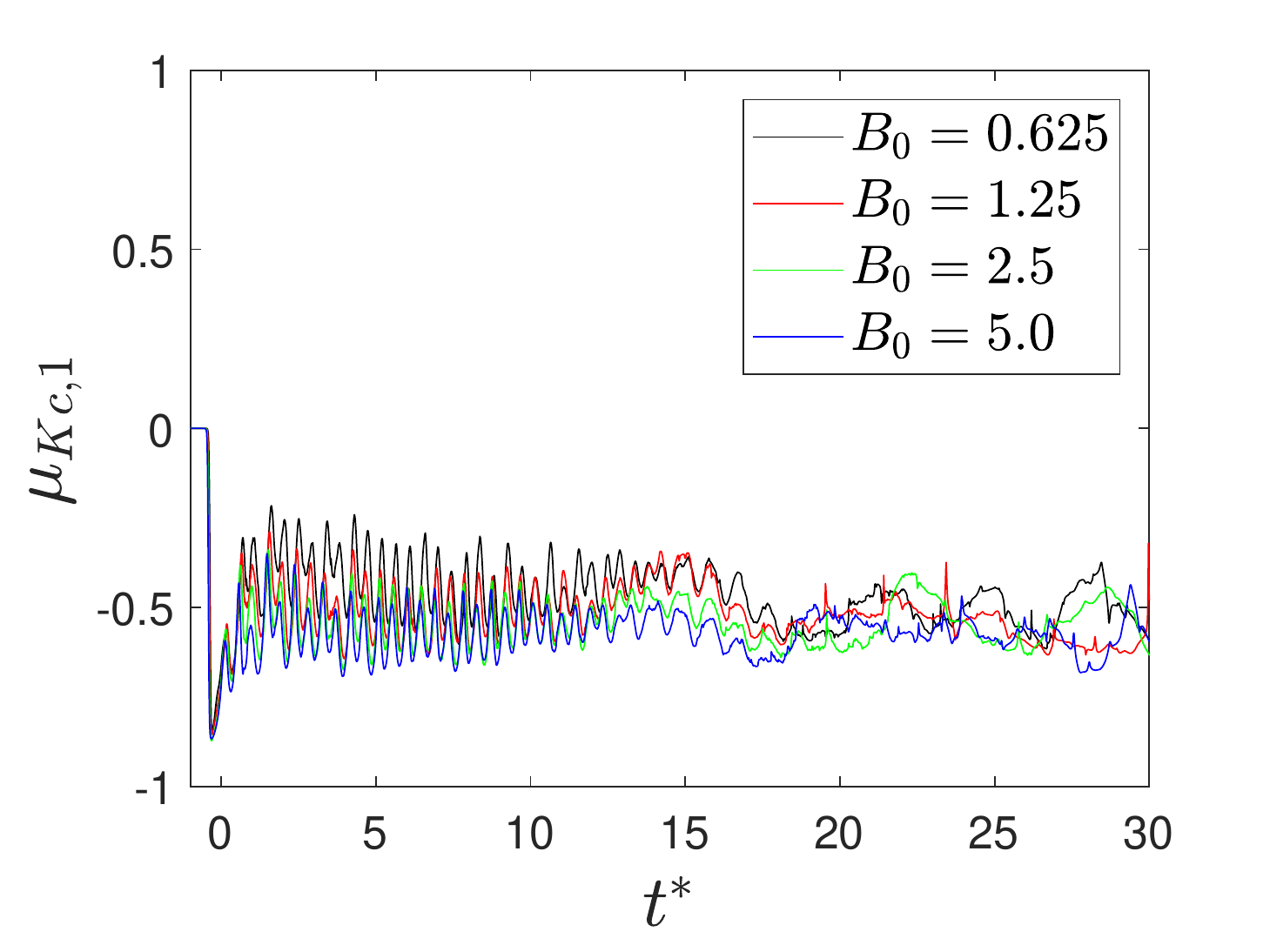}}
  \caption{Evolution of the moment of volume, $\mu_{Vc,1}$, (top row) and the kinetic energy, $\mu_{Kc,1}$, (bottom row) in the simulated clouds. Results of (a,c) monodisperse and (b,d) polydisperse clouds are shown.}
   \label{fig:focus_moment} 
\end{figure}
For quantification of the anisotropic structure, we compute evolutions of the moments of volume and the kinetic energy in each cloud during the course of simulations.
Figure \ref{fig:focus_moment} shows the result.
In all the clouds, the moment of volume oscillates around -0.25 during the passage of the wave until around $t^*=17$ then grows back to zero. This suggests that the size of the proximal bubbles are larger than the distal bubbles for all $t^*$ and the structural anisotropy is the most significant around at $t^*=17$.
After the initial transient, the moment of kinetic energy oscillates between -0.25 and -0.5 for all $t^*$.
This indicates that the proximal bubbles experience a larger amplitude of pressure excitation and oscillate more actively than the proximal bubbles.

The results above indicate that the bubble dynamics are relatively insensitive to both the population and initial polydispersity of the clouds.
Therefore, the anisotropic structure is expected to be observed over a wide range of the nuclei distribution and population.

\section{Parametric simulations using plane ultrasound waves}
\label{section:parametric}
\subsection{Setup}
In the setup considered in the previous section, bubbles are forced by the pressure wave with a complex waveform generated by a specific transducer.
This hinders further generalization of the obtained results, including the anisotropic structure and the bubble-induced kinetic energy, to the bubble cloud dynamics excited in other geometries of pressure fields.
For generalization, analysis using a wider range of parameters, but with a simpler geometry of acoustic source is desirable.
To this end, as an idealized problem, we conduct parametric simulations of bubble cloud dynamics excited by plane ultrasound waves of various amplitudes.

\begin{table}
 \begin{center}
  \begin{tabular}{cc}
    Run  & $A$ \\[3pt]
       A1v[1-4] & $10^{-1.5}$\\
       A2v[1-4] & $10^{-1.0}$\\
       A3v[1-4] & $10^{-0.5}$\\
       A4v[1-4] & $1$        \\
       A5v[1-4] & $10^{0.5}$ \\
       A6v[1-4] & $10$       \\
\end{tabular}
\caption{List of parameters used in the parametric study. The numbering after the symbol v denotes values of nuclei densities, corresponding to $B_0=0.625, 1.25, 2.5, 5$, respectively. For each set of $(A,B_0)$, 5 bubble clouds with distinct initial coordinates of bubbles are simulated.}{\label{tab:kd}}
 \end{center}
\end{table}
The set of parameters addressed in the simulations is summarized in table \ref{tab:kd}.
The radius of clouds and variations of $B_0$ follow the previous section.
It is realistic to assume that the radial distribution of bubbles is polydisperse rather than monodisperse in practical conditions.
Thus we assume that the distribution of the initial radius of nuclei follows a log normal distribution with ${R}_{b0,ref}=10$ $\mu$m and $\sigma=0.7$, but we expect only small differences with monodisperse clouds in the present cases.

The mesh size follows the previous section.
We excite 10 cycles of a plane, sinusoidal pressure wave from a source plane located at $x=-20$ mm to the positive $x$ direction, that gives the pressure at the origin, without bubble cloud, of
\begin{equation}
p_a=p_0[1+H(10-t^*)A\mathrm{sin}(2\pi t^*)],
\end{equation}
where $H$ is the Heaviside step function.
The frequency of the wave is $f=300$ Hz, thus the wavelength is 4.9 mm and approximately equal to the diameter of the bubble clouds.
In order to assess the variability of the bubble cloud dynamics due to spatial distribution of bubbles, with each set of $(A,B_0)$ we simulate $N_s=5$ clouds with distinct, random spatial distributions of nuclei.
In what follows, we denote quantities obtained by averaging $N_s$ bubble clouds with the same set of $(A,B_0)$ as those of {\it{ensemble averaged cloud}}.
We denote the ensemble average of arbitrary quantity $f$ as
\begin{equation}
f_{ens}=\frac{1}{N_s}\sum_{i=1}^{N_s}f_i,
\end{equation}
where $f_i$ is obtained from $i$-th realization of the bubble cloud.
In the present simulations, $N_s=5$ is sufficient to obtain ensemble averaged quantities.

\subsection{Anisotropic structure}
\label{subsec:anis}
Here we analyze the volumetric evolution and the anisotropic structure of the clouds.
We begin by looking at the highest amplitude case, $A=10$, in detail.

\begin{figure}
  \center
  \subfloat[]{\includegraphics[width=70mm,trim=0 0 0 0, clip]{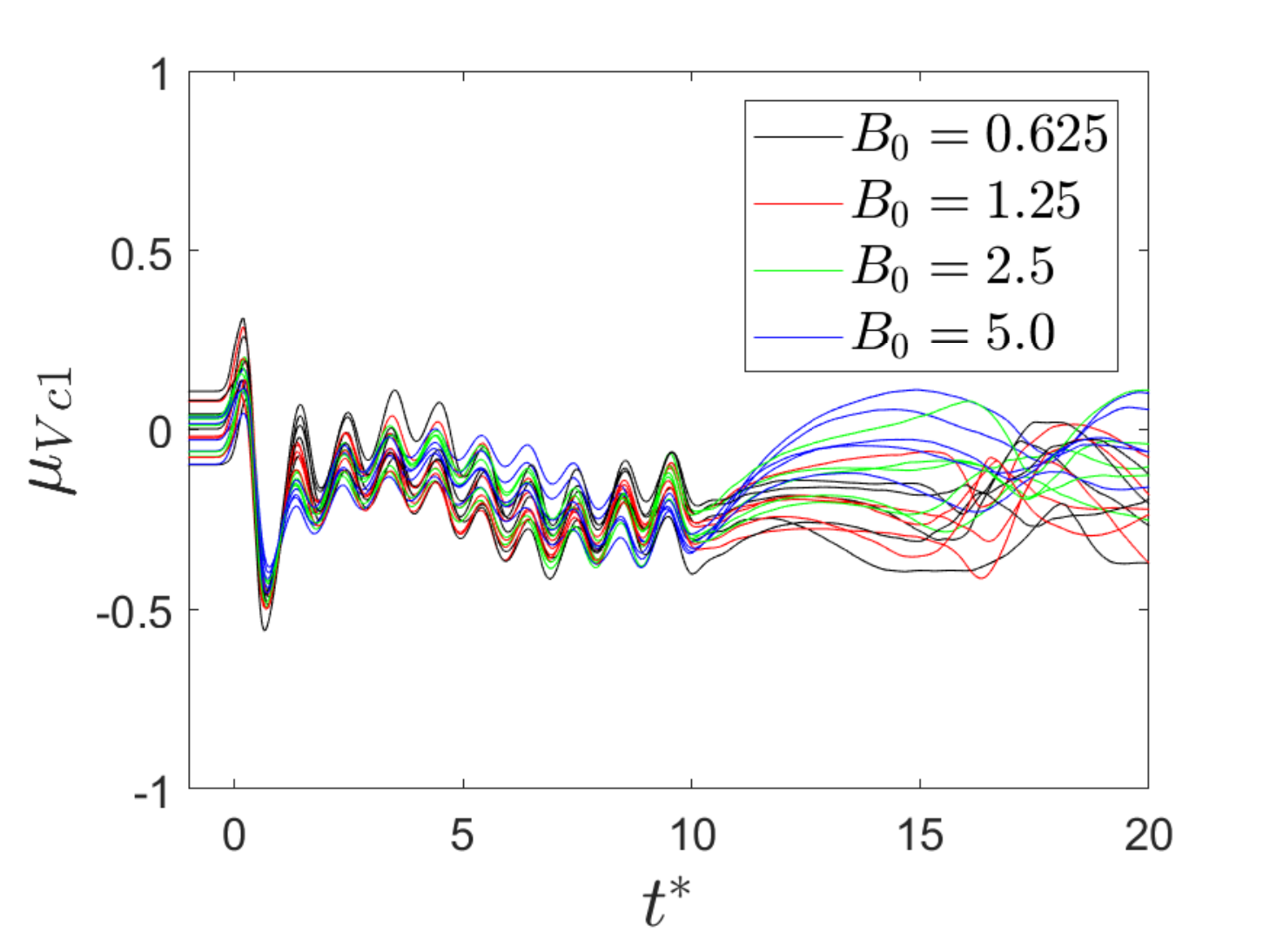}}
  \subfloat[]{\includegraphics[width=70mm,trim=0 0 0 0, clip]{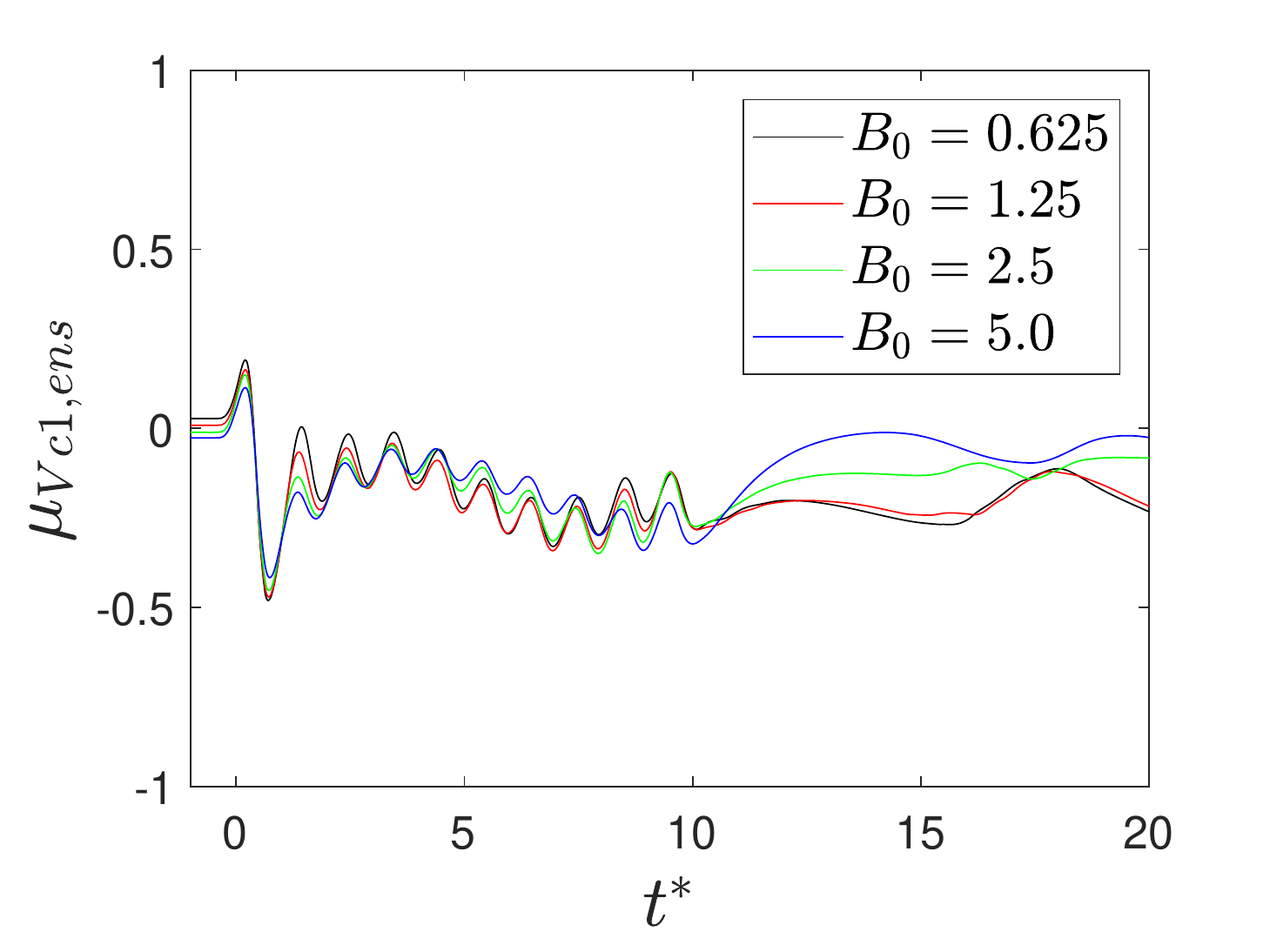}}
  \caption{Evolution of the moment of volume with $A=10$ in (a) each realization (b) ensemble averaged clouds.}
   \label{fig:xv_cent}
\end{figure}
Figure \ref{fig:xv_cent} (a) shows evolutions of the moment of volume of bubble clouds from run A6v during the course of simulation.
The moment of volume oscillates between -0.3 and 0 for all values of $B_0$ after initial transient until $t^*=10$. After $t^*=10$ the range of moment takes on a wider spread in values.
In order to assess variability associated with the random position of bubbles, figure \ref{fig:xv_cent} (b) shows the same quantities of the ensemble averaged clouds.
The similarity of the moments in the two plots indicates small incoherence among the the dynamics of bubble clouds of distinct realizations.
The clouds share the same anisotropic structure regardless of the initial population and spatial distribution of nuclei.

\begin{figure}
  \center
  \subfloat[]{\includegraphics[width=33mm,trim=120 0 120 0, clip]{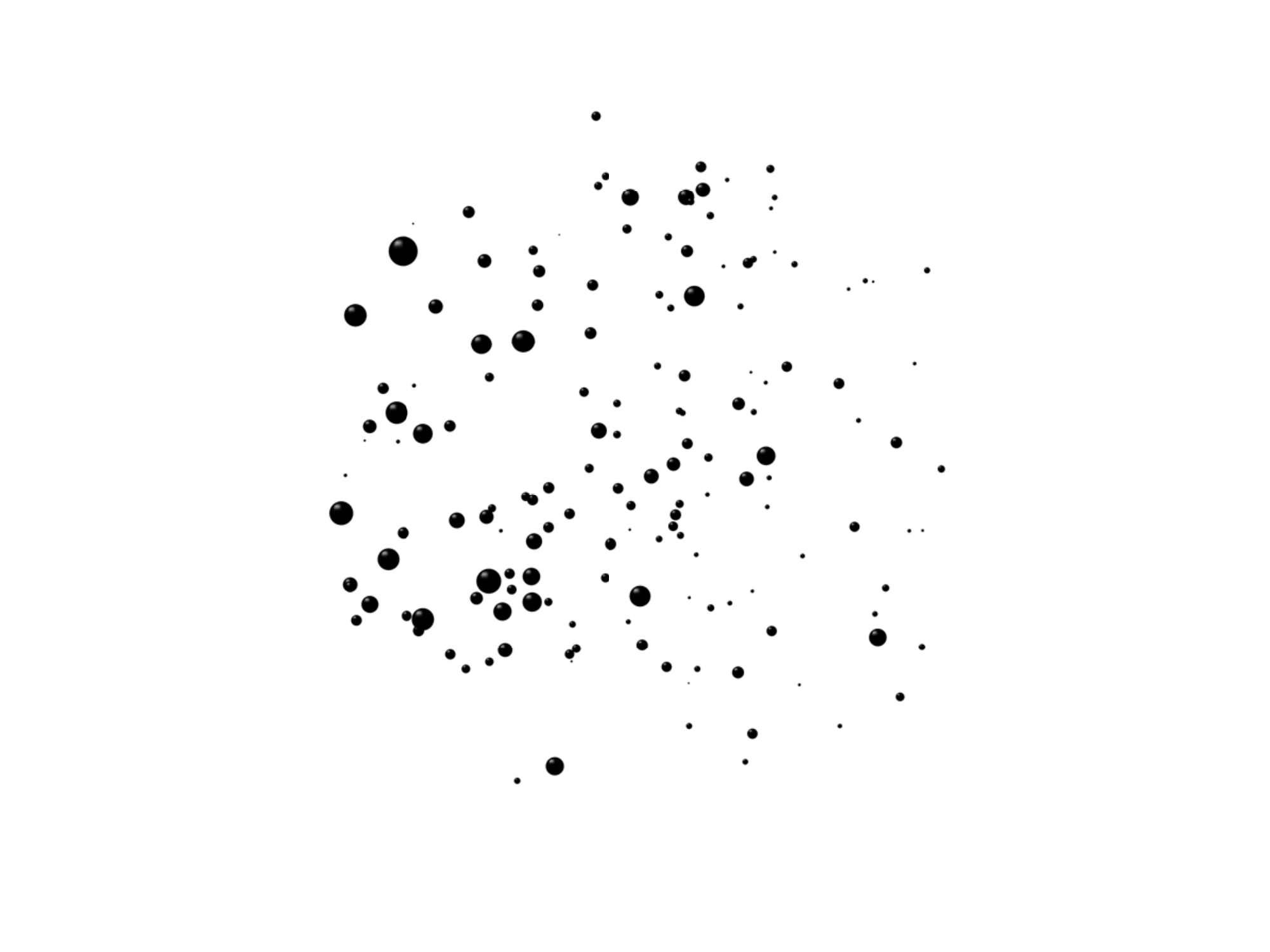}}
  \subfloat[]{\includegraphics[width=33mm,trim=120 0 120 0, clip]{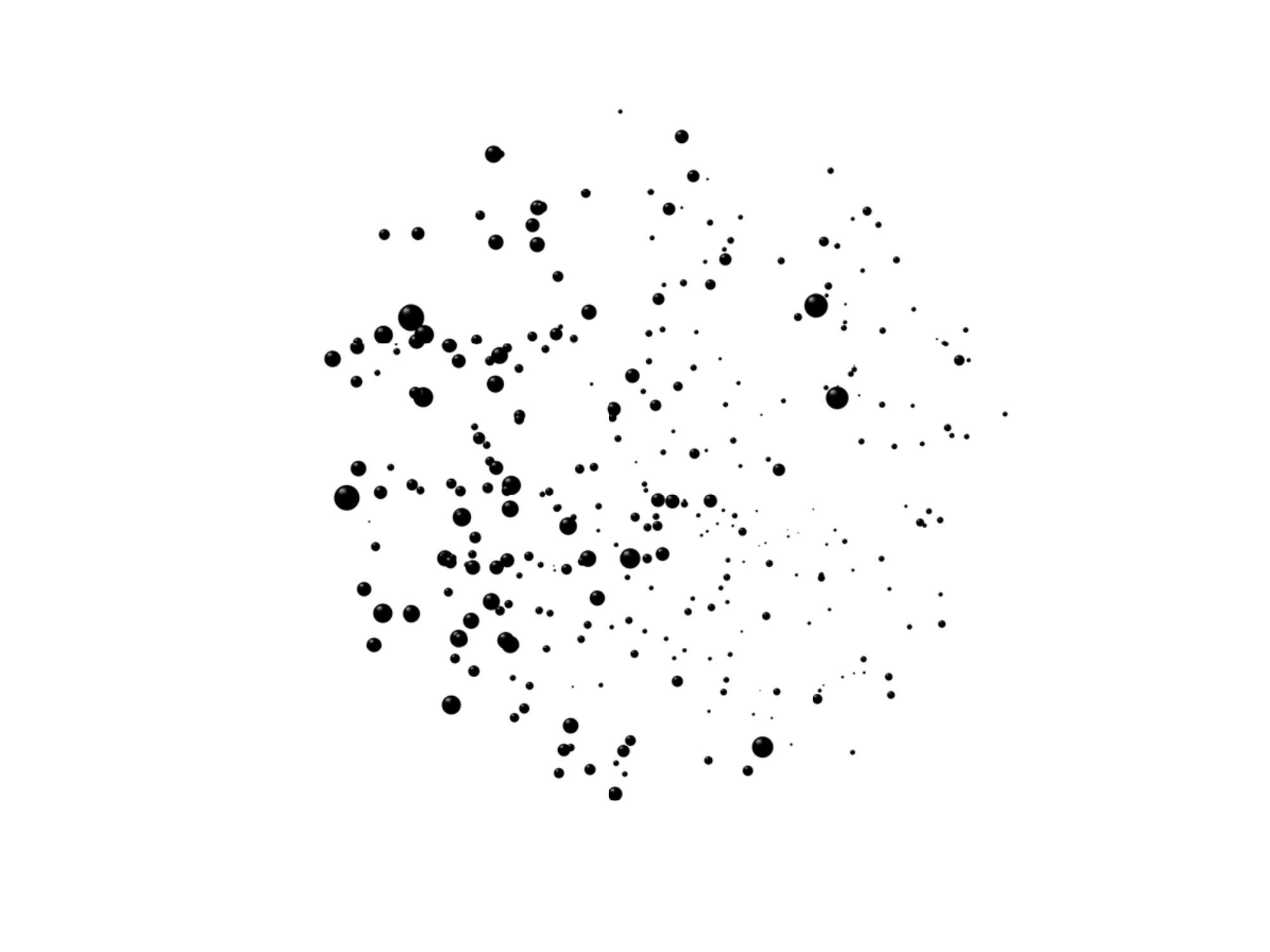}}
  \subfloat[]{\includegraphics[width=33mm,trim=120 0 120 0, clip]{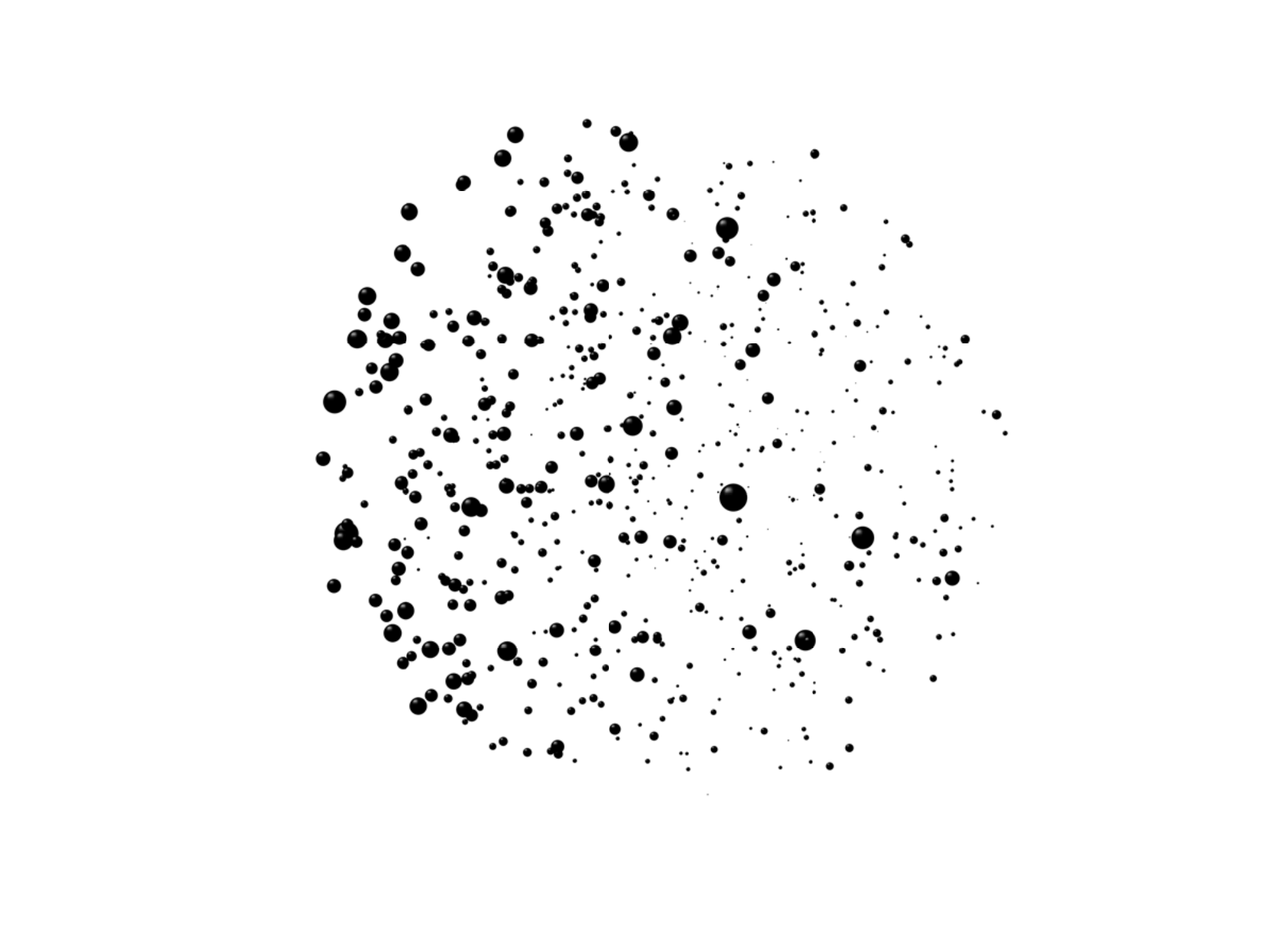}}
  \subfloat[]{\includegraphics[width=33mm,trim=120 0 120 0, clip]{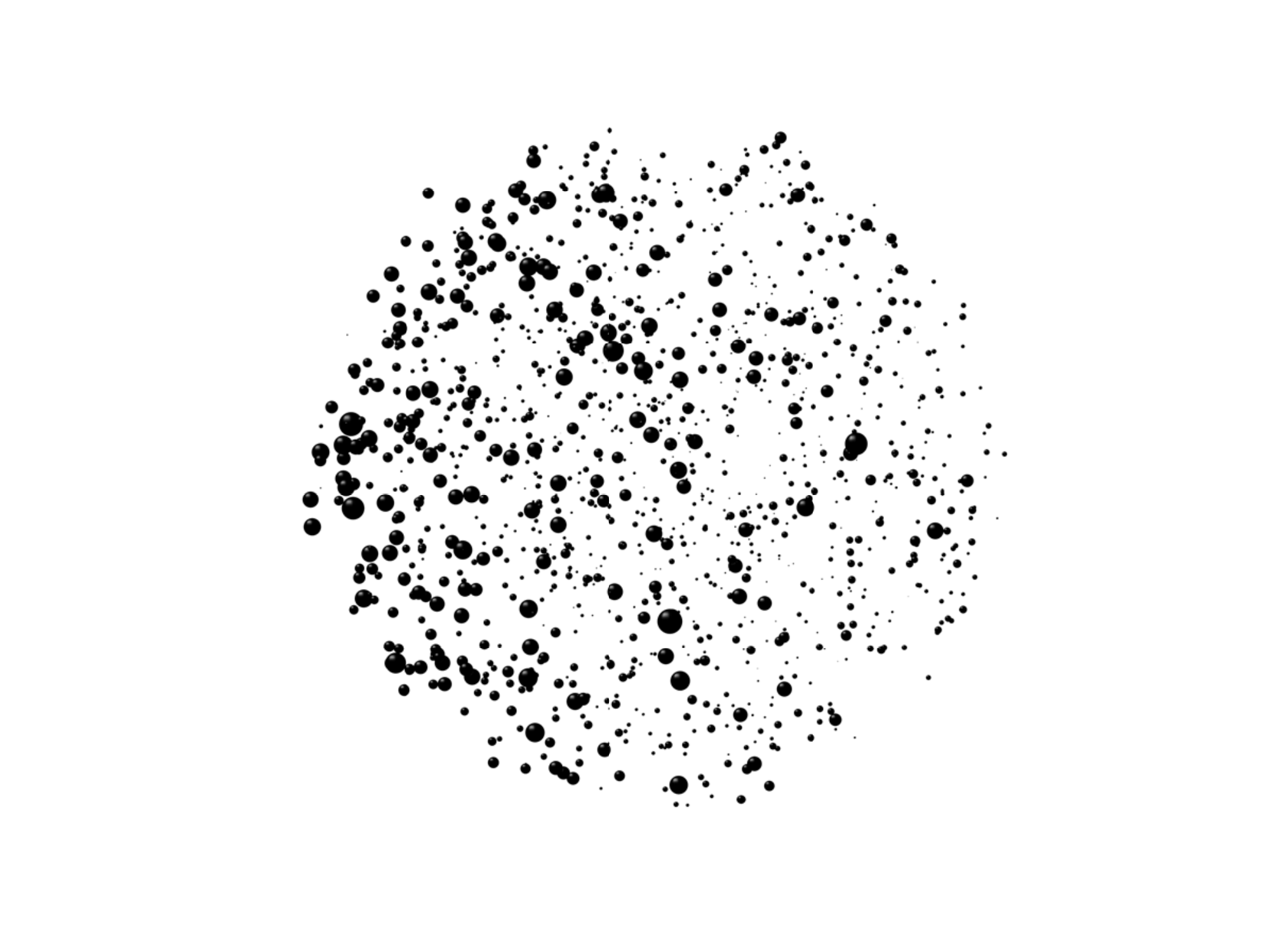}}
  \caption{Images of the bubble clouds with various values of $B_0$ at $t^*=5.7$ from runs (a) A6v1, (b) A6v2, (c) A6v3, and (d) A6v4.
   }
   \label{fig:snap_f300}
\end{figure}
Figure \ref{fig:snap_f300} shows images of bubble clouds at $t^*=5.7$, obtained from one of the realizations from runs A6v1 - A6v4.
As expected, the anisotropic structure is similar to the clouds excited by HIFU.

\begin{figure}
  \center
  \includegraphics[width=70mm,trim=0 0 0 0, clip]{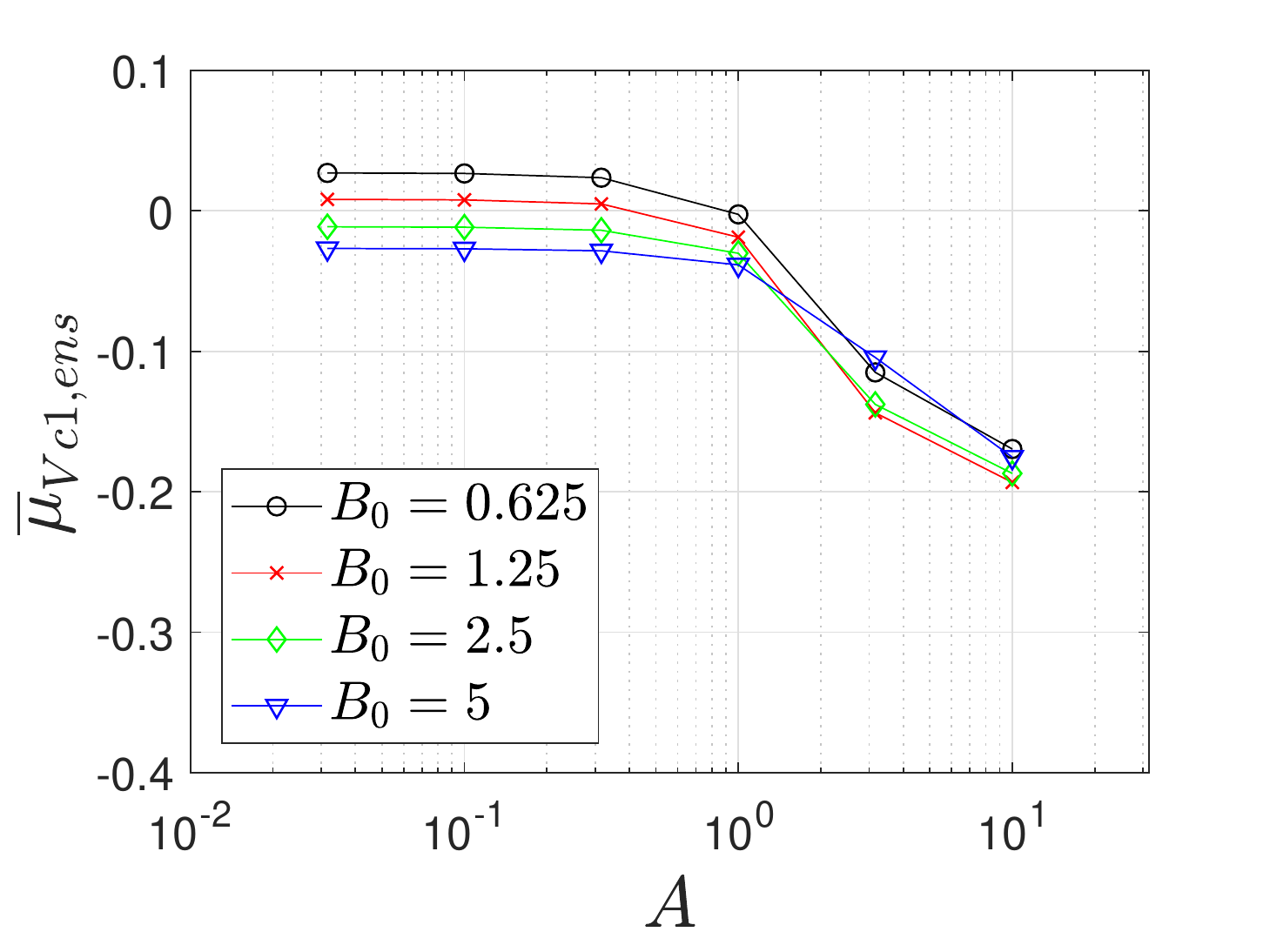}
  \caption{Correlations of the time averaged moment of volume of ensemble averaged clouds and the normalized amplitude of the incident wave.}
   \label{fig:p_xvc} 
\end{figure}
We now consider the effect of varying the excitation amplitude, $A$.
In order to simplify the discussion, we concentrate on the dynamics during the excitation phase where $2 < t^* < 9$, and time-average (denoted by $\overline{(\cdot)}$) the corresponding moments.
Figure \ref{fig:p_xvc} shows the time averaged moment of volume plotted against the normalized amplitude of the incident wave.
Regardless of $B_0$, the moment of volume is small and nearly constant with $A$ up to around $A=1$. For $A >1$ the moment decreases, indicating larger anisotropy.  Thus anisotropy is observed with high amplitude excitation.

\begin{figure}
  \center
  \subfloat[]{\includegraphics[width=70mm,trim=95 200 100 200, clip]{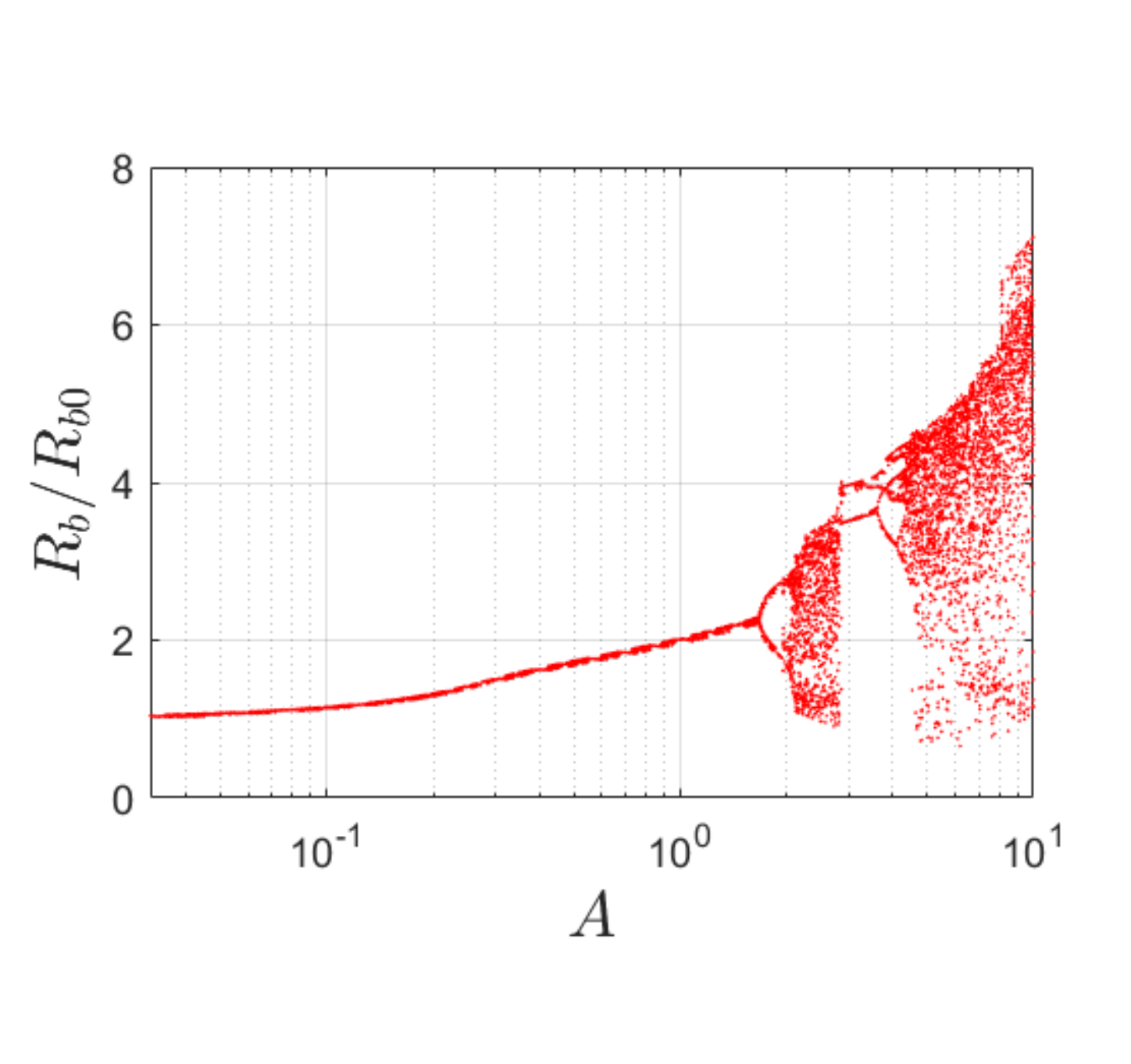}}
  \subfloat[]{\includegraphics[width=70mm,trim=95 200 100 200, clip]{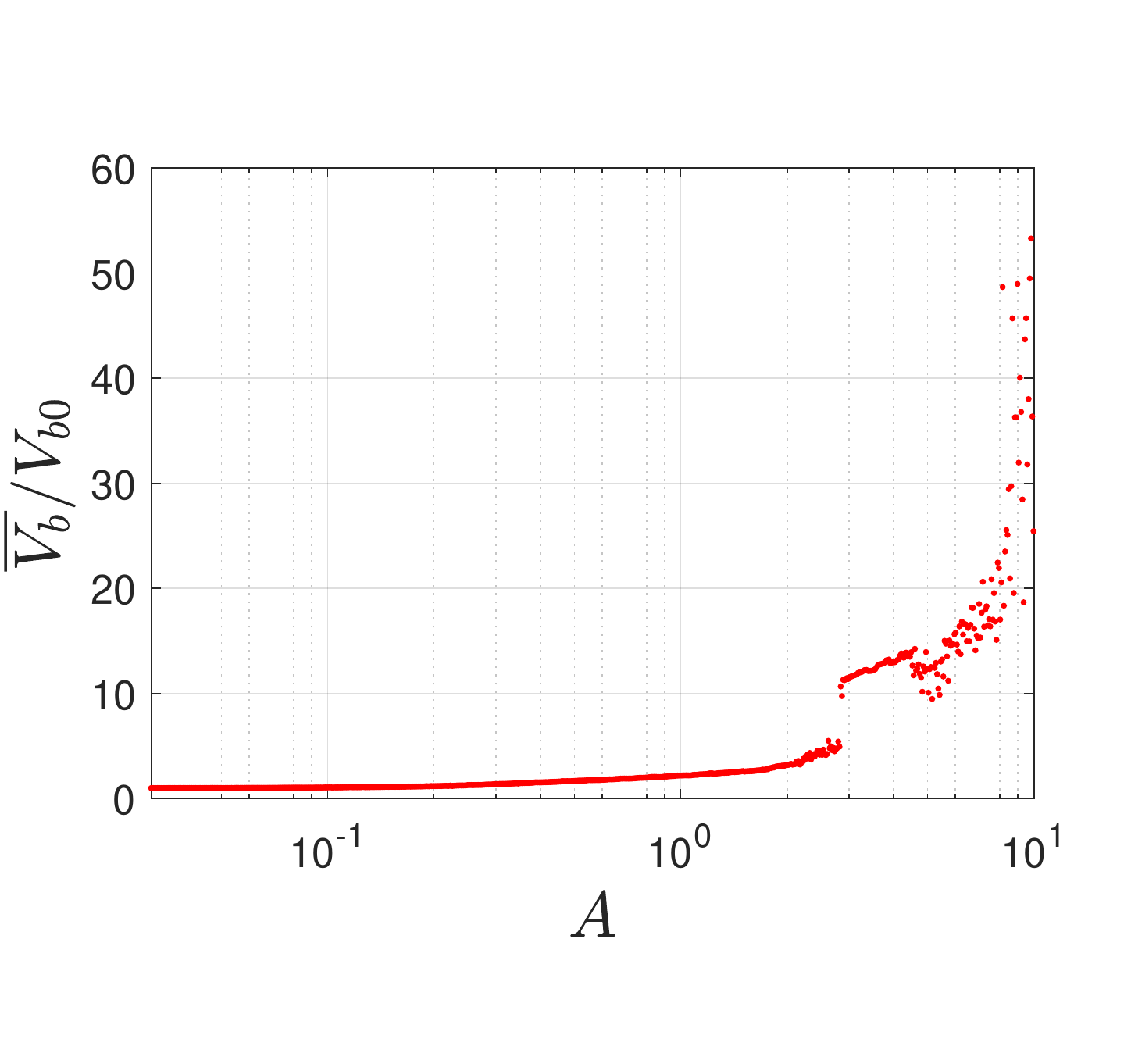}}
  \caption{(a) Bifurcation diagram of the bubble radius and (b) averaged volume of a single, spherical bubble under periodic pressure excitation with varying amplitude.
   }
   \label{fig:bif}
\end{figure}
To understand this dependency of the structure on the pressure amplitude, we use the Keller-Miksis equation to examine the nonlinear response of a single, isolated spherical bubble with an initial diameter of 10 $\mu$m under periodic far-field pressure excitation with a frequency of 300 kHz.
Figure \ref{fig:bif} (a) shows a bifurcation diagram of the radius of the bubble sampled at every period of forcing pressure with a slowly increasing forcing amplitude within the range addressed in the parametric study.
The computed radius monotonically grows with $A$ and experiences a sub-harmonic bifurcation at $A\approx1.65$, then transits to a chaotic regime with a growing amplitude of radius. The bifurcation diagram in this range of the excitation amplitude was also reported by \citet{Preston07}.
At $A\approx2.85$, the radius returns to a quasi-periodic behavior, then at $A\approx4$ it re-transits to a chaotic regime with an amplitude growing with $A$.

Figure \ref{fig:bif} (b) shows the time averaged volume of the same bubble during the period of forcing.
The growth of the averaged volume follows a similar trend to that of the radius, but with a larger slope.
The volume smoothly grows to $\overline{V}_b/V_{b0}\approx5$ with $A$ then discontinuously grows to $\overline{V}_b/V_{b0}\approx12$ at $A\approx2.85$.
Then it grows with much faster rate with $A$, toward $\overline{V}_b/V_{b0}\approx50$ at $A=10$.
The nonlinear growth of the volume with $A>1$ corresponds to cavitation.

\begin{figure}
  \center
  \subfloat[]{\includegraphics[width=70mm,trim=0 0 0 0, clip]{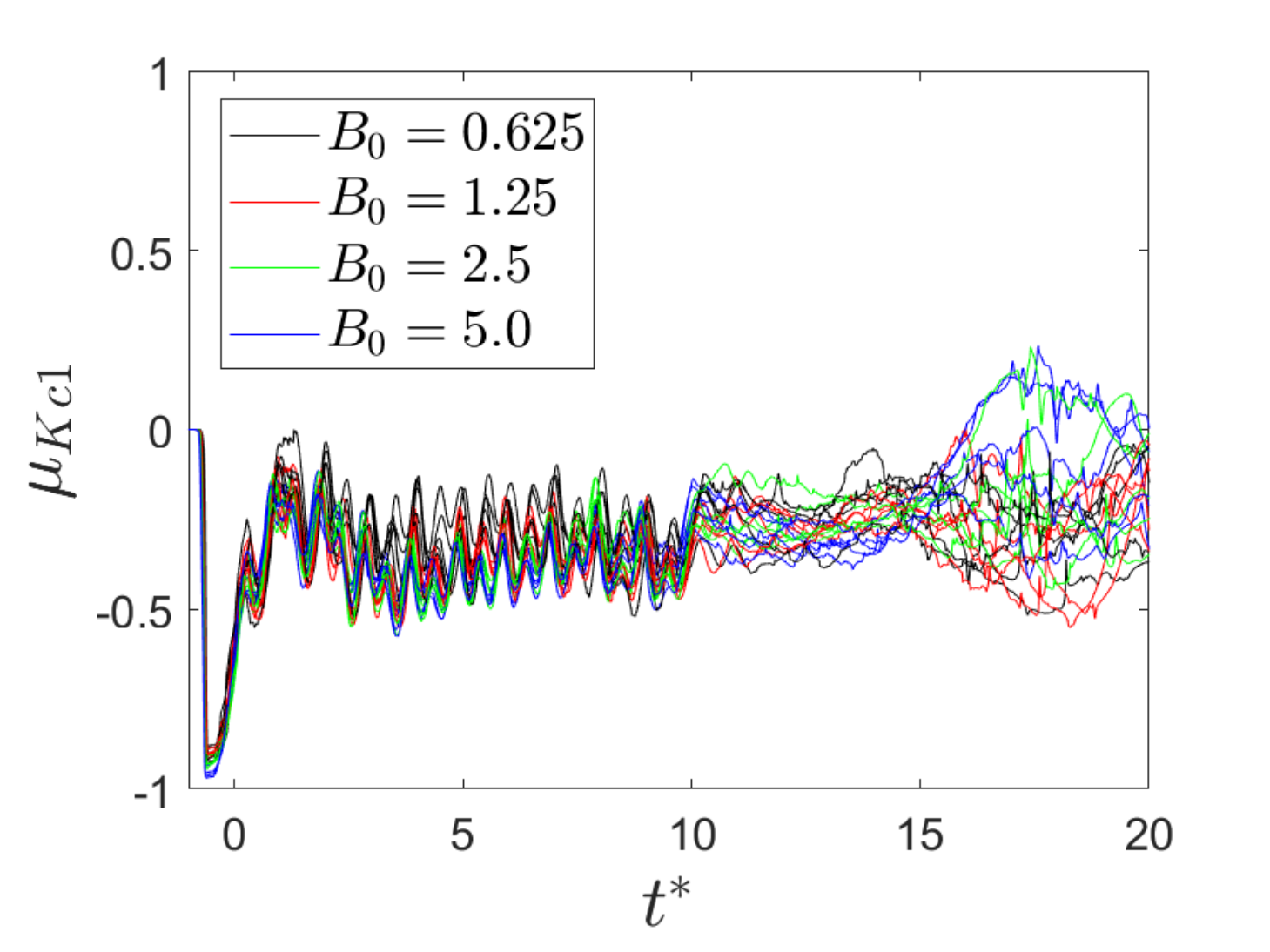}}
  \subfloat[]{\includegraphics[width=70mm,trim=0 0 0 0, clip]{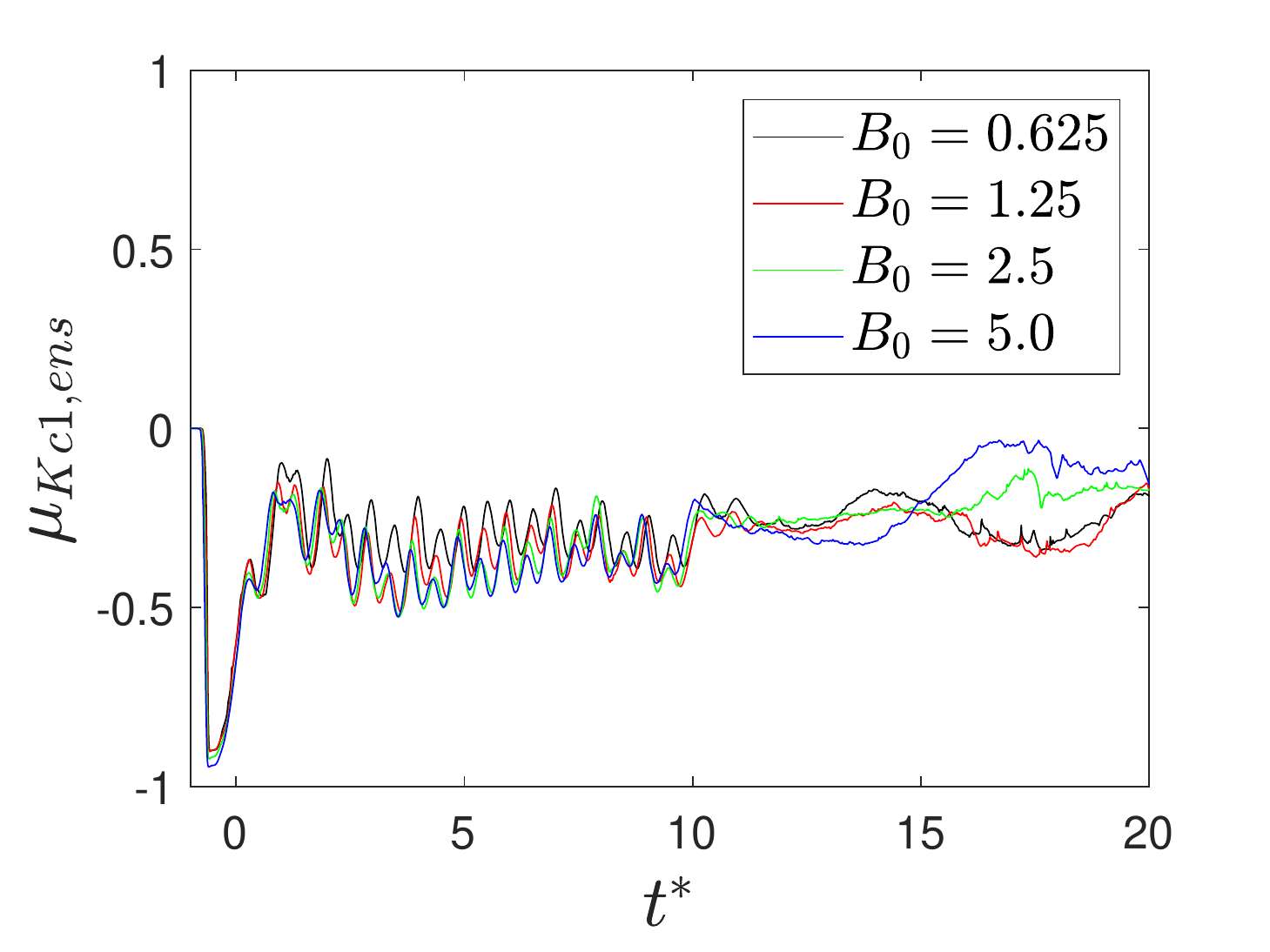}}
  \caption{Evolution of the moment of kinetic energy for (a) all the clouds and (b) ensemble averaged clouds.}
   \label{fig:xk_cent}
\end{figure}
Figure \ref{fig:xk_cent} (a) shows the evolution of the moment of kinetic energy of clouds A6v. After the initial transient until $t^*=10$, the moments of kinetic energy oscillate between -0.5 and 0 around an approximately constant level, after which oscillations with larger amplitudes occur. Figure \ref{fig:xk_cent} (b) shows the same quantities of the ensemble-averaged clouds. The result is similar to figure \ref{fig:xk_cent} (a), further confirming that the trend of the moment results from the coherent dynamics of the cloud.
The plots indicate that the oscillations of proximal bubbles are more energetic than the distal bubbles during the course of excitation, regardless of the initial nuclei population.
Since the moment of volume and the moment of kinetic energy reach quasi-stationary states during the 10 cycles of pressure excitation, increasing the number of cycles of the pressure excitation may not largely affect the structure of the clouds.
When the number of cycle is as small as 1, however, the bubble dynamics do not reach the stationary state, as shown in $t^*\in[0,1]$ in figure \ref{fig:xk_cent}, and the structure may not be observed.

The results of single bubble dynamics in figure \ref{fig:bif} and the moment of energy in figure \ref{fig:xk_cent} may explain the mechanism of the anisotropic structure.
The bubbles nearest the source are exposed to an incoming pressure wave, while the distal bubbles experience smaller amplitudes of pressure fluctuations due to the scattering of the wave by the proximal bubbles. This results in larger amplitude of oscillations of bubbles locally in the region near the proximal surface of the cloud, seen as the bias in the moment of kinetic energy.
With a pressure amplitude larger than $A>1$, the proximal bubbles can grow to much larger radius than the distal bubbles due to local cavitation, which results in the bias in the center of volume, and becomes visible as the anisotropic structure.

\subsection{Dynamic cloud interaction parameter}
\label{section:dynamicB}
In order to further quantify the bubble cloud dynamics, we seek to generalize the definition of the cloud interaction parameter introduced by DB.
The critical difference in the bubble cloud dynamics in the present study and those considered by DB lies in the wavelength and the amplitude of the pressure excitation.
As discussed in $\S$ \ref{section:metrics}, the original interaction parameter can be interpreted as a scaling parameter of the global kinetic energy of liquid induced by a small amplitude oscillations of bubble cluster under weak pressure excitation with long wavelength.
Meanwhile, in the bubble clouds considered in the present study, the wavelength is as small as the size of a cloud.
Due to the strong amplitude of the pressure, bubbles experience cavitation growth and their radii can deviate from their initial values.
The radius of bubbles can also vary in space.

\begin{figure}
  \center
  \includegraphics[width=70mm,trim=0 0 0 0, clip]{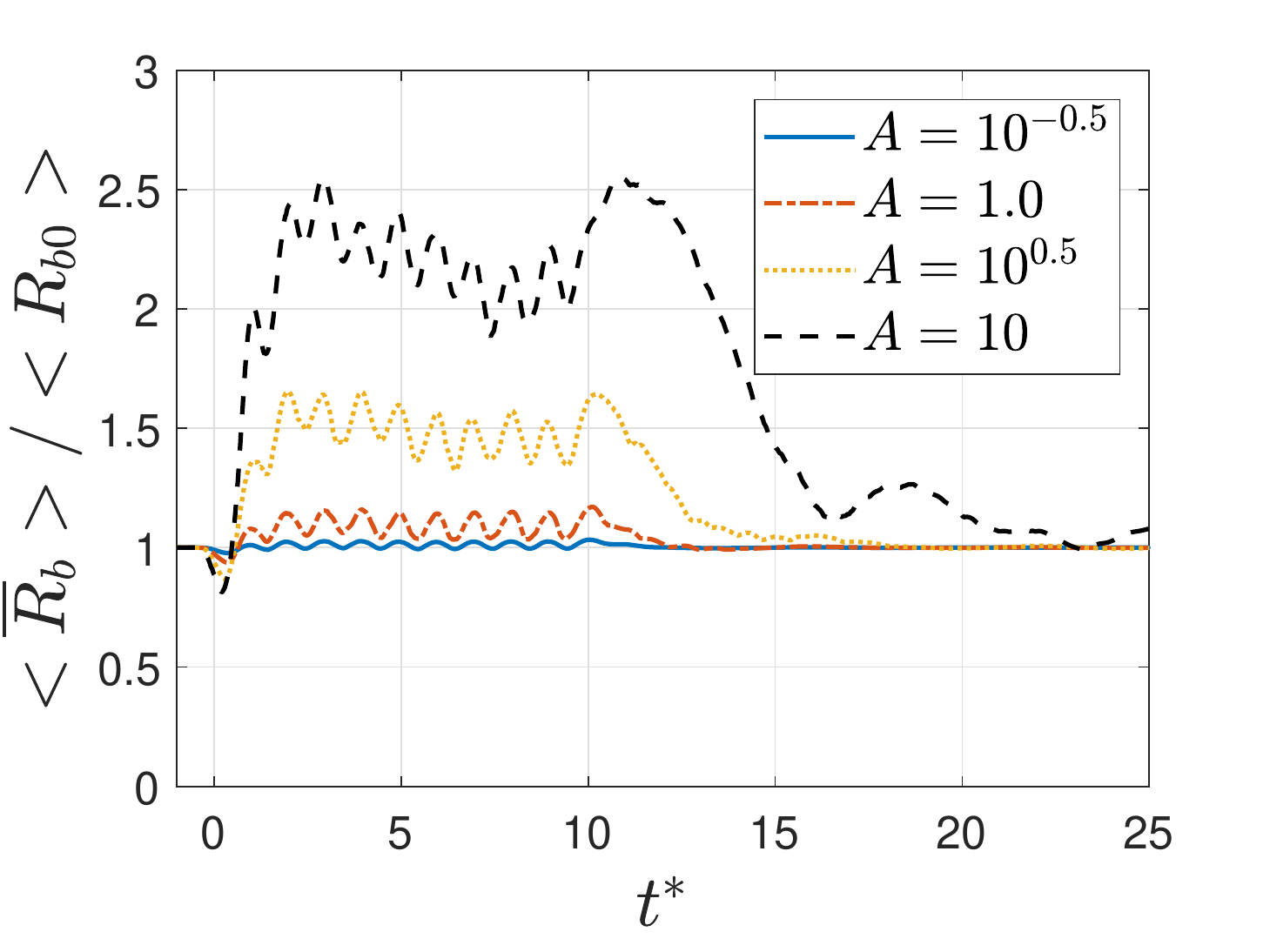}
  \caption{Evolution of the mean radius of bubbles in the cloud.}
   \label{fig:meanRv2} 
\end{figure}
Figure \ref{fig:meanRv2} shows the evolutions of the spatial mean of the radius of bubbles in the cloud with $B_0=1.25$, normalized by its initial value, with various values of the excitation amplitude: $A=[10^{-0.5},1.0,10^{0.5},10]$.
For $A>1$, the mean radius grows rapidly on arrival of the wave, then oscillates around an approximately constant value larger than 1 until $t^*=10$, while with $A<1$ the mean radius oscillates around 1.
After $t^*>10$ the radius decays to the initial value in all cases.
This indicates that the spatial mean of the bubble radius oscillates around their quasi-stationary equilibrium whose value is unique to the pressure amplitude during the course of excitation.

Motivated by this result, we extend the definition of the cloud interaction parameter as
\begin{equation}
B=\frac{N_b<\overline{R}_b>}{R_{c,L}},
\end{equation}
where
$\overline{R}_{b}$ is the time averaged radius of bubble during the pressure excitation.
Hereafter we denote this parameter as {\it{dynamic cloud interaction parameter}}.
A detailed discussion motivating the specific form of $B$ is given in appendix B.

\begin{figure}
  \center
  \includegraphics[width=70mm,trim=0 0 0 0, clip]{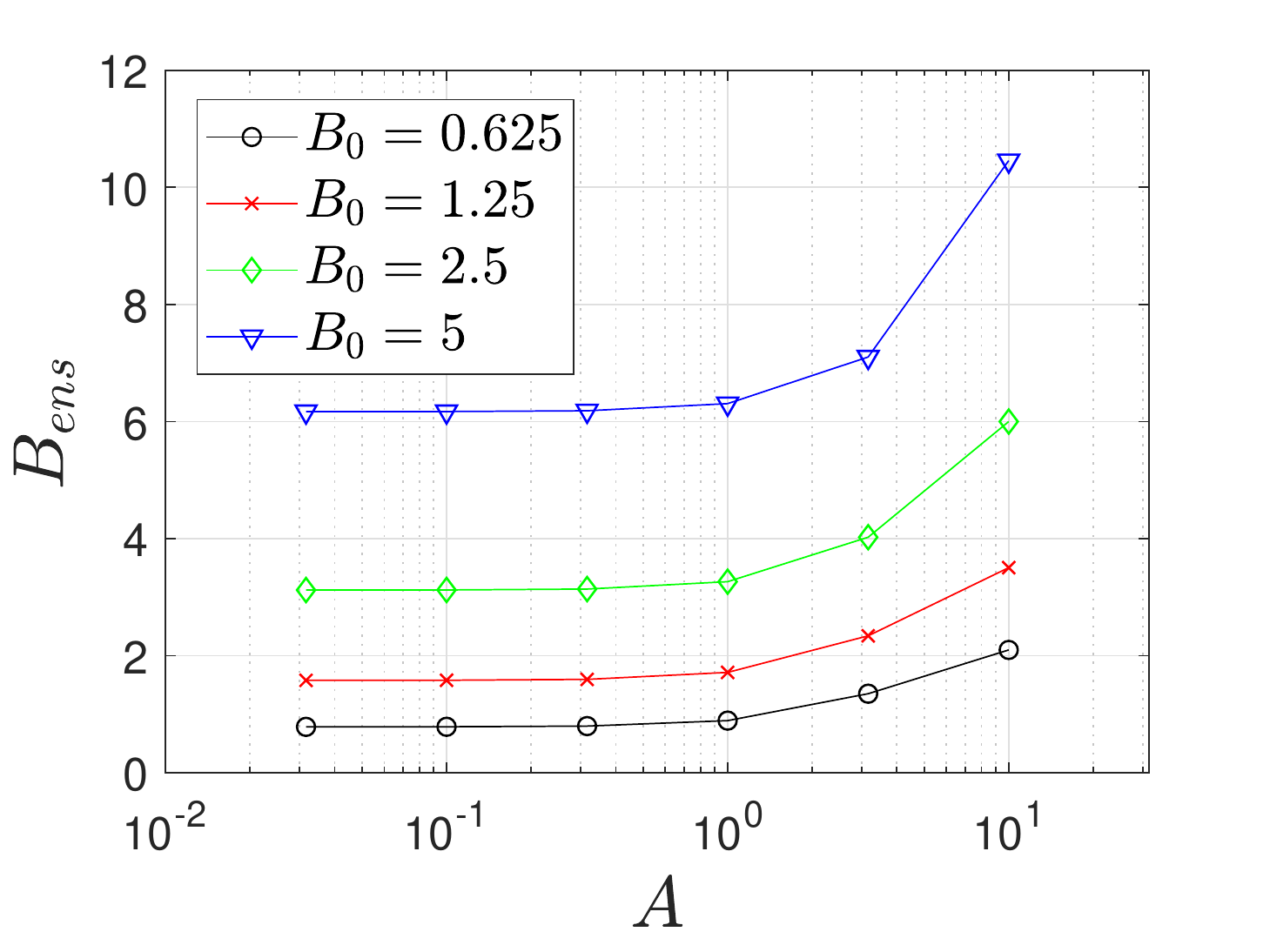}
  \caption{Dynamic interaction parameter plotted against the amplitude of pressure excitation for various values of $B_0$.}
   \label{fig:P_B} 
\end{figure}
In figure \ref{fig:P_B}, we plot the average value of $B$ (over the ensemble) obtained for all the runs in Table 1.  
For all the values of $B_0$, $B$ monotonically grows and deviates from $B_0$ for $A>1$.
This is due to deviation of $<\overline{R}_b>$ from $R_{b,ref}$ and thus can be associated with the cavitation growth of the mean bubble radius in the cloud shown in figure \ref{fig:P_B} with $A>1$.

\subsection{Scaling of the moment of kinetic energy}
\label{section:moment}
The dynamic interaction parameter is proposed as an appropriate scaling parameter for bubble cloud dynamics excited in the short wavelength regime.
In this section, to examine the extent to which $B$ controls the dynamics, we correlate the moment of kinetic energy against both $B$ and the original cloud interaction parameter, $B_0$, and compare the results.

\begin{figure}
  \center
  \subfloat[]{\includegraphics[width=70mm,trim=0 0 0 0, clip]{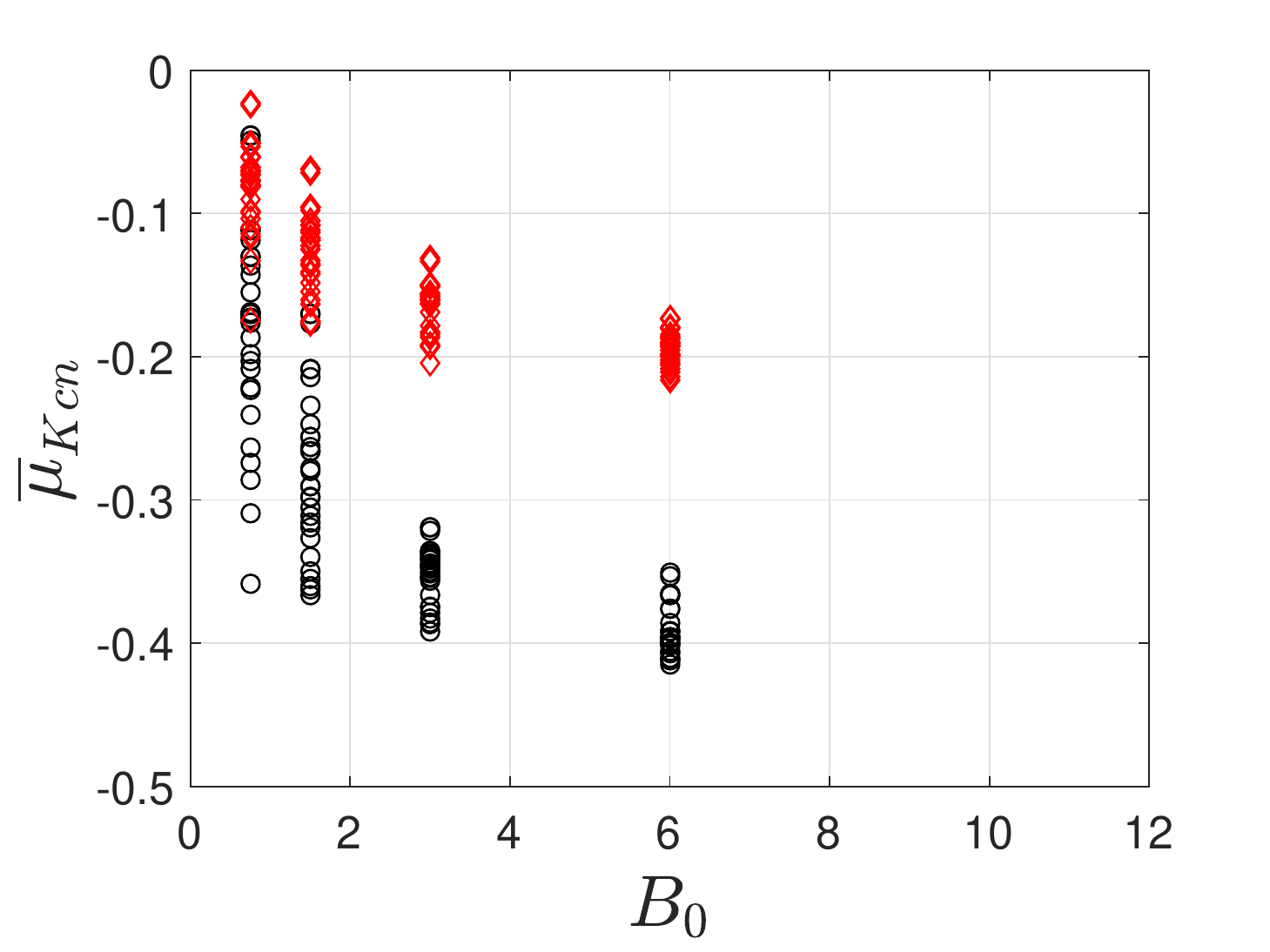}}
  \subfloat[]{\includegraphics[width=70mm,trim=0 0 0 0, clip]{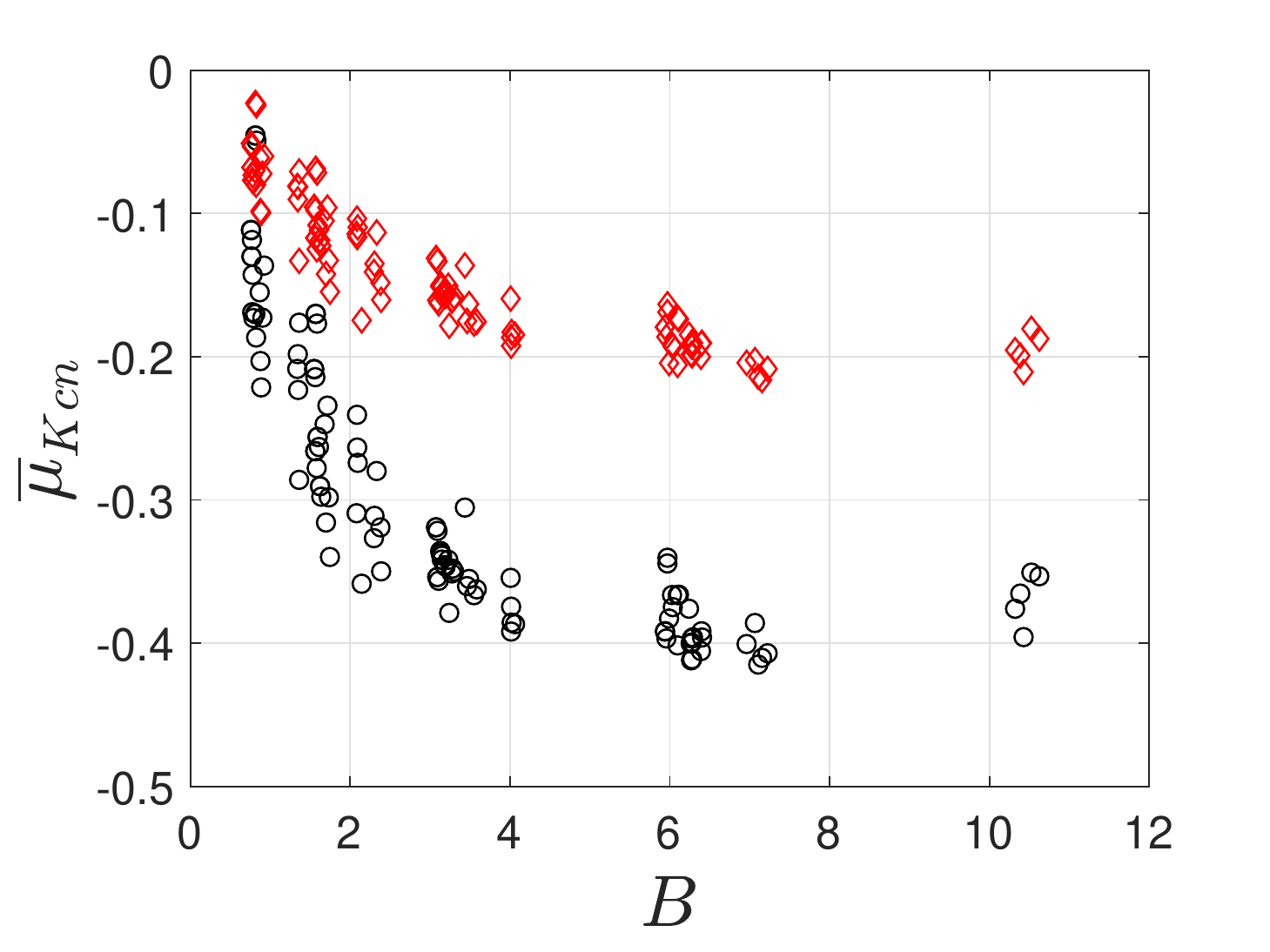}}\\
  \subfloat[]{\includegraphics[width=70mm,trim=0 0 0 0, clip]{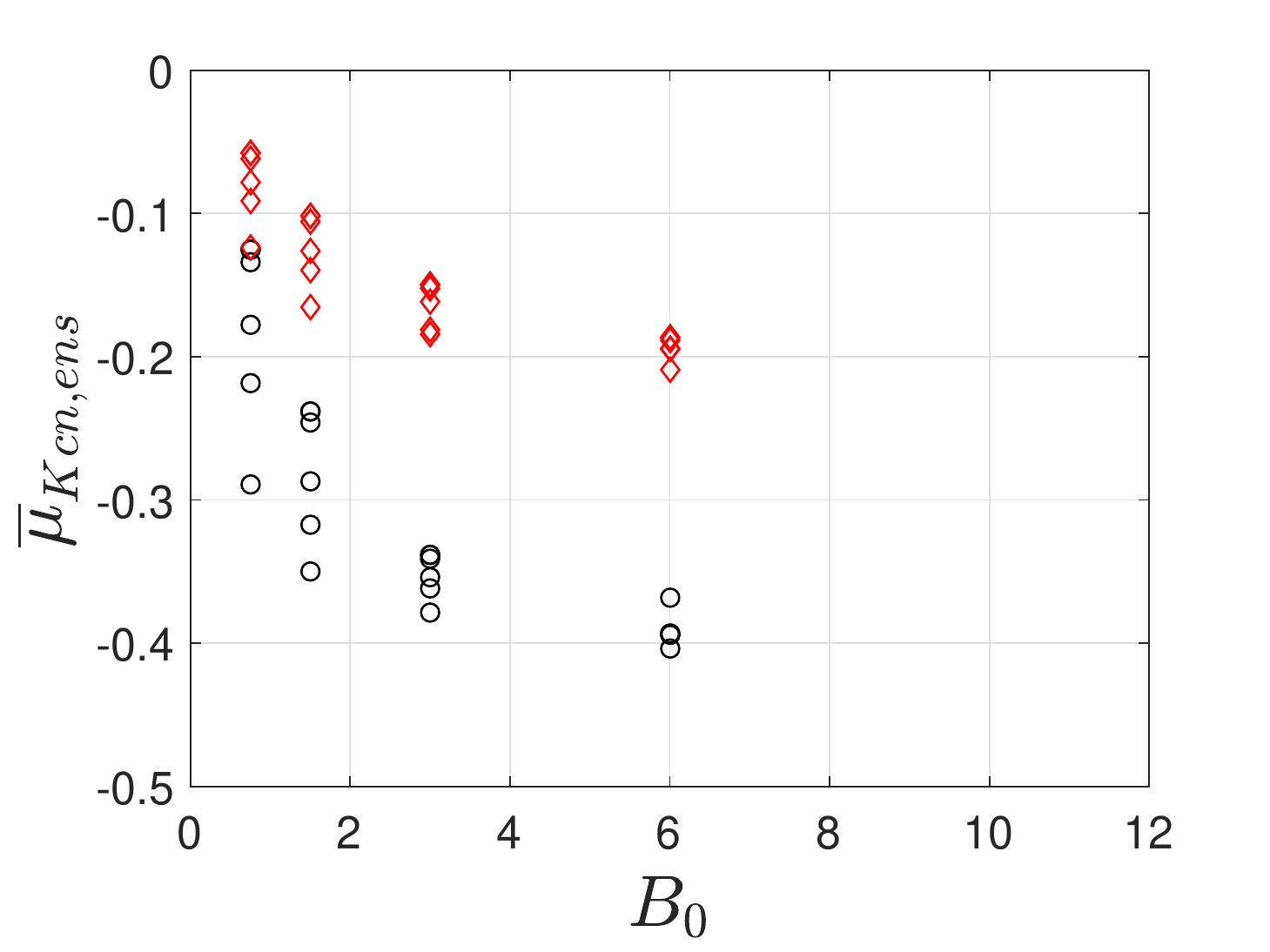}}
  \subfloat[]{\includegraphics[width=70mm,trim=0 0 0 0, clip]{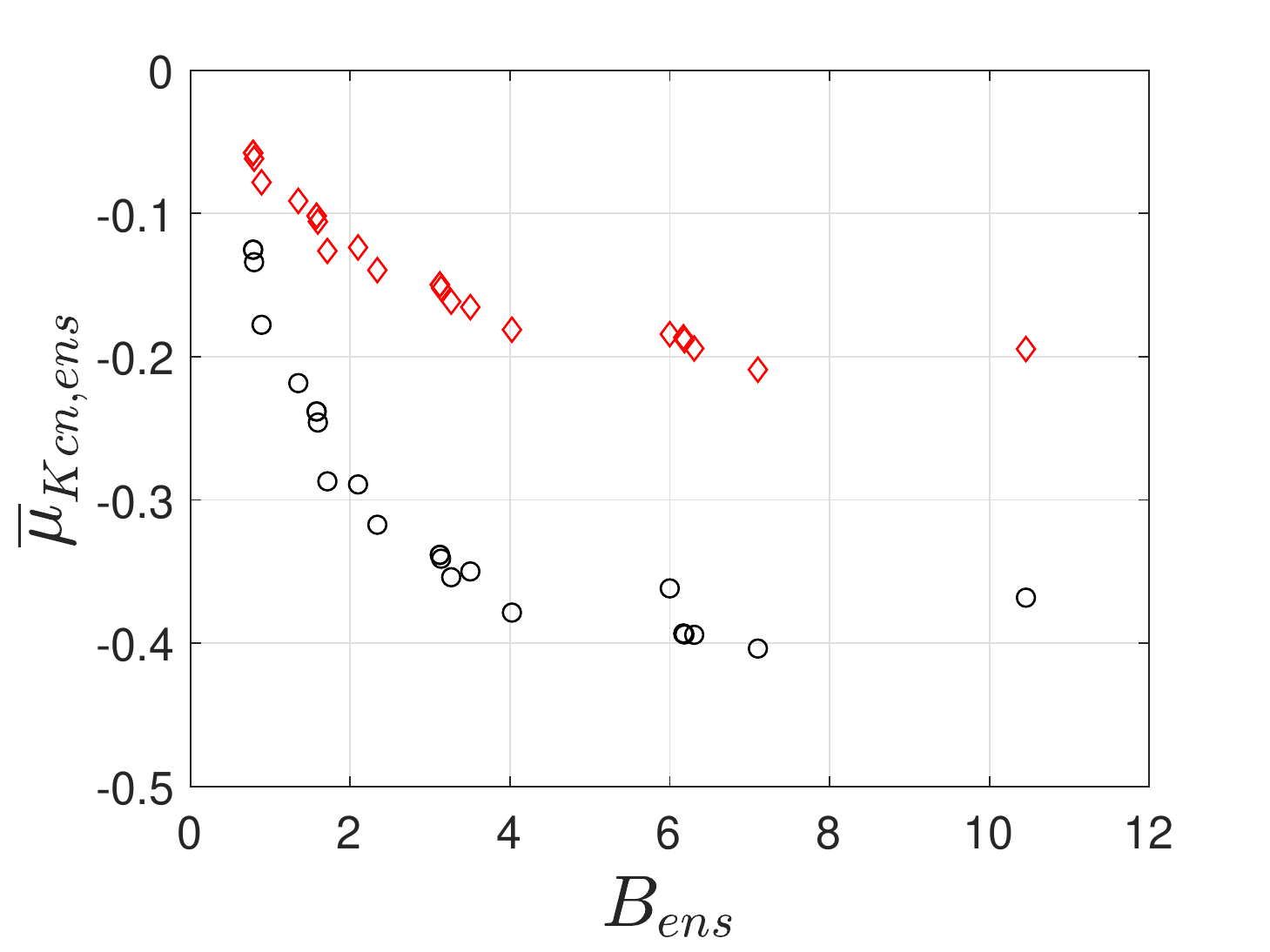}}
  \caption{Scattered plots of the time averaged moments of kinetic energy against $B_0$ and $B$. The top row show all realizations while the bottom row shows the ensemble-averaged values. Circle and diamond denote the first and the third moments, respectively.
   }
   \label{fig:B_xkc}
\end{figure}
Figure \ref{fig:B_xkc}(a) and (b) show scatter plots of the time-averaged, first and third moments of kinetic energy against $B_0$ and $B$, respectively.
The 1st and 3rd moments show negative correlations against both $B_0$ and $B$, while the data points are vertically more spread against $B_0$ than against $B$.
The dynamic parameter does a somewhat better job of collapsing the dynamics than the original one.
However, once we ensemble-average the data in figure 16 (c) and (d), we see that much of the variation is associated with the randomized positions of the bubbles, and, in general, the dynamic interaction parameter collapses the moment of kinetic energy of the clouds.
This confirms that the moments of kinetic energy can been seen as monotonic, decreasing functions of $B$.
Overall, the results indicate that $B$ is more appropriate parameter to scale the moments than $B_0$.
The similarity of figure \ref{fig:B_xkc} (b) and (d) indicates small variability of the spatial bias in the energy due to initial spatial distribution of bubbles.

The moments in figure \ref{fig:B_xkc} (d) approach zero for small $B$, which confirms that in the limit of $B=0$ inter-bubble interactions are negligible and the resulting spatial bias in the mean kinetic energy is statistically zero, since bubbles experience the same amplitude of pressure excitation at any location in the cloud.
The plots also indicate that as $B$ increases, the slope of the curve monotonically decreases and thus the moment saturates. 
This indicates that the distribution of energetic bubbles in the cloud becomes more localized in the proximal side of the cloud with increasing the pressure amplitude, while the magnitude of energy localization eventually becomes invariant to the amplitude.

Overall, the results of the parametric simulation further elucidate the underlying mechanism of the anisotropic structure.
When the inter-bubble interaction becomes dominant, the energy localization occurs to the cloud, and this happens regardless of the amplitude of pressure excitation. Meanwhile, the anisotropic structure becomes visible only when the energetic, proximal bubbles cavitate and reach a large radius is a nonlinear function of the amplitude of pressure excitation.
It is notable that, in fact, the moment of volume is not collapsed by the dynamic cloud interaction parameter.
\begin{figure}
  \center
  \subfloat[]{\includegraphics[width=70mm,trim=0 0 0 0, clip]{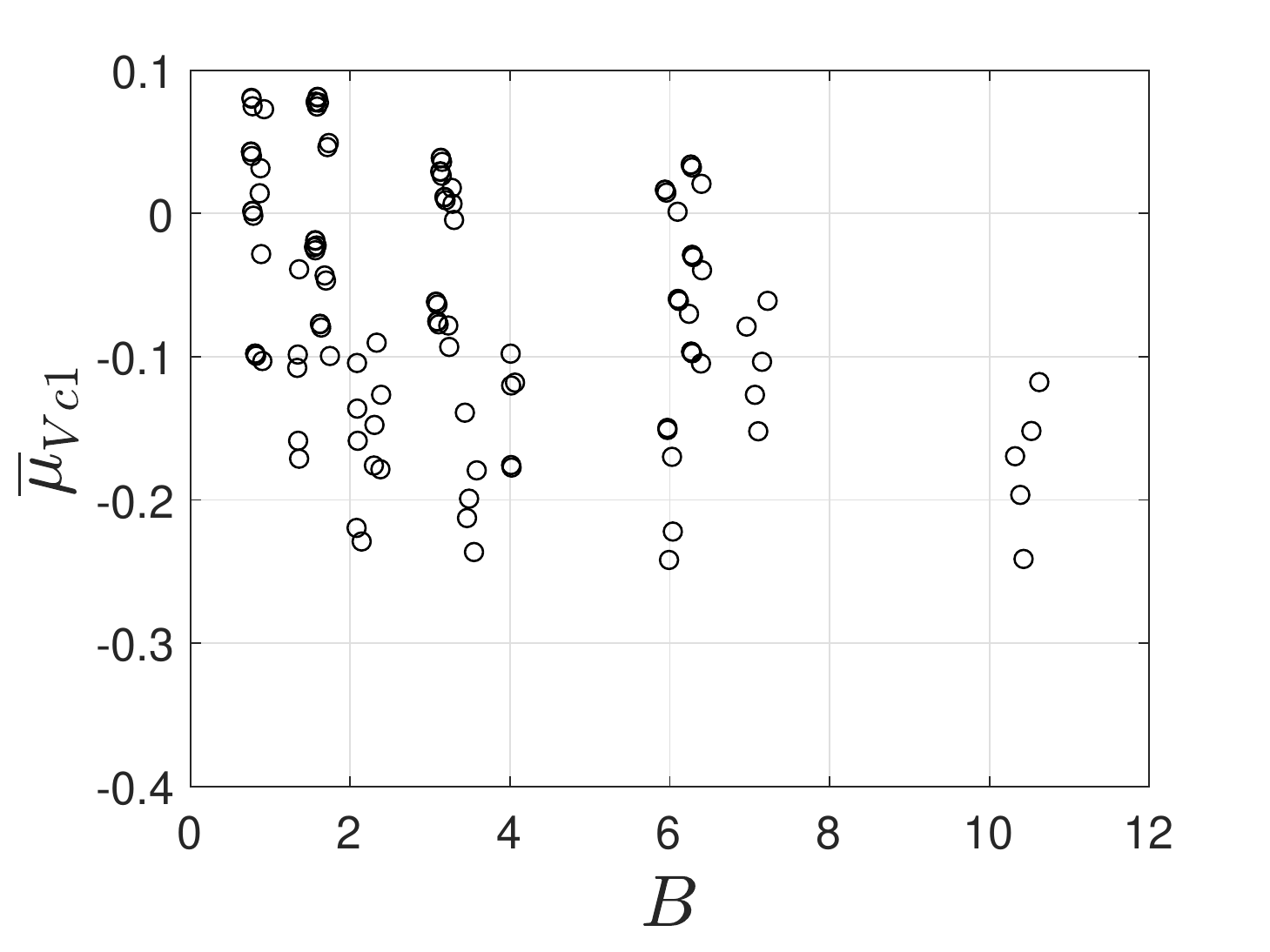}}
  \subfloat[]{\includegraphics[width=70mm,trim=0 0 0 0, clip]{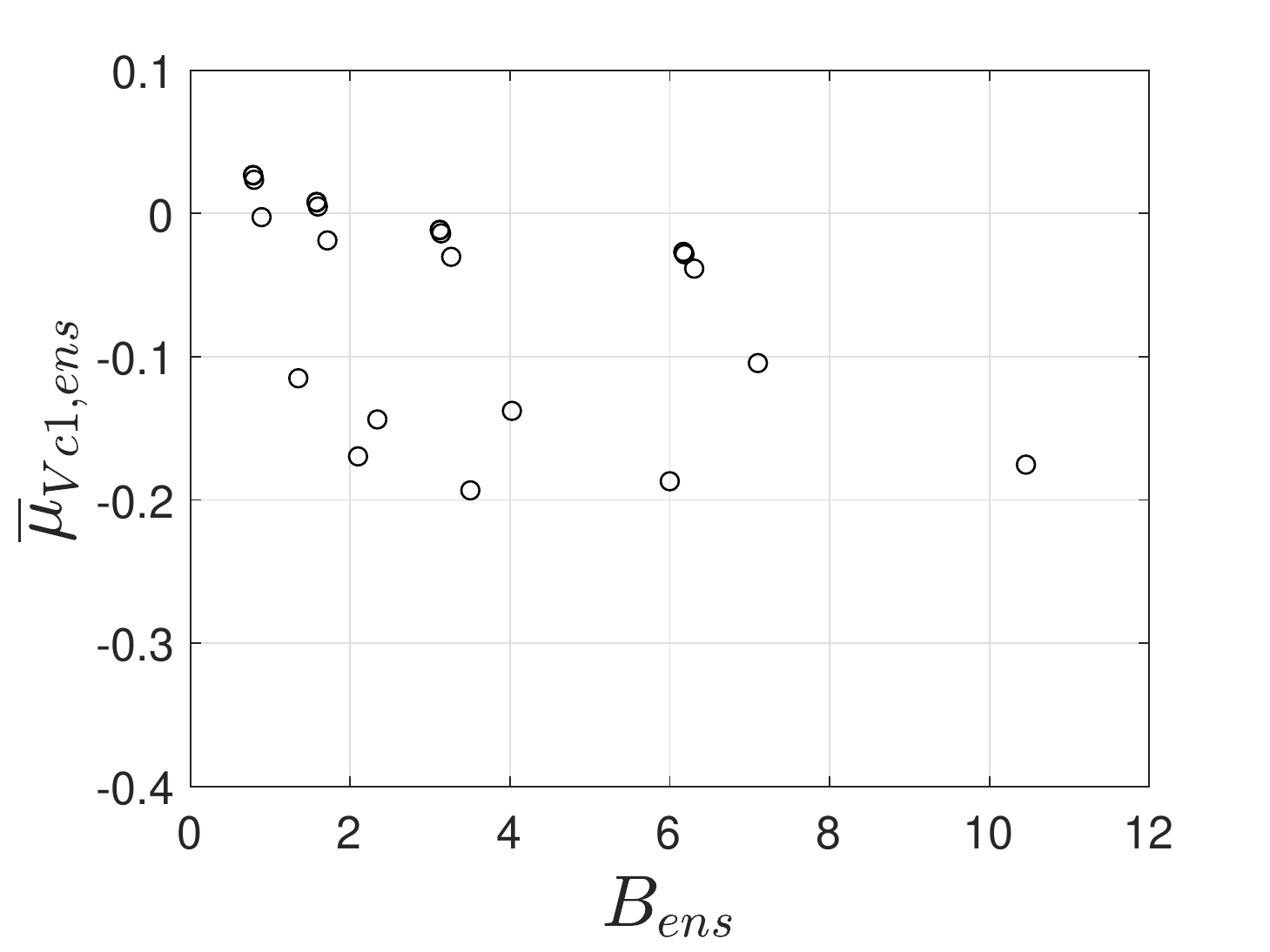}}
  \caption{Scattered plots of the time averaged moments of volume against $B$. (a) shows all realizations while (b) shows the ensemble-averaged values.
   }
   \label{fig:B_xvc}
\end{figure}
Figure \ref{fig:B_xvc} shows scatter plots of the time-averaged moment of volume against the dynamic cloud interaction parameter. For the entire range of $B$, the moment is scattered between -0.3 and 0.1 for both all realizations and the ensemble-averaged values.

\subsection{Scaling of the far-field, bubble-scattered acoustics}
\label{section:acoustics}
Given the successful scaling of the moments of kinetic energy in terms of the dynamic interaction parameter, we are motivated to explore scaling of the far-field, bubble-scattered acoustics that result from the bubble cloud dynamics.

\begin{figure}
  \center
  \subfloat[]{\includegraphics[width=70mm,trim=0 0 0 0, clip]{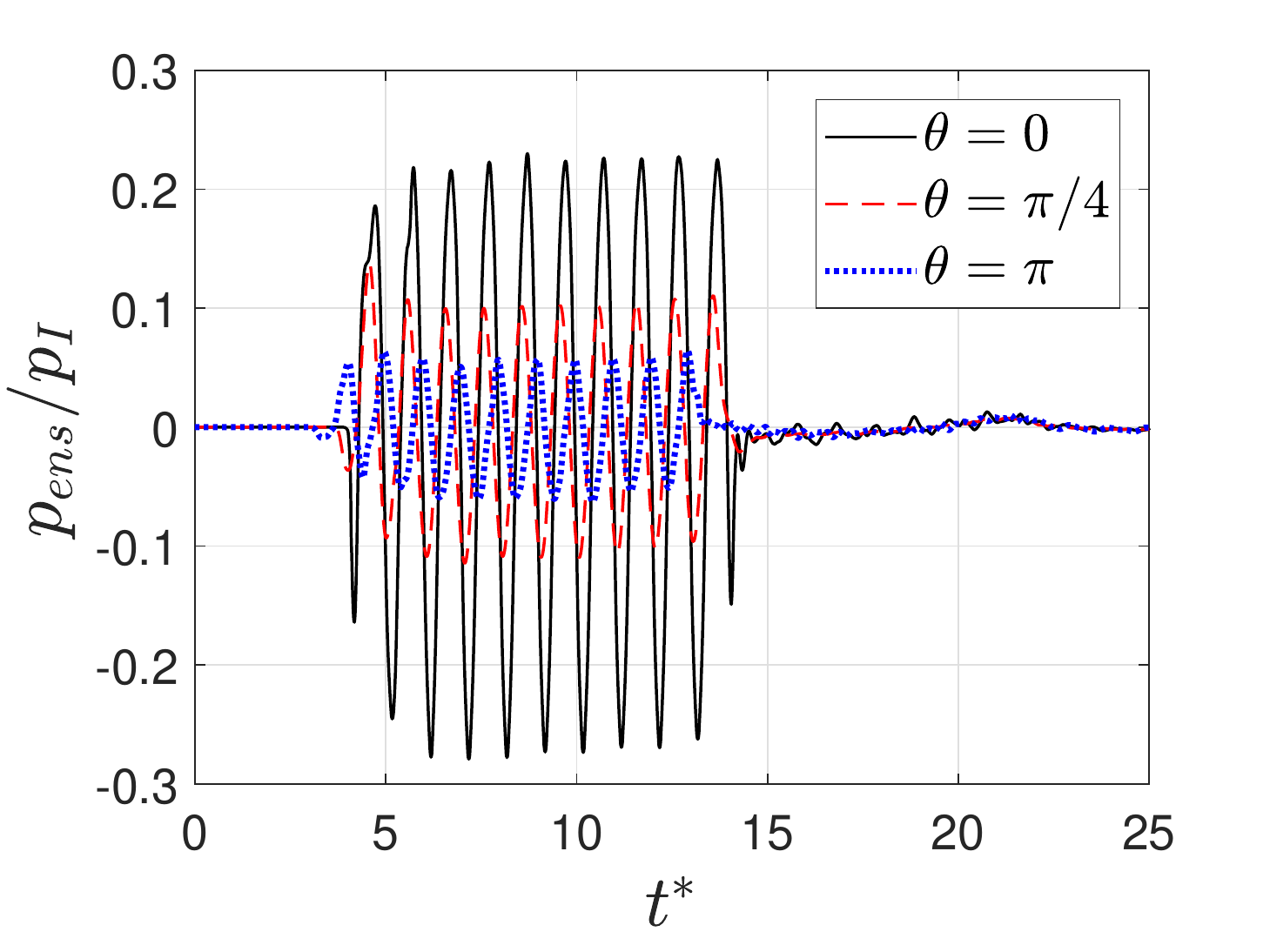}}
  \subfloat[]{\includegraphics[width=70mm,trim=0 0 0 0, clip]{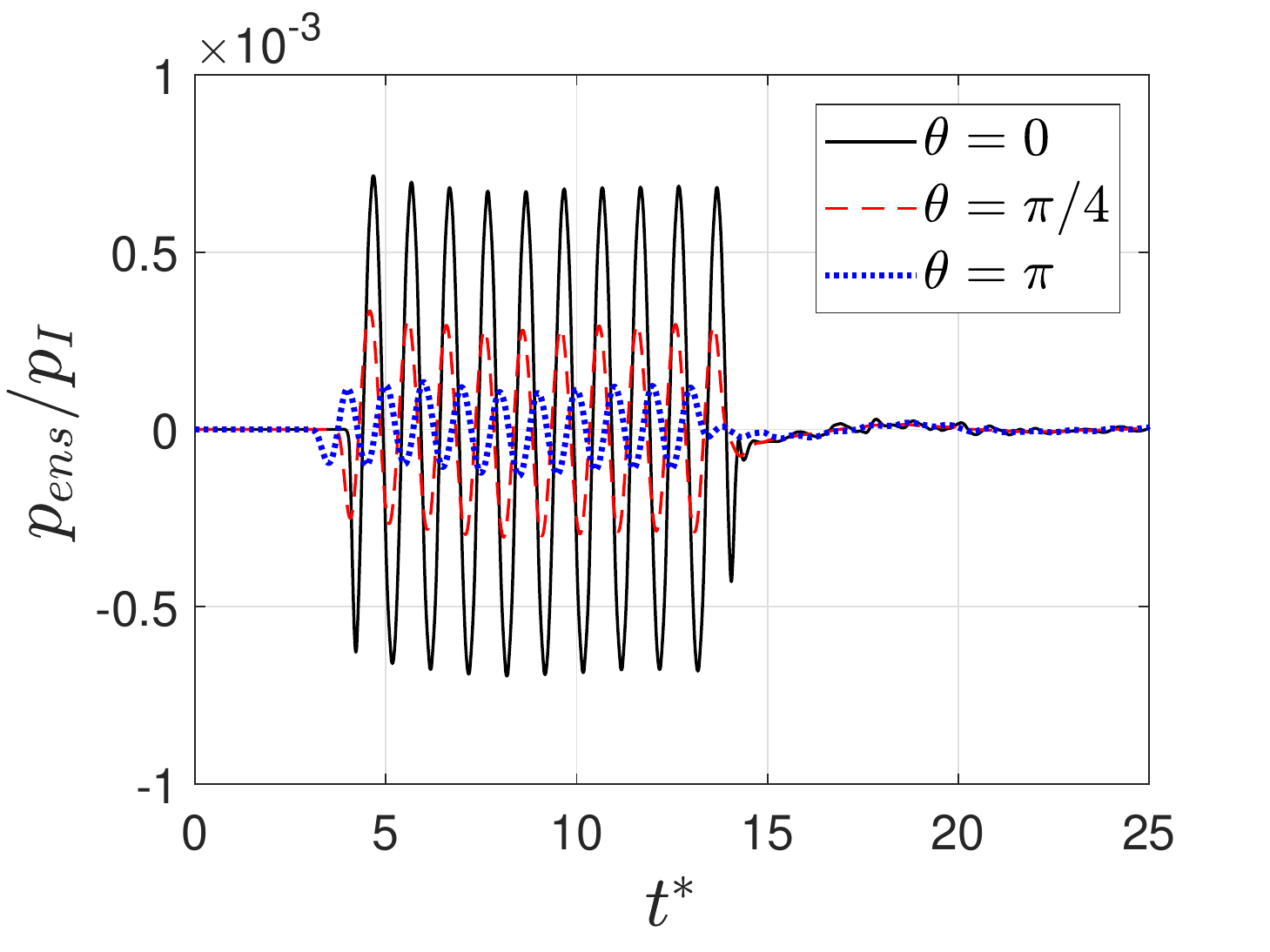}}
  \caption{Evolution of the scattered pressure field at a distance $r=8R_c$ from ensemble averaged clouds from runs (a) A6v4 and (b) A1v1.
  }\label{fig:t_p} 
\end{figure}
Figure \ref{fig:t_p} shows the evolution of the far-field sound at different angles to the direction of incident radiation.  The pressure has been normalized by the amplitude of the incident wave.  These are plotted for two cases: figure \ref{fig:t_p} (a) shows the densest cloud excited by the highest amplitude wave (thus obtaining the largest value of $B$), while figure \ref{fig:t_p} (b) shows the most dilute cloud with the lowest amplitude of excitation (lowest value of $B$). 
The scattered pressure shows sinusoidal oscillations at a retarded time associated with the incident wave scattered to the sampling location.
The amplitude of the scattered pressure from the dense cloud is an order of magnitude larger than than the dilute one for all $t^*$.
The small, rapid fluctuations at large $t^*$ are due to the bubble oscillations after the passage of the incident wave. The small amplitude of these fluctuations indicate the absence of a strong, coherent cloud collapse.

\begin{figure}
  \center
  \subfloat[]{\includegraphics[width=80mm,trim=30 42 24 30, clip]{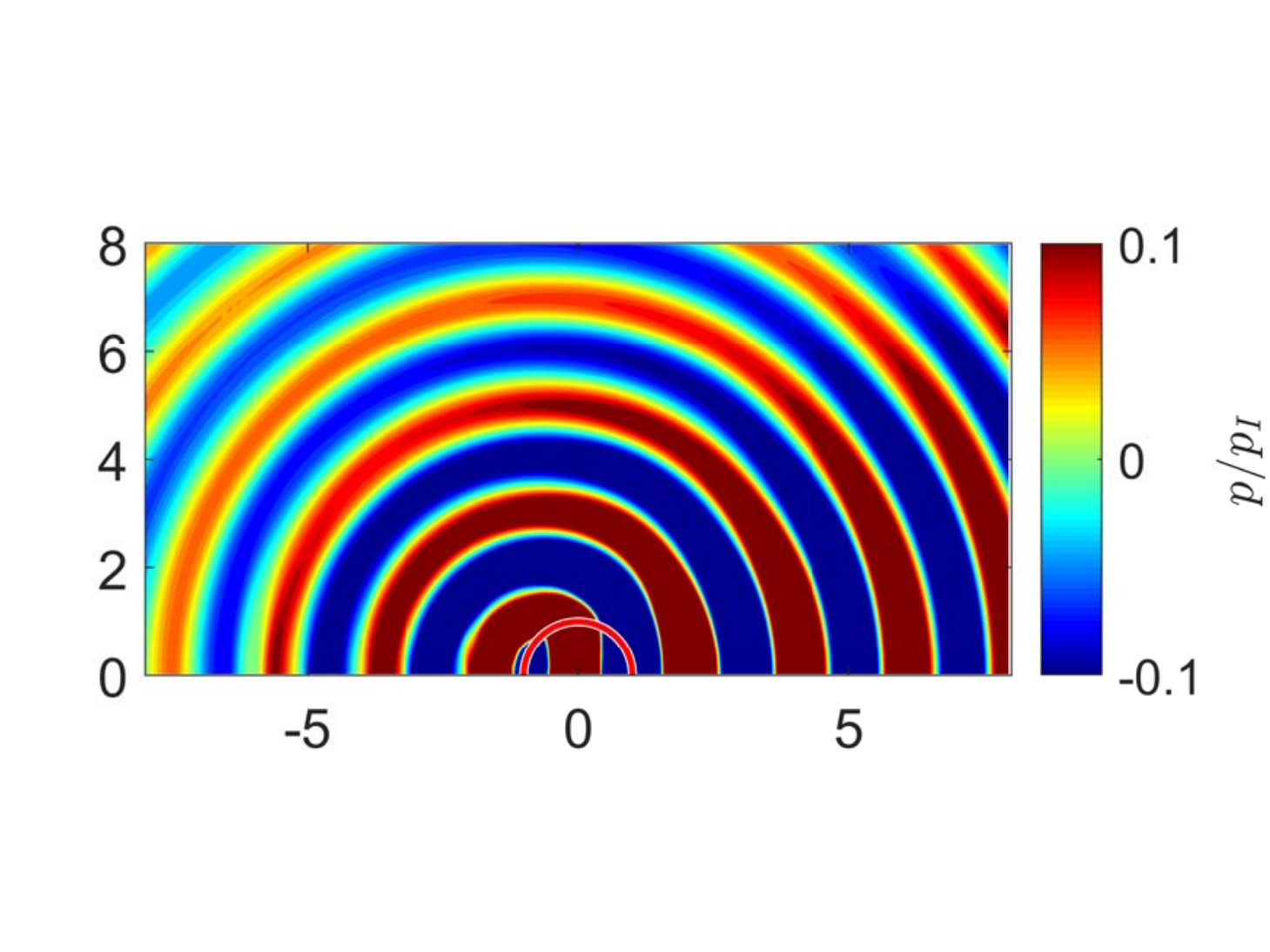}}
  \subfloat[]{\includegraphics[width=55mm,trim=-10 0 0 0, clip]{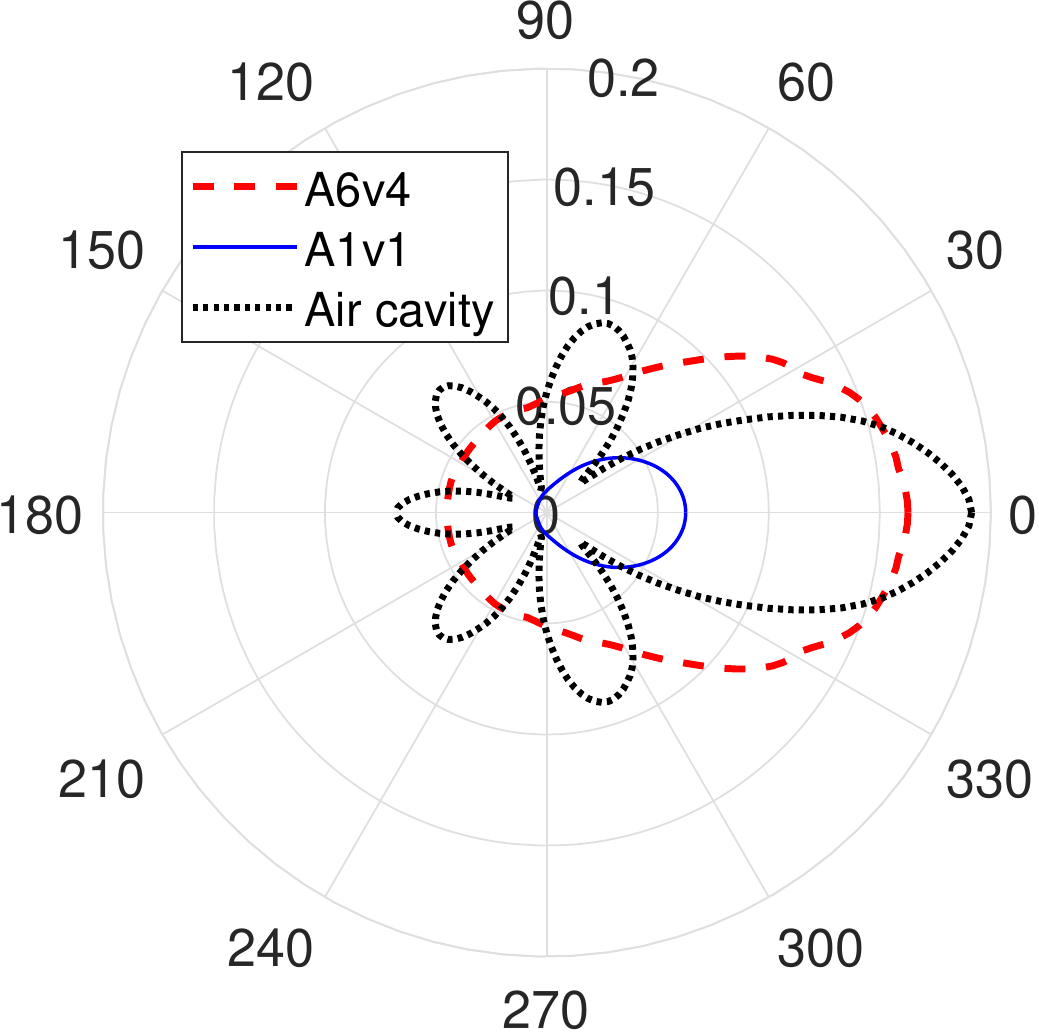}}
  \caption{(a) Contours of the scattered pressure field at $t^*=9.6$, normalized by the amplitude of the incident wave, obtained from run A6v4. The length unit is $R_c=2.5$ mm. (b) Polar plots of the root-mean-square pressure sampled on spherical surface with $r/R_c=8$. Results from runs A6v4 and A1v1 are compared to the scattering of a single spherical air bubble of the same size as the cloud.}
   \label{fig:pres_contour} 
\end{figure}
Figure \ref{fig:pres_contour} (a) shows an contour plot of the bubble-scattered component of the pressure field at $t^*=9.6$ from the dense cloud. The scattered component is obtained by subtracting the contribution of the incident pressure wave from the total pressure field. The scattered wave propagates radially outward from the bubble cloud.
Figure \ref{fig:pres_contour} (b) shows a polar plot of the scattered waves from both clouds averaged over the period of direct scattering.
The linear scattering from a single spherical air cavity with the same radius as the clouds is also shown for reference.
With both clouds, scattering is dominant over angles in the forward direction.
The amplitude of scattering is larger at all angles from the dense cloud than the dilute one.

\begin{figure}
  \center
  \subfloat[]{\includegraphics[width=70mm,trim=0 0 0 0, clip]{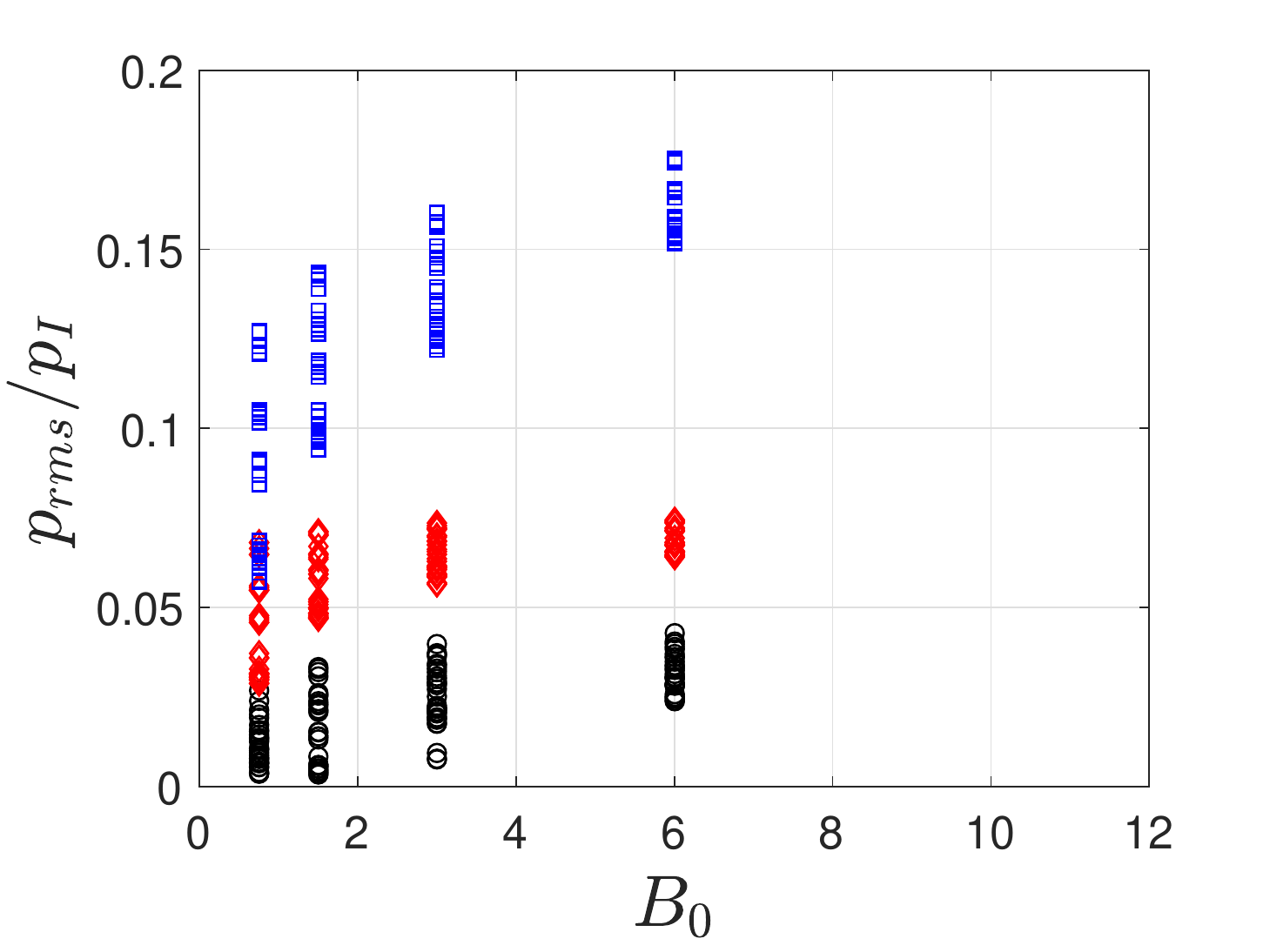}}
  \subfloat[]{\includegraphics[width=70mm,trim=0 0 0 0, clip]{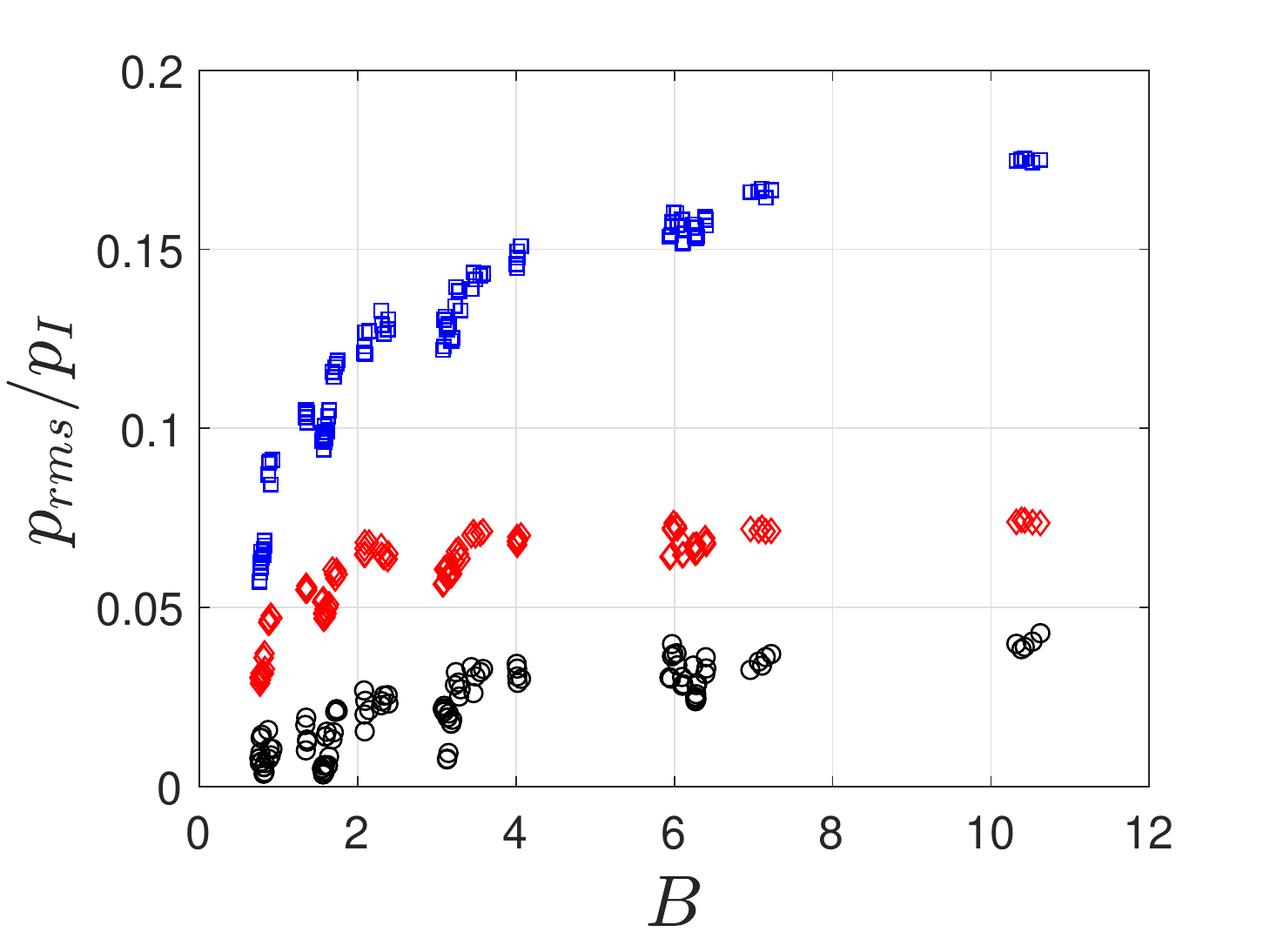}}\\
  \subfloat[]{\includegraphics[width=70mm,trim=0 0 0 0, clip]{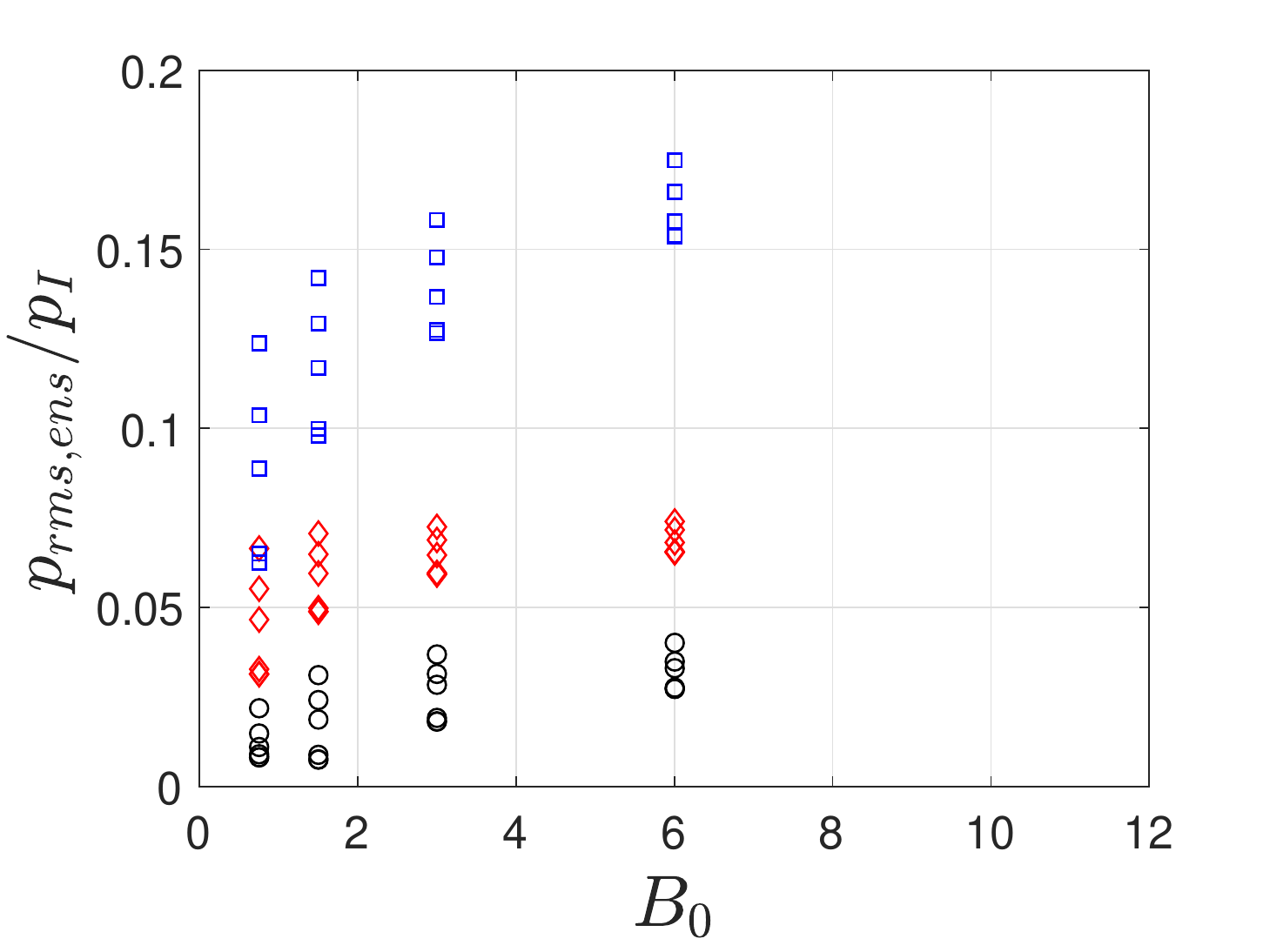}}
  \subfloat[]{\includegraphics[width=70mm,trim=0 0 0 0, clip]{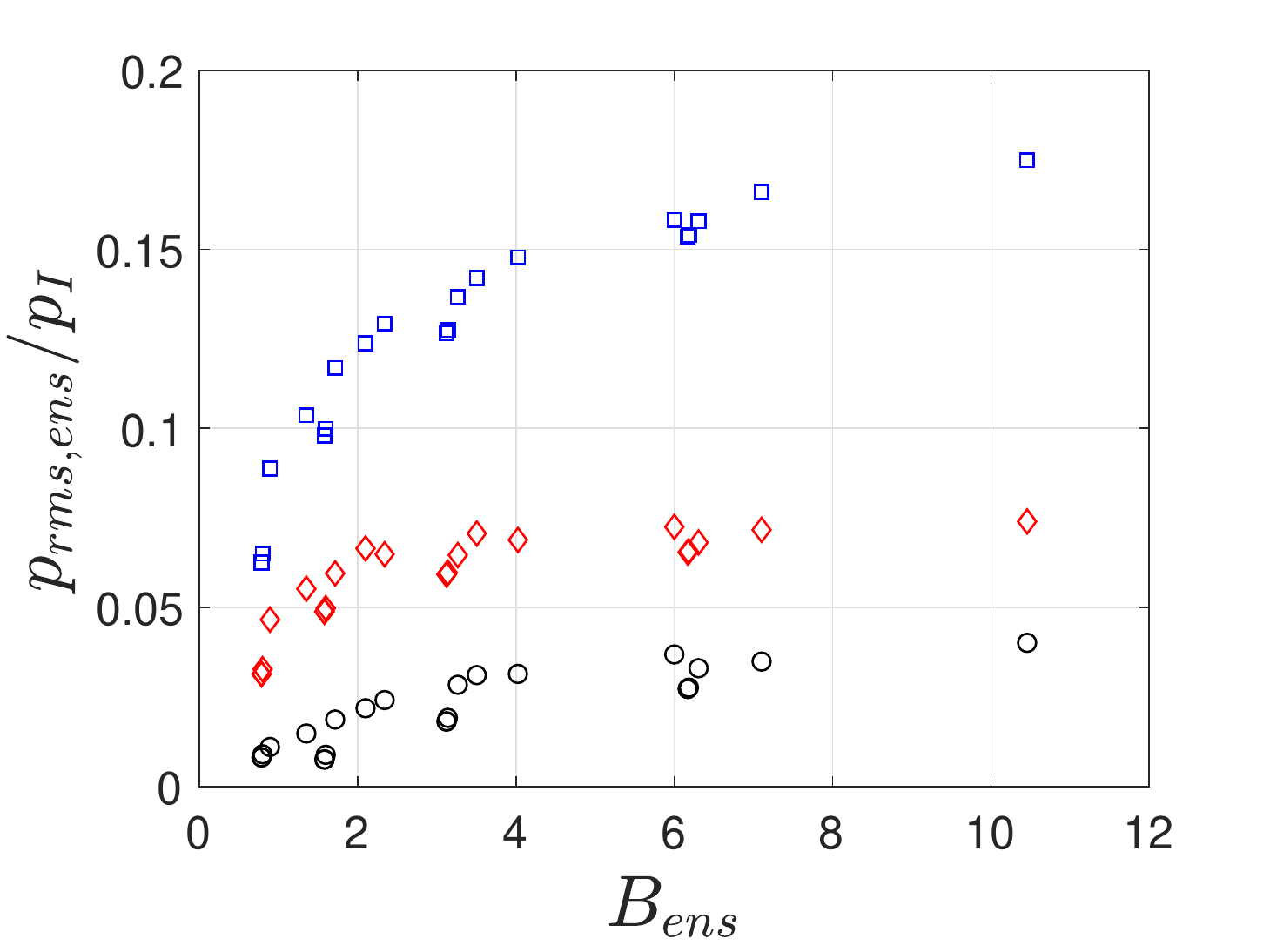}}
  \caption{Scatter plots of the root-mean-square pressure against $B_0$ and $B$. The top row show all realizations while the bottom row shows the ensemble-averaged values.
   }
   \label{fig:B_prms} 
\end{figure}
Figure \ref{fig:B_prms} is the analog to figure \ref{fig:B_xkc} but with the root-mean-square pressure plotted versus the original and dynamic cloud interaction parameters.  Shown by the different colors are the 3 scattering angles considered in figure \ref{fig:t_p}.
The scattered pressure shows positive correlations with both $B_0$ and $B$.
The data points are widely spread against $B_0$, but collapse better with $B$.
As was the case with the kinetic energy moment, ensemble averaging of the clouds remove additional scatter associated with the randomized bubble positions and distribution.
Overall, the results confirm that the proposed interaction parameter scales the amplitude of the bubble-scattered acoustics better than the original parameter.

The polar plots shown in figure \ref{fig:pres_contour} may help explain the saturation of both the moments of kinetic energy and the amplitude of bubble-scattered acoustics with a large value of the dynamics interaction parameter.
Due to the large mismatch in the acoustic impedance across the air-water interface, a cavity can scatter the most portion of the incident wave energy.
The dense cloud gives a similar magnitude of scattering as a single large bubble.
A subsequent increase in either the excitation amplitude or cloud volume fraction (thus increasing $B$), yields no further effect; the scattered acoustics saturate at a level similar to a single bubble of the cloud dimension.
The smaller directionality of the scattered acoustics by the cloud than by the air cavity is associated with the spatially random distributions of bubbles.
In multiple scattering theory, scatters with a random, disordered distribution may act as a rough surface and result in randomized angles of scattering of the incoming wave, compared to a smooth surface like that of the air cavity \citep{Ishimaru78}.

Overall, it has been shown that the dynamic interaction parameter scales both the amplitude of the scattered acoustic field, as well as the moment of kinetic energy. Furthermore, this indicates direct correlations between the far-field acoustics and the moment.
\begin{figure}
  \center
  \subfloat[]{\includegraphics[width=70mm,trim=0 0 0 0, clip]{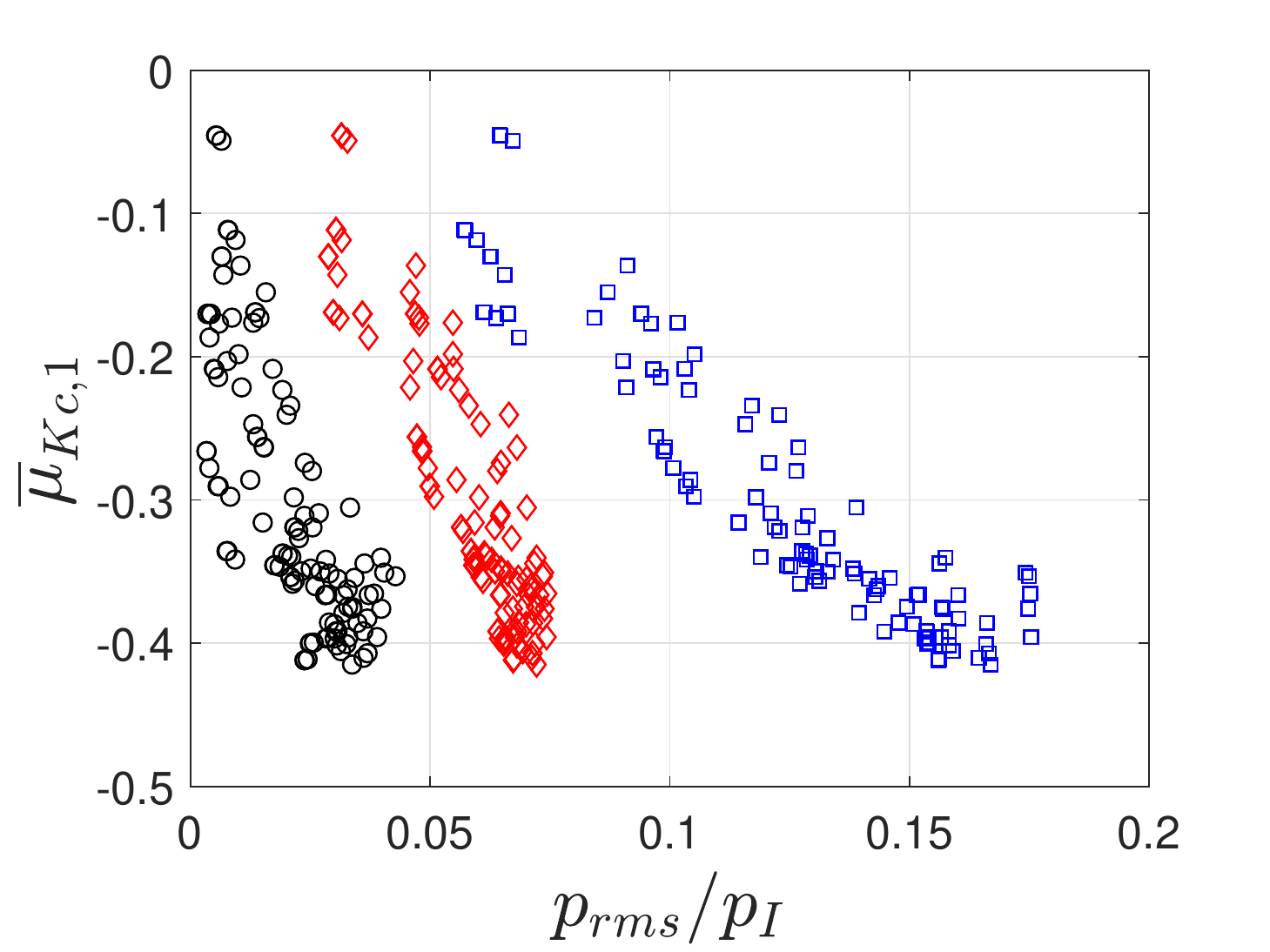}}
  \subfloat[]{\includegraphics[width=70mm,trim=0 0 0 0, clip]{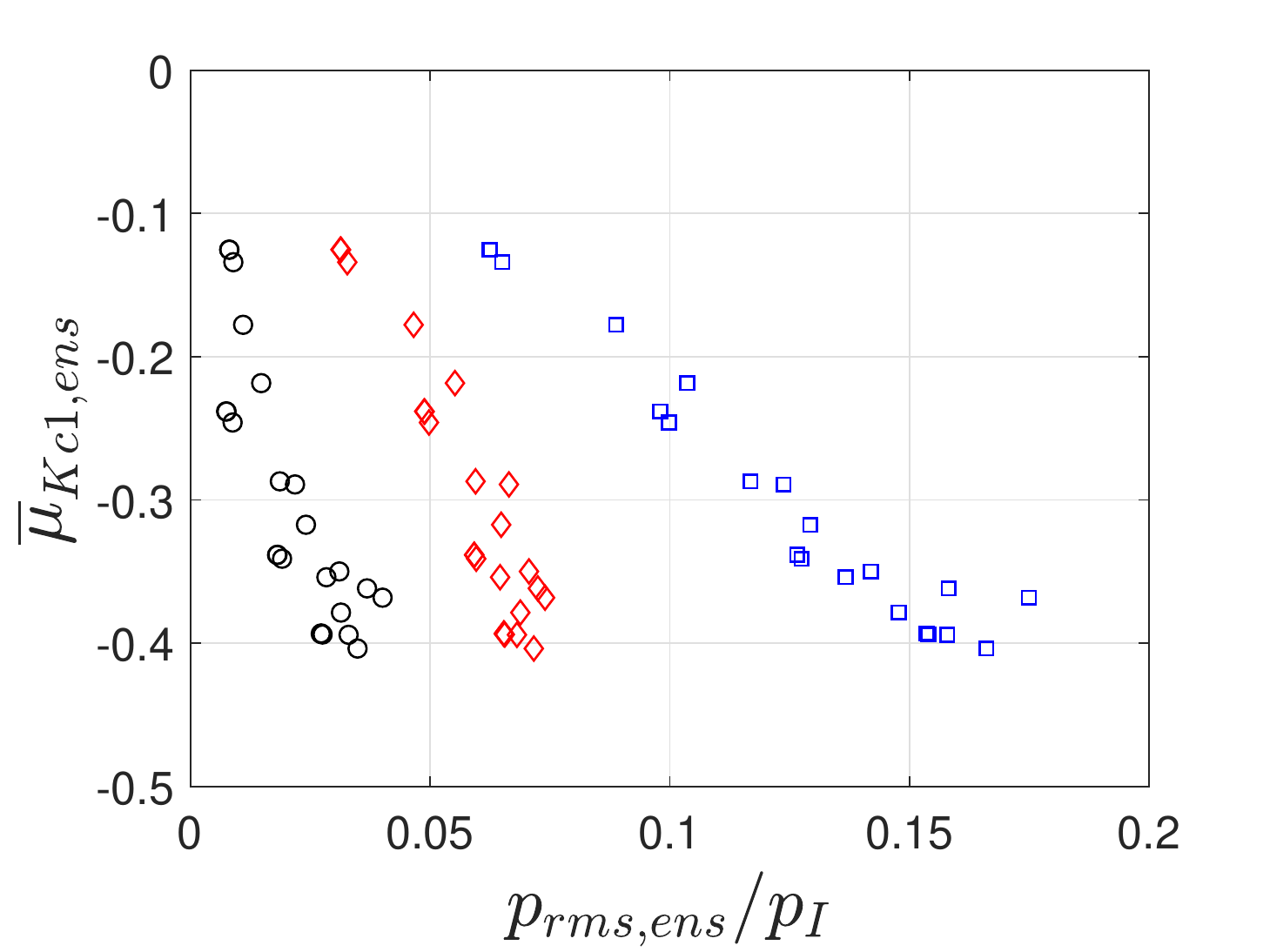}}
  \caption{Scatter plots of the moments of kinetic energy against $p_{rms}$ for (a) all clouds and (b) ensemble averaged clouds.}
   \label{fig:p_xk} 
\end{figure}
Figure \ref{fig:p_xk} (a) and (b) show data for all clouds considered at the 3 observer angles, with and without ensemble averaging, resepectively. 
For applications, the result indicates that the measurement of the far-field, bubble-scattered pressure waves can be used as a surrogate for the magnitude of the energy localization in the bubble cloud as well as a means to estimate the value of $B$.

\section{Implications for cavitation in lithotripsy}
\label{section:lith}
As the central application and motivation of the present study, it is worth discussing implications of the present results of numerical experiments to HIFU-based lithotripsy.

In ESWL, bubbles in a cloud experience a nearly identical amplitude of pressure excitation during the passage of the tensile component of the wave since the tensile tail typically has a much larger width than the cloud size.
The dynamics of bubble cloud consist of spherically symmetric structures, similar to what was shown in figure \ref{fig:snap_f50}.
The inward-propagating shockwave causes violent cloud collapse, that results in erosion of surrounding materials.
This injurious effect has been seen as a major disadvantage of ESWL \citep{McAteer05,Bailey06}.

The use of HIFU for lithotripsy has been proposed as an alternative to ESWL due to their potentials for safer and more efficient stone comminution \citep{Ikeda06,Yoshizawa09,Maxwell15}.
Cavitation bubble clouds in the HIFU-based lithoripsy, by contrast to those in ESWL, have a cloud size commensurate with the ultrasound wavelength.
The resulting energy localization of the cloud and scattering of the incoming waves identified in the previous sections indicate that the bubble clouds with a size at an order of the incident pressure wavelength can result in strong scattering of the incident wave, with strong implications for HIFU-based lithotripsy.

\begin{figure}
  \center
  \subfloat[]{\includegraphics[width=90mm,trim=20 80 0 85, clip]{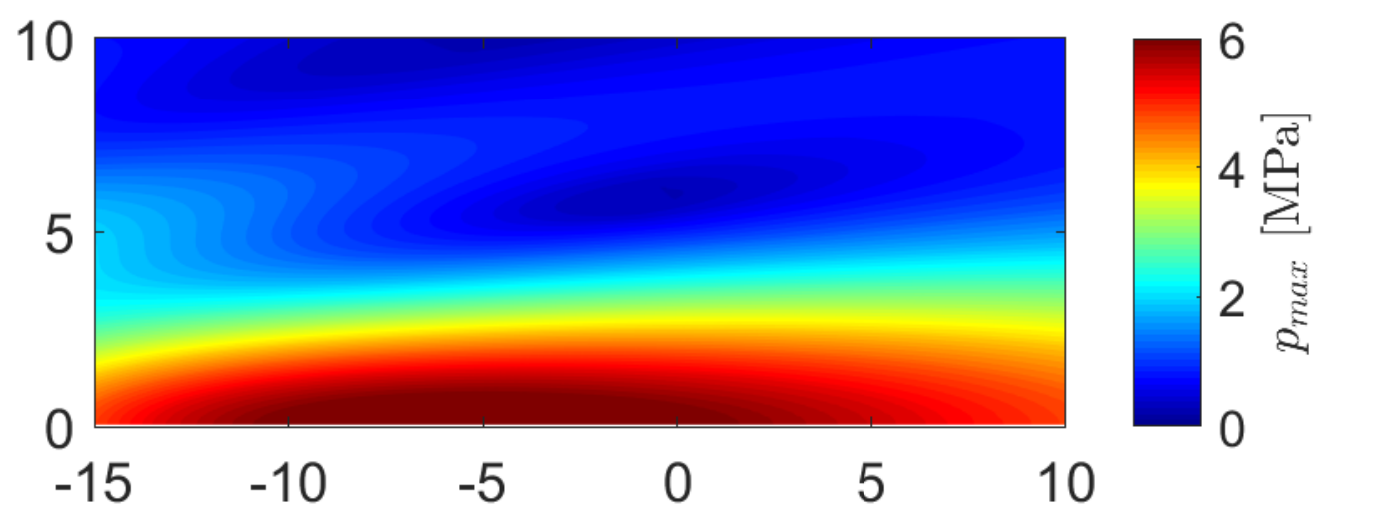}}\\
  \subfloat[]{\includegraphics[width=90mm,trim=20 80 0 85, clip]{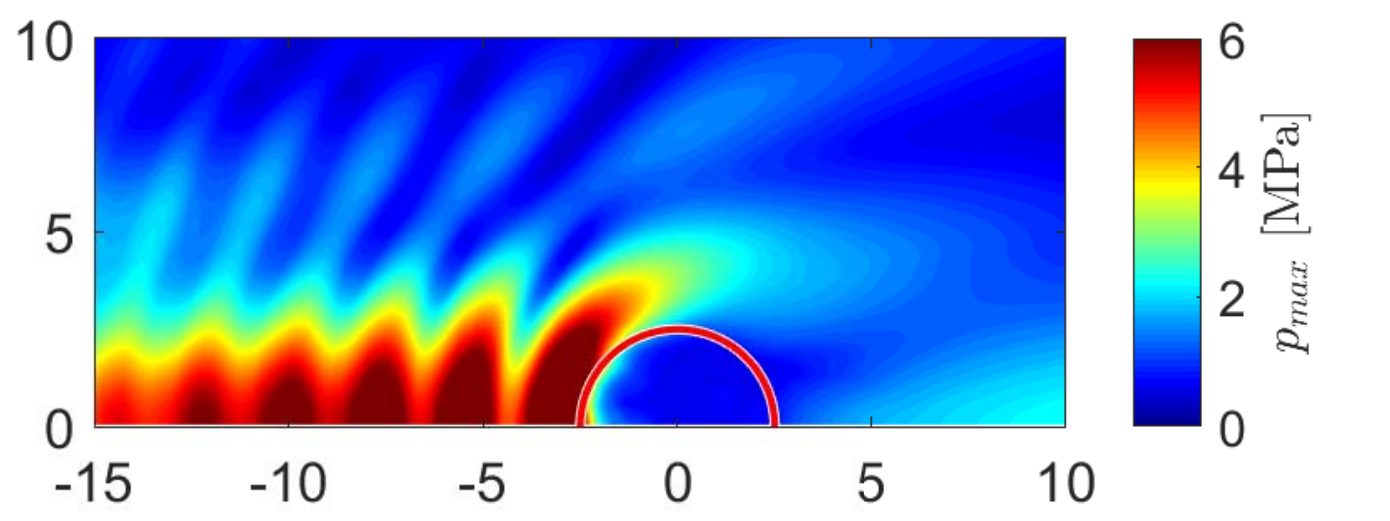}}
  \caption{Contours of the maximum pressure over the course of the simulations (a) without and (b) with the bubble cloud from run F8. The length unit is mm.
   }
   \label{fig:cont_max} 
\end{figure}
Figure \ref{fig:cont_max} compares contours of the maximum pressure on the cross plane over the course of the simulations with and without the bubble cloud.
Without the bubble cloud, the region of high-pressure ($>$5 MPa) is localized to the focal region of $x\in[-10,5]$ mm, while with the bubble cloud, the region of high pressure does not penetrate into the cloud except near the proximal surface.
This can be interpreted that the energetic proximal bubbles scatter the incoming wave to prevent the wave from penetrating into the cloud and suppress excitation and oscillation of the distal bubbles. There exists {\it{energy shielding}}.
The small values of the maximum pressure within the cloud shown in figure \ref{fig:cont_max} (b) also indicate that strong cloud collapse does not occur during or after passage of the wave.
This agrees with the absence of strong acoustic signals from bubble clouds after the passage of an incident plane wave confirmed in the numerical experiment, shown in figure \ref{fig:P_B}.
The results are reminiscent of a bubble screen\citep{Carstensen47,Commander89} that provides a similar shielding effect.

In practical conditions of HIFU-based lithotripsy, cavitation bubble clouds can be nucleated on the surface of a kidney stone. It can be conjectured that such bubble clouds may have both positive and negative effects on outcomes of the therapy; they can be less injurious due to the absence of violent cloud collapse, but they could reduce lower the efficacy of stone comminution by scattering the incident radiation.
Meanwhile, it is apparent that a presence of kidney stones may complicate the resulting bubble cloud dynamics. For instance, non-spherical bubble collapse may occur on the surface of a stone to cause erosion \citep{Tomita86,Johnsen09}, an effect not considered in the present study. For future research, simulations of bubble cloud dynamics in the presence of a stone are desirable.

\section{Conclusion}
\label{section:conc}
We investigated the dynamics of cavitation bubble clouds excited by strong ultrasound waves in a regime where the cloud size is similar to the ultrasound wavelength.
In a first set of simulations, we excite bubble clouds by a focused ultrasound wave to mimic the laboratory setup of HIFU-based lithtoripsy.
An anisotropic cloud structure was observed in both experiments and simulations.
The proximal bubbles grow to larger radius than the distal bubbles.
In a second series of simulations, we elucidated the underlying mechanisms leading to the anisotropy of the observed structure and dynamics. In these simulations, we varied the amplitude of (plane-wave) excitation, the number density of bubbles, and we considered an ensemble of five runs for each case with different locations and populations of bubbles.
Based on the kinetic energy of liquid induced by oscillations of a bubble cluster, we proposed a new scaling parameter, namely a dynamic cloud interaction parameter, that scales the observed anisotropy and dynamics.
The parameter is generalized from the cloud interaction parameter introduced by \citet{dAgostino89} for linearized bubble cloud dynamics in the long wavelength regime.
We likewise showed that the scattered acoustic field collapses with the same dynamic interaction parameter, and thus can serve as a surrogate measure for the extent of energy localization in the cloud.
This correlation may be of use to diagnose {\it in situ}, via acoustic monitoring, the state of cavitation during ultrasound therapy.

\section*{Acknowledgements}
The authors thank Adam Maxwell, Wayne Kreider and Michael Bailey for their support in the companion experiments. K.M acknowledges the Funai Foundation for Information Technology, for the Overseas Scholarship. This work was supported by the National Institutes of Health under grant P01-DK043881 and ONR Grant N00014-17-1-2676. The computations presented here utilized the Extreme Science and Engineering Discovery Environment, which is supported by the National Science Foundation grant number CTS120005.

\appendix
\section{Experimental setup}
\begin{figure}
  \center
  \subfloat[]{\includegraphics[width=75mm,trim=0 0 0 0, clip]{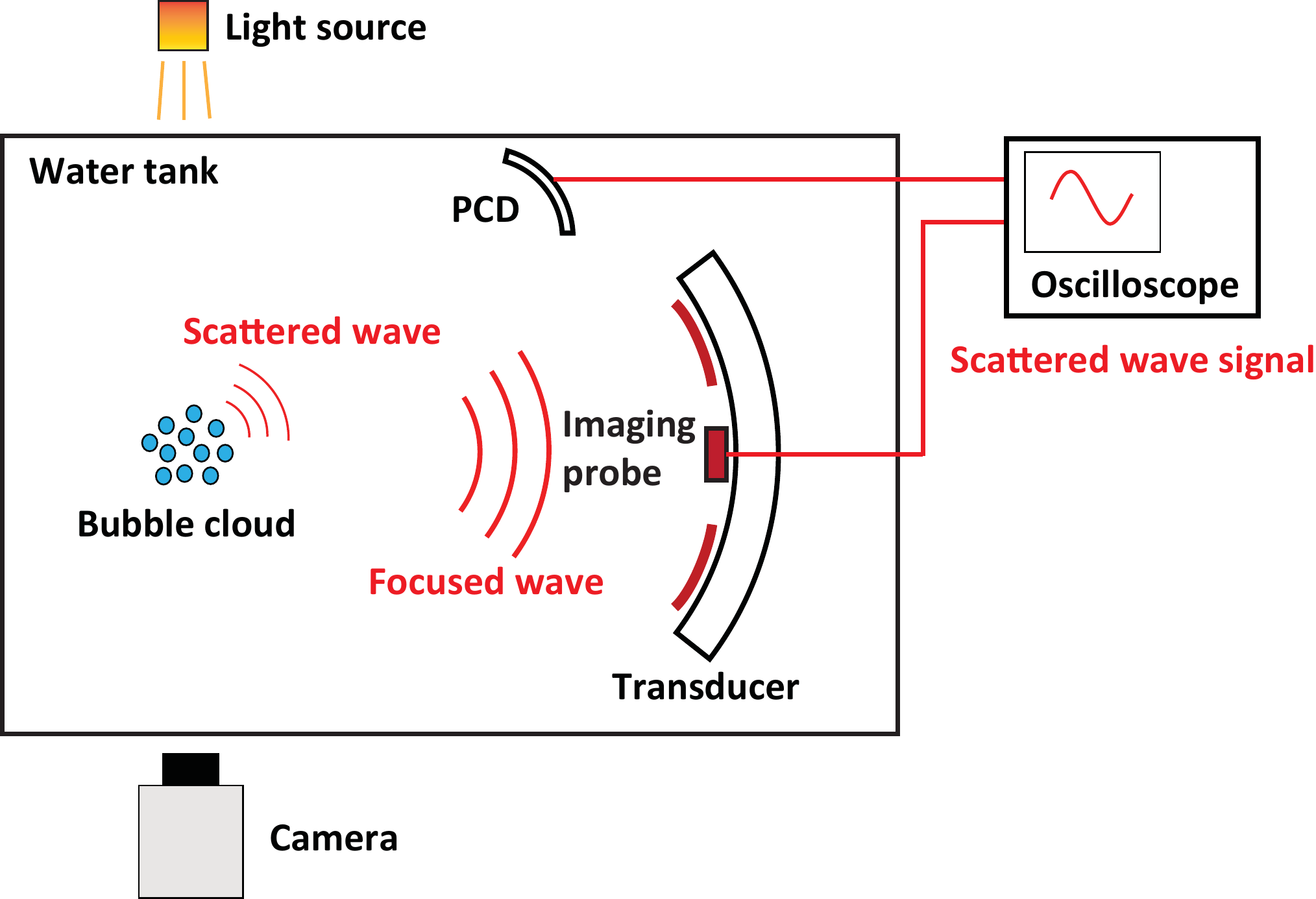}}
  \subfloat[]{\includegraphics[width=45mm,trim=-200 -400 -200 -70, clip]{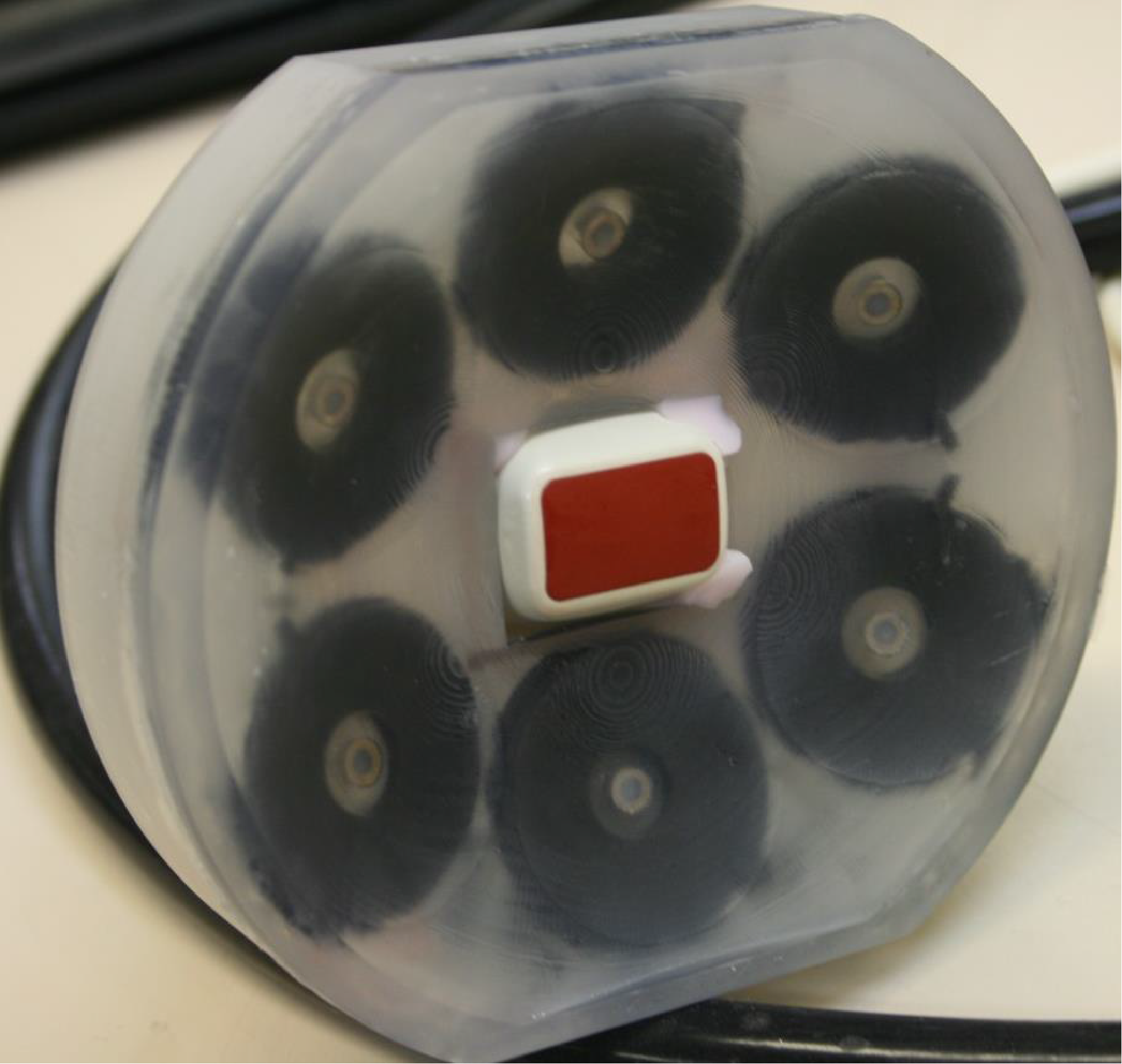}}
  \caption{(a) Schematic of the experimental setup. (b) Piezo-ceramic medical transducer used in the experiment.
   }
   \label{fig:exp} 
\end{figure}
In this section we describe the experimental setup used to obtain images of bubble cloud shown in figure \ref{fig:imacon}.
Figure \ref{fig:exp} (a) shows the schematic of the setup.
The setup is designed to capture the evolution of a single, isolated cavitation bubble clouds excited in a focused, traveling ultrasound wave.
The temperature and pressure are ambient.
The water is degassed by a vacuum pump to realize the oxygen level of 75\%.
A medical transducer composed of six piezo-ceramic array elements (figure \ref{fig:exp} (b)) is immersed in water.
The transducer has an aperture of $110\times104$ mm, and a focal length of 120 mm.
An imaging probe is attached at the center of the transducer.
We excite burst waves at the transducer with a pulse-repetition-frequency (PRF) of 200 Hz.
A high-speed camera captures a rectangular region with a dimension of $15.3\times12.5$ mm around the focal point of the transducer.
The camera captures 14 consecutive frames with a frame rate of 6 $\mu$s and an exposure time of 50 ns, with a resolution of $1200\times980$ pixels.
A focused passive cavitation detector (PCD) with Polyvinylidene difluoride (PVDF) membrane, with ROC of 150 mm and an aperture of 50mm, is positioned confocal to the transducer. We use acoustic signals captured by the imaging probe and the PCD to map the location of cavitation site to confirm that the bubble cloud captured by the camera is isolated and no other cloud is present outside the window of the camera.
Note that all the high-speed images presented in this paper are vertically reflected for consistency with the simulations.

With the input of $N_c$ cycles of a sinusoidal voltage, the output of the transducer is modeled by the following formula:
\begin{equation}
p_{trans} = p_a\mathrm{cos}(2\pi ft)[(1-e^{-t/\tau_u})-(1-e^{-(t-N_c/f)/\tau_d})H(t-\frac{N_c}{f})],
\end{equation}
where $\tau_u$ and $\tau_d$ are the ring-up and ring-down time, respectively.
In the simulations of focused waves, we excite this expression of the pressure at the source plane, with $\tau_u=4.0$ and $\tau_u=8.0$ $\mu$s.
The validity of acoustic source model is confirmed by comparing the focal pressure with experimental measurement as shown in figure \ref{fig:waveform}.

\section{Local cloud interaction parameter}
In this appendix we provide a rationale for defining the dynamic cloud interaction parameter.
To treat a bubble cloud in that bubbles experiences non-uniform forcing pressure due to the short wavelength, we introduce the notion of {\it{local}} kinetic energy.
The local energy is defined as the kinetic energy of liquid induced by bubbles in a spherical region around a coordinate $\vector{x}$ with a radius of $R_{c,L}$ that experiences approximately uniform pressure excitation, namely local-cloud:
\begin{align}
K_{Local}(\vector{x})
&= 2\pi\rho\left[\sum^{N_{b,L}}_iR_i^3\dot{R}_i^2+\sum^{N_{b,L}}_i\sum^{N_{b,L}}_j\frac{R_i^2R_j^2\dot{R}_i\dot{R}_j}{r_{i,j}}+O\left(\frac{R^7\dot{R}^2}{r^4}\right)\right]\\
&\sim
N_{b,L}R_L^3(\vector{x})\dot{R}_L^2(\vector{x})(1+\frac{N_{b,L}R_L}{R_{c,L}}).\label{eqn:Ksub}
\end{align}
As discussed in $\S$3, with a strong pressure excitation, the mean bubble radius can largely deviate from its initial value.
In case of periodic pressure excitation, a natural choice of $R_{L}$ can be its time averaged value during the course of excitation:
\begin{equation}
R_{L}\approx\overline{R}_{b,L}.
\end{equation}
This leads us to define the following {\it{local}} cloud interaction parameter:
\begin{equation}
B_{L}(\vector{x})\approx\frac{N_{b,L}\overline{R}_{b,L}(\vector{x})}{R_{c,L}}.
\end{equation}
The local interaction parameter characterizes the kinetic energy of the {\it{local}} cloud.
We take a summation of this parameter over all the local clouds:
\begin{equation}
\sum^{N_c}_{i=1}B_L(\vector{x})
=
N_c<B_L>
=
\frac{N_{b}<\overline{R}_{b}>}{R_{c,L}},
\end{equation}
where $<\cdot>$ denotes the spatial average over the global bubble cloud.
By multiplying a factor $R_{c,L}/R_c$, we obtain
\begin{equation}
\frac{N_cR_{c,L}}{R_c}<B_L>
=
B.
\end{equation}
The relation indicates that if the spatial distribution of $B_L$ in distinct clouds are identical, the clouds also possess the same $B$ and show a similar dynamic response.
Conversely, if distinct clouds possess the same value of $B$, we could expect a similar spatial distribution of $B_L$, thus that of the local kinetic energy, though this is not a necessary condition since clouds with the same value of $B$ may have different distributions of $B_L$.

The scaling of the kinetic energy (\ref{eqn:Ksub}) implies that the bubble cloud with larger values of $B$ tends to induce a larger amount of kinetic energy in the liquid, with the same values of $(R_L,\dot{R}_L)$.
Conversely, bubble clouds with a larger value of $B$ need smaller values of $(R_L,\dot{R}_L)$ to induce the same amount of kinetic energy. 
This qualitatively implies that, with the same amplitude of pressure excitation, bubbles in a bubble cloud with a larger $B$ (or $B_0$) tend to grow less, compared to a single, isolated bubble and bubble clouds with smaller values of $B$.
The suppression of bubble/bubble cloud growth with a large value of $B$, thus inter-bubble interactions, has been observed in numerical simulation by \citet{Wang99} and in experiment by \citet{Bremond06}, and also qualitatively agrees with the results of the present simulations, in that bubble clouds present smaller differences in the moments, compared to differences in their values of $B_0$ (figs. \ref{fig:focus_moment}, \ref{fig:xv_cent} and \ref{fig:xk_cent}).

\bibliography{jfm-references}
\bibliographystyle{jfm}

\end{document}